\documentclass[prb,superscriptaddress,twocolumn,showpacs,floatfix,amsmath,amssymb,longbibliography]{revtex4-2}
\usepackage{xcolor}
\usepackage[utf8]{inputenc}
\usepackage{amsmath}
\usepackage{amssymb}
\usepackage{mathtools}
\usepackage{siunitx}
\DeclareSIUnit\bar{bar}
\DeclareSIUnit\ryd{Ry}
\usepackage[version=4]{mhchem}
\usepackage{graphicx}
\usepackage{booktabs}
\usepackage{svg}
\usepackage{gensymb}
\usepackage{graphics}
\usepackage[T1]{fontenc}
\usepackage{soul}
\usepackage{hyperref}
\usepackage[margin=1in]{geometry}
\usepackage{graphicx}
\usepackage{subcaption}
\usepackage{adjustbox}
\usepackage{braket}
\begin{document}

\title{\textit{Ab Initio} Spinor Kadanoff-Baym Approach to Nonequilibrium Electron, Phonon and Magnon Dynamics in Itinerant Ferromagnets}
\author{Giovanni Marini}
\affiliation{Department of Physics, University of Trento, Via Sommarive 14, 38123 Povo, Italy}
\begin{abstract}
This work introduces a theoretical framework based on the Kadanoff-Baym equations in spinor space to study ultrafast magnetization dynamics in itinerant ferromagnetic systems from first principles. By incorporating spin-orbit coupling into the \textit{ab initio} Hamiltonian and generalizing the self-energies to include terms beyond the charge sector, I derive scattering integrals within the Markov approximation and quasiparticle renormalizations for electrons and phonons in the presence of spin-dependent effective interactions within a many-body perturbation theory approach. I explicitly discuss how magnons emerge in this framework, and derive, through suitable approximations, a closed and tractable set of equations for coupled electron, phonon and magnon dynamics that can be solved from first principles. I discuss the equations of motion within a minimal ferromagnetic model, where the coupled equations are solved numerically. This approach allows one to treat coherent and incoherent magnetic dynamics on an equal footing, paving the way for a first principles understanding of ultrafast magnetic dynamics and demagnetization, and opening to a truly predictive theory of femtomagnetism in periodic systems.
\end{abstract}

\maketitle

\section{Introduction}

Femtosecond laser pulses can induce magnetization changes hundreds of times faster than any conventional magnetic switching technique~\cite{RevModPhys.82.2731}, offering a pathway toward next-generation magnetic devices with unprecedented speed and energy efficiency~\cite{Kimel2019,Stanciu2007,10.1126/science.1253493}. The landmark observation of sub-picosecond demagnetization by Beaurepaire and coworkers~\cite{Beaurepaire1996} established that all-optical control of magnetic states is possible at the femtosecond timescale, opening a rich research direction that has since led to the discovery of numerous intriguing phenomena: all-optical switching~\cite{Stanciu2007,Graves2013,Mangin2014,Ignatyeva2019,Xu2019,Kimel2019}, ultrafast magnetization reversal, light-induced magnetic phase transitions~\cite{PhysRevLett.93.197403,Radu2011}, light-enhanced magnetism~\cite{PhysRevLett.98.217401,PhysRevLett.125.267205,Lu2024}, optically induced intersite spin transfer~\cite{doi:10.1021/acs.nanolett.7b05118,Siegrist2019,doi:10.1126/sciadv.aay8717,doi:10.1126/sciadv.adn4613}, light-induced magnetization of intrinsically non-magnetic materials~\cite{Cheng2020,PhysRevB.86.100405,Hennecke2024,Basini2024}. 

Despite these advances, a comprehensive theoretical understanding of ultrafast magnetic dynamics remains elusive, owing to the complex interplay of multiple microscopic mechanisms. The fundamental role of spin-orbit coupling (SOC) in demagnetization was recognized early~\cite{Zhang2000}, yet after three decades no consensus has been reached on the relative importance of the concomitant processes involved, including SOC-mediated transfer of angular momentum to the orbital degrees of freedom~\cite{Krieger2015,PhysRevB.95.024412,Krieger_2017} (subsequently dissipated via phonon scattering~\cite{Tauchert2022}), superdiffusive spin transport~\cite{Battiato2010,Balaz2023}, spin-mediated electron-phonon coupling~\cite{Koopmans2005,Koopmans2010,PhysRevB.102.054442}, and hot magnon generation~\cite{https://doi.org/10.1002/apxr.202300103,7vr4-57mk}. While each of these mechanisms captures certain aspects of ultrafast demagnetization, a truly unified theoretical framework is still missing. 

Other theoretical efforts have focused on understanding how ultrafast laser excitation can induce magnetic order in non-magnetic systems: for example, a magnetic instability has been predicted in the V$_2$O$_5$ semiconducting compound following above-gap photoexcitation~\cite{PhysRevB.105.L220406} employing constrained density-functional theory, a quasi-equilibrium approach capable of identifying laser-induced instabilities after above-gap pumping in semiconductors~\cite{Tangney1999,Tangney2002,Murray2007,Marini2021,Murray2015,Mocatti2023,Furci2024}, while a mechanism exploiting spin transfer from a ferromagnetic layer was proposed to induce ferromagnetism in MoSe$_2$ monolayers interfaced with MnSe$_2$~\cite{Junjie2022}. Further theoretical studies have shown that magnetic dynamics can also be initiated by off-resonant pumping through time-local explicit time-reversal symmetry breaking~\cite{Neufeld2023}, even with linearly polarized light, or through a carrier-envelope phase that explicitly breaks time-reversal symmetry~\cite{75tm-3t9b}. A large part of these works is grounded in real-time time-dependent density functional theory (TDDFT), and there is broad consensus that SOC plays a pivotal role in the coherent magnetic dynamics simulated within TDDFT~\cite{10.21468/SciPostPhys.20.2.048}. TDDFT and \textit{ab initio} (or \textit{ab initio} informed) molecular dynamics have  also been widely applied to ultrafast demagnetization~\cite{Krieger2015,PhysRevB.95.024412,Krieger_2017,Zhang2023}, Hubbard $U$ renormalization~\cite{Tancogne-Dejean2018}, photoinduced structural dynamics~\cite{Lian2020,Wen-Hao2022,Chen2018,Liu2022,Corradini2025}, and hot-carrier relaxation~\cite{Zheng2023,Lively2024}. It is important to note that TDDFT-based approaches generally suffer from a fundamental limitation: they lack intrinsic dephasing and relaxation mechanisms, which can however be in principle incorporated through Lindbladian extensions~\cite{TEMPEL2011130}. 

Nonequilibrium Green's function (NEGF) methods based on the Kadanoff-Baym ansatz~\cite{Lipavsky1986,Kadanoff2018} and their generalization to electron-boson systems~\cite{Karlsson2021,Stefanucci2023}, as well as density matrix formulations~\cite{Rossi2002,PhysRevB.90.125140}, have emerged as alternative powerful tools for studying non-equilibrium electronic, optical, and lattice dynamics~\cite{Karlsson2021,Pavlyukh2022,Pavlyukh_Perfetto2022,Perfetto_Pavlyukh2022,Perfetto2023}, also in the context of correlated systems~\cite{RevModPhys.86.779}. Often, Kadanoff-Baym equations and their electron-phonon generalization~\cite{Stefanucci2024} are solved within the Markov approximation~\cite{Schafer2002,Haug2009,Haug2010,Kira2011} to reduce their prohibitive scaling.  Within this class of methods, photocarrier-induced modifications to electronic, optical~\cite{Steinhoff2014,Schmidt2016,Erben2018} and lattice dynamics~\cite{Girotto_Novko2023}, as well as certain aspects of photoexcited carrier dynamics~\cite{Butscher2007,Malic2011,Breusing2011,Steinhoff2016}, have been analyzed, often in the semiclassical Boltzmann limit~\cite{Marini2013,Bernardi2014,Sangalli2015,Bernardi2016,Sadasivam2017,OMahony2019,Tong2021,Chen2022,Maliyov2024,Sjakste2025,Caruso2021,Caruso2022,Emeis2024}.  Recently, a first principles framework that simultaneously treats electron-electron, electron-phonon, and phonon-phonon interactions in real time has also been proposed in this context~\cite{Mocatti2026}.

The study of spin dynamics with NEGF methods is relatively less frequent and has regarded few selected problems. Spin Bloch equations have been employed to model spin dynamics in non-magnetic semiconductors~\cite{PhysRevB.61.2945} and Rashba SOC~\cite{WANG2018395}, while first principles spin relaxation calculations have only recently appeared in literature, either with the inclusion of electron-phonon and impurity scattering within a Lindbladian formalism~\cite{PhysRevB.104.184418} or through the explicit calculation of spin-spin correlation function~\cite{PhysRevLett.129.197201}. Conversely, spin dynamics has been extensively investigated theoretically within the Kadanoff-Baym formalism applied to non-equilibrium quantum field theory in the context of relativistic high-energy physics~\cite{PhysRevD.66.043502,Yang2020,PhysRevD.104.016029}.

Considering all of the above, extending the many-body NEGF formalism to treat non-equilibirium magnetic dynamics in crystalline materials would represent a desirable step forward, and give access to a theoretical framework where coherent and incoherent magnetization dynamics could be addressed on equal footing. To this end, this work presents a practical approach based on suitable approximations of the Kadanoff–Baym equations to describe magnetization dynamics in ferromagnetic crystals. Starting from the first principles noncollinear Hamiltonian, I derive the Kadanoff-Baym equations for electrons in spinor space and identify practical approximations for the electron self-energy, discussing the role of SOC and their coupling to phonons and other microscopic mechanisms relevant to magnetization dynamics. After explicitly showing how magnons emerge from transverse contributions to the electron self-energy, I outline a strategy to close the dynamics and obtain a numerically treatable set of equations for electrons,phonons and magnons within the Markov approximation. I identify and discuss the most important contributions to scattering integrals and quasiparticle renormalization, including electron-electron, electron-phonon, and electron-magnon channels, and outline a practical route for future numerical implementations. I then study the numerical solution of the resulting set of equations in a simple one-dimensional ferromagnetic model, and qualitatively discuss the most notable findings. The present framework provides a unified description of coherent and incoherent magnetic dynamics, opening to a rigorous and comprehensive treatment of magnetization dynamics and ultrafast demagnetization from first principles.\\

\section{The $ab~initio$ Hamiltonian}

While most aspects of the formalism discussed below are general, I will at times explicitly refer to the prototypical case of a photoexcited metallic system with a pre-existing ferromagnetic state, where ultrafast demagnetization was first observed. Rydberg atomic units are employed across the manuscript, i.e. $\hbar=2m_e=e^2/2=4\pi\epsilon_0=1$, unless explicitly stated. I'll work in the framework of out-of-equilibrium many-body theory, and apply the Kadanoff-Baym equations formalism to the electron-phonon problem~\cite{Stefanucci2023} in the noncollinear magnetic case. It is important to remark that the inclusion of noncollinearity in the formalism, SOC in particular, is strictly necessary to have a qualitatively correct discussion of demagnetization processes, even for collinear systems; this has already been discussed extensively elsewhere~\cite{Krieger2015,PhysRevB.95.024412,Krieger_2017,Elliott2018,Zhang2000} and will become even more evident later on in this work. Thus, a fully noncollinear treatment is necessary even in a collinear system. Due to the nontrivial internal spin structure of the Hamiltonian in the presence of noncollinear terms, magnetic systems can host a very complex out-of-equilibrium spin dynamics. To understand why, I start from the $ab~initio$ relativistic Hamiltonian for electrons and nuclei: 

\begin{equation}
    \hat{H} = \hat{H}^{0,e} + \hat{H}_{0,ph}+\hat{H}_{e-e}+\hat{H}_{e-ph} \label{H_fp}
\end{equation}

Each term is defined in analogy to Ref.~\cite{Stefanucci2024}: $\hat{H}_{0,ph}$ is the bare phonon Hamiltonian, $\hat{H}_{e-e}$ is the electron-electron Hamiltonian and $\hat{H}_{e-p}$ is the electron-phonon Hamiltonian, and they are all defined as in Ref.~\cite{Stefanucci2024}; here though, the electronic creation and annihilation operators in Bloch space will be labeled as $\hat{c}^\dagger_{\mathbf{k}\mu}$ (or $\hat{c}^\dagger_{\mathbf{k}n\alpha}$), and $\hat{c}_{\mathbf{k}\mu}$ (or  and $\hat{c}_{\mathbf{k}n\alpha}$), where the index $\mu$ is a cumulative band and spin index $\mu = \{n,\alpha\}$, and $\alpha$ can assume either spin-$1/2$ value, $\{\uparrow,\downarrow\}$. I additionally include relativistic terms in the single particle electron Hamiltonian $\hat{H}^{0,e}$ (see Appendix A for details and notation), at variance with previous works on the $ab~initio$ dynamics of electrons and nuclei~\cite{Stefanucci2023,Stefanucci2024,Mocatti2026}, where they were neglected in the study of nonmagnetic dynamics. 

Among the relativistic terms, SOC plays a crucial role in magnetic dynamics. When the SOC contribution, $\hat{H}_{SO}\propto \hat{\sigma}^x \hat{L}_x+\hat{\sigma}^y \hat{L}_y+\hat{\sigma}^z \hat{L}_z$, is included in the Hamiltonian, two qualitative differences arise with respect to the collinear case. First, no spin projection is a good quantum number anymore since the Hamiltonian does not commute with any Pauli matrix, $[\hat{H}_{SO},\hat{\sigma}^{x,y,z}]\neq0$, as one can infer from the commutation rules of the Pauli matrices. This implies that  spin evolution can be nontrivial, even in the absence of magnetism. Second, the Hamiltonian acquires a more complex spin structure, possessing $\hat{\sigma}^{x,y,z}$ components.

Let me now consider the concrete case of a non-equilibrium magnetic dynamics initiated by ultrafast laser light. This is described in the present formalism by an explicitly time-dependent light-matter interaction term $\hat{H}_{drive}(t)$, which is added to $\hat{H}$, and which I assume to couple only to electronic degrees of freedom. Its form follows from the minimal-coupling substitution, and in the Coulomb gauge and within the dipole approximation, is written in second quantization as:
\begin{equation}
\label{eq:electronic_drive_hamiltonian}
    \hat{H}_{drive}(t) = \sum_{\mathclap{\mathbf{k},\mu,\mu'}} \bar{\mathcal{F}}_{\mathbf{k}\mu\mu'}(t) \hat{c}^\dagger_{\mathbf{k}\mu} \hat{c}_{\mathbf{k}\mu'}
\end{equation} 
where I introduce the bare Rabi frequency matrix
\begin{equation}
    \bar{\mathcal{F}}_{\mathbf{k}\mu\mu'}(t) = \sqrt{2}\mathbf{E}(t) \cdot \mathcal{D}_{\mathbf{k}\mu\mu'}
\end{equation}

and the electric-dipole matrix elements $\mathcal{D}_{\mathbf{k}\mu\mu'}$:

\begin{equation}
    \mathcal{D}_{\mathbf{k}\mu\mu'} = i \langle u_{\mathbf{k}\mu} |\nabla_{\mathbf{k}} | u_{\mathbf{k}\mu'} \rangle_{\Omega}\label{dipole}
\end{equation}
where the matrix element is evaluated over the unit cell $\Omega$ between the  periodic parts $u_{\mathbf{k}\mu}$ of the Bloch states $\psi_{\mathbf{k}\mu} = e^{i\mathbf{k}\cdot\mathbf{r}} u_{\mathbf{k}\mu}$, and $i\nabla_{\mathbf{k}}$ 
is the gradient representation of the position operator acting on Bloch states in a periodic solid.
This operator has a purely ``charge'' character, being proportional to $\hat{\sigma}^0$ in spinor space. The coupling between matter and the magnetic field component of the laser is assumed to be negligible. The presence of the explicitly time-dependent $\hat{H}_{drive}(t)$ brings the system out of equilibrium, and the out-of-equilibrium dynamics will be investigated here within the Kadanoff-Baym formalism. \\

\section{Kadanoff-Baym equations in spinor space}

Having established the Hamiltonian, I now focus on the system's real-time dynamics. The starting point is the electronic Green's function in real space. Due to the presence of SOC, the electronic Green's function and self-energy generally acquire non-zero components over all four $\hat{\sigma}^I$ channels. The time-ordered electronic Green's function $\bf{G}$ is a $2\times2$ matrix in spinor space:

\begin{equation}
    \mathbf{G}(1,2) = \begin{pmatrix} G_{\uparrow \uparrow}(1,2) & G_{\uparrow \downarrow}(1,2) \\ G_{\downarrow \uparrow}(1,2) & G_{\downarrow \downarrow}(1,2) \end{pmatrix}
\end{equation}

where $\uparrow$ and $\downarrow$ represent the two spin components with respect to a chosen quantization axis, conventionally $\hat{z}$ (for a ferromagnetic system, the magnetization axis). I employ the shorthand coordinate notation $1=\{\mathbf{r}_1,z_1\}$ to indicate a generic space-time point at position $\mathbf{r}_1$ and complex time $z_1$ lying on the Schwinger-Keldysh contour. The components of $\mathbf{G}$ are defined as:

\begin{equation}
    G_{\alpha \beta}(1,2) = -i \langle T\hat{\Psi}_\alpha(1)\hat{\Psi}^\dagger_\beta(2)\rangle \label{prop}
\end{equation}

where $\hat{\Psi}_\alpha(1)$ represents the electronic field operator  that annihilates an electron with spin $\alpha$ at $1=\{\mathbf{r}_1,t_1\}$ and $T$ is the time-order operator on the Schwinger-Keldysh contour. Since in this paper I consider lattice-periodic systems, it is useful to keep in mind the expansion of electronic field operators in the complete orthonormal Bloch basis :

\begin{equation}
   \hat{\Psi}_{\alpha}(\mathbf{r}_1) = \sum_{\mathbf{k}n} \psi_{\mathbf{k}n}^{\alpha}(\mathbf{r}_1) \hat{c}_{\mathbf{k}n\alpha} \label{bloch_fields}
\end{equation}

where $\psi_{\mathbf{k}n}^{\alpha}$ is the Bloch wavefunction of the single particle Bloch state labeled by $\{\mathbf{k},n,\alpha\}$.

 A formal derivation of the Kadanoff-Baym equations in spinor space is given in Appendix B. The result is an equation for the lesser propagator, Eq.~\ref{density_mat}, whose equal-time form is an equation of motion for the one-body electronic density matrix, defined as $\rho^<_{\alpha \beta}(t) = -i G_{\alpha \beta}^<(t,t)$. In periodic systems, Eq.~\ref{density_mat} can be recast in Bloch space, as a matrix in the band index space, as~\cite{Stefanucci2023,Stefanucci2024}: 

\vspace{2.5pt}\begin{widetext}
\begin{equation}
   i\partial_{t} \rho^\mathbf{<,k}_{\alpha \beta}(t)+[{h}^{qp}(\mathbf{k},t),\rho^\mathbf{<,k}(t)]_{\alpha \beta} = -~\int_0^t[\Sigma^>_{\mathbf{k},\alpha \gamma}(t,t')G^<_{\mathbf{k},\gamma \beta}(t',t)-\Sigma^<_{\mathbf{k},\alpha \gamma}(t,t')G^>_{\mathbf{k},\gamma \beta}(t',t)]  + h.c. \label{density_mat_rec}
\end{equation}
\end{widetext}

where I have defined the matrix $h^{nn'qp}_{\alpha\beta}(\mathbf{k},t) = \bra{\psi_{\mathbf{k}n}^\alpha}\hat{H}^{qp}(t)\ket{\psi_{\mathbf{k}n'}^\beta}  $ and the greater/lesser electronic Green's function matrix in Bloch space:

\begin{equation}
\begin{gathered}
    G^{<nm}_{\mathbf{k},\alpha \beta}(t_1,t_2) = i \langle \hat{c}^\dagger_{\mathbf{k}m\beta}(t_2)~\hat{c}_{\mathbf{k}n\alpha}(t_1)\rangle\\
        G^{>nm}_{\mathbf{k},\alpha \beta}(t_1,t_2) = -i \langle \hat{c}_{\mathbf{k}n\alpha}(t_1)~\hat{c}^\dagger_{\mathbf{k}m\beta}(t_2)\rangle
        \end{gathered}
\end{equation}

where $\hat{c}^\dagger_{\mathbf{k}n\alpha}(t_1)$ and $\hat{c}_{\mathbf{k}n\alpha}(t_1)$ are in the Heisenberg picture, and $\rho^{<,\mathbf{k}}(t_1)$ is the $t_2 \to t_1$ limit of $G^{<}_{\mathbf{k}\alpha \beta}(t_1,t_2)$ . The spin density matrix equation Eq.~\ref{density_mat_rec} is formally equivalent to the one obtained in the spinless case. However, it generally describes a non trivial spin evolution, both in the coherent and collisional part, since both the electron density matrix and self-energy are generally non-diagonal in spinor space. Due to the presence of the electron-phonon Hamiltonian term in Eq.~\ref{H_fp}, the electron self-energy contains terms that couple the evolution of electrons and phonons. For this reason, an additional equation for the (one-body) phonon density matrix is required for the propagation of interacting electrons and phonons. Following the discussion of Ref.~\cite{Karlsson2021}, the phonon propagator is expressed in terms of a two-dimensional vector of fluctuation operators:

\begin{equation}
\begin{pmatrix}
\delta\hat{\phi}^1_{\mathbf{Q}\nu}(t) \\
\delta \hat{\phi}^2_{\mathbf{Q}\nu}(t) \\
\end{pmatrix} = 
\begin{pmatrix}
\hat{\phi}^1_{\mathbf{Q}\nu}(t)-\langle\hat{\phi}^1_{\mathbf{Q}\nu}(t)\rangle \\
\hat{\phi}^2_{\mathbf{Q}\nu}(t)-\langle\hat{\phi}^2_{\mathbf{Q}\nu}(t)\rangle \\
\end{pmatrix}
\end{equation}
where

\begin{equation}
\begin{pmatrix}
\hat{\phi}^1_{\mathbf{Q}\nu}(t) \\
\hat{\phi}^2_{\mathbf{Q}\nu}(t) \\
\end{pmatrix} = 
\begin{pmatrix}
\frac{\hat{a}_{\mathbf{-Q}\nu}^\dagger(t)+\hat{a}_{\mathbf{Q}\nu}(t)}{\sqrt2} \\
i\frac{\hat{a}_{\mathbf{-Q}\nu}^\dagger(t)-\hat{a}_{\mathbf{Q}\nu}(t)}{\sqrt2} \\
\end{pmatrix}
\end{equation}

and $\hat{a}_{\mathbf{Q}\nu}(t)$ and $\hat{a}_{\mathbf{Q}\nu}^\dagger(t)$ are the phonon annihilation and creation operators for the equilibrium Born-Oppenheimer (BO) phonon mode $\nu$ and momentum $\mathbf{Q}$. The phononic greater and lesser Green's functions in reciprocal space are defined from the fluctuation operators as in Ref.~\cite{Stefanucci2024} :

\begin{equation}
\begin{gathered}
\mathcal{L}^{ij,>}_{\mathbf{Q},\nu\nu'}(t_1,t_2) = -i\langle \delta\hat\phi^i_{\mathbf{Q}\nu}(t) \delta\hat{\phi}^j_{-\mathbf{Q}\nu'}(t') \rangle \\
\mathcal{L}^{ij,<}_{\mathbf{Q},\nu\nu'}(t_1,t_2) = -i\langle \delta\hat\phi^j_{-\mathbf{Q}\nu'}(t') \delta\hat{\phi}^i_{\mathbf{Q}\nu}(t) \rangle 
\end{gathered}
\end{equation}

and the one-body phonon density matrix equation in matrix notation reads~\cite{Stefanucci2024}:

\begin{equation}
\begin{gathered}
   \partial_{t} \gamma_\mathbf{Q}^<(t)+i(\mathcal{J}h^{ph}_{qp}(\mathbf{Q},t)\gamma_\mathbf{Q}^<(t)-\gamma_{\mathbf{Q}}^<(t)h^{ph}_{qp}(\mathbf{Q},t)\mathcal{J})\\
   =\mathcal{J} \int_0^t dt'[\Xi_\mathbf{Q}^>(t,t')\mathcal{\mathcal{L}}_\mathbf{Q}^<(t',t)-\Xi_{\mathbf{Q}}^<(t,t')\mathcal{\mathcal{L}}_{\mathbf{Q}}^>(t',t)]+h.c.
   \label{density_mat_ph}
   \end{gathered}
\end{equation}

where $\Xi_{\mathbf{Q}}^<(t,t')$ is the lesser phonon self-energy, $\gamma^{<,ij}_{\mathbf{Q}\nu}(t) = i \mathcal{L}_{\mathbf{Q}\nu\nu}^{ij,<}(t,t) 
$ is the one-body phonon density matrix, while the commutator metric $\mathcal{J}$ acts in the two-dimensional space as $\mathcal{J}_{\nu\nu'}=-\hat{\sigma}^y\delta_{\nu\nu'}$~\cite{Stefanucci2024}, and the quasiparticle phonon Hamiltonian $h^{ph}_{qp}(\mathbf{Q},t)$ is defined as : 

\begin{equation}
\begin{gathered}
        h_{qp}^{ph}(\mathbf{Q},t) = h_0^{ph}(\mathbf{Q})+\Xi_{\mathbf{Q}}^{\delta}(t)   \\
        h_{0,\nu\nu'}^{ph}(\mathbf{Q})= \begin{pmatrix}K_{\nu\nu'}(\mathbf{Q}) &0 \\0 & \delta_{\nu\nu'}\end{pmatrix}
        \label{hqp}
\end{gathered}
\end{equation}
with $K_{\nu\nu'}(\mathbf{Q})$ the bare elastic tensor and $\Xi^{\delta}_{\mathbf{Q}}(t)$ being the singular part of the time-dependent phonon self-energy. The equation is identical to the nonmagnetic collinear case~\cite{Stefanucci2024}, the only difference being that in the noncollinear case the dressed electron-phonon coupling, appearing in the phonon self-energy in many approximated expressions, can  generally be non-diagonal in spinor space, as I will discuss in Sec.~\ref{sec:selfen}. 

\section{Spin structure of the electron density matrix}



When compared to the spinless case~\cite{Stefanucci2024}, the evolution of the electron density matrix, Eq.~\ref{density_mat_rec}, presents the extra complication of the internal spin structure. Some particularly relevant aspects in this regard are discussed below. First, the coherent evolution of the density matrix is described by the commutator in the left hand side of Eq.~\ref{density_mat_rec}. The coherent evolution in the noncollinear magnetic case discussed here is generally non-trivial in spin, since the single particle Hamiltonian $\hat{H}^{qp}$ is non-diagonal in spinor space due to the presence of SOC. This means that the total spin magnetization is generally not conserved in time, and spin and orbital degrees of freedom can exchange angular momentum. This coherent process has been discussed in depth in the framework of TDDFT~\cite{Krieger2015,PhysRevB.95.024412,Krieger_2017}, and is a source of coherent demagnetization in the presence of light-matter coupling. The right-hand side of Eq.~\ref{density_mat_rec} contains instead the collision integral term, which describes incoherent dynamics, and can include electron-boson or electron-electron scattering contributions~\cite{Karlsson2021,Stefanucci2024,Mocatti2026}. Due to the presence of off-diagonal terms in spinor space in the electron Green's function, even a non-magnetic interaction vertex can cause a spin flip, both through electron-electron and electron-phonon processes, through the Elliott-Yafet mechanism~\cite{PhysRev.96.266,YAFET19631,Kiss2016,7vr4-57mk}. Conversely, in the non-magnetic collinear case, each scattering event conserves spin. A very useful alternative perspective on Eq.~\ref{density_mat_rec} is obtained by introducing a different single particle orthonormal representation $\{\mathbf{k},\lambda\}$, defined by the unitary transformation $U^{\lambda\mathbf{k}}_{\alpha n}$ that diagonalizes the quasiparticle Hamiltonian at $t=0$, $h^{qp,nn'}_{\alpha \beta}(\mathbf{k},t=0)$:

\begin{equation}
    h^{qp,nn'}_{\alpha \beta}(\mathbf{k},t=0)= \sum_{\lambda}~U^{\mathbf{k}\lambda}_{n \alpha} \varepsilon^0_{ \mathbf{k}\lambda} (U^\dagger)^{\mathbf{k}\lambda}_{n' \beta} \label{eq:basis_change}
\end{equation}

In this alternative $\{\mathbf{k},\lambda\}$ representation, one can identify quasiparticle eigenvalues $\varepsilon^0_{\mathbf{k}\lambda}$ and eigenvectors $\psi_{\mathbf{k}\lambda}$. The generic eigenvector $\psi_{\mathbf{k}\lambda}$ is non-diagonal in the spin basis, $i.e.$

\begin{equation}
    \psi_{\mathbf{k}\lambda} = a_{\mathbf{k}\lambda} \ket{\uparrow}+b_{\mathbf{k}\lambda }\ket{\downarrow} 
\end{equation}

where $a_{\mathbf{k}\lambda }$ and $b_{\mathbf{k}\lambda }$ are the orbital components.
In this basis, it is possible to rewrite Eq.~\ref{density_mat_rec} as:

\vspace{2.5pt}\begin{widetext}
\begin{equation}
   i\partial_{t} \rho^{<,\mathbf{k}}_{\lambda \lambda'}(t)+[h^{qp}(\mathbf{k},t),\rho^{<,\mathbf{k}}(t)]_{\lambda \lambda'} = -~\int_0^t dt'~[\Sigma^{>}_{\mathbf{k},\lambda \lambda''}(t,t')G^<_{\mathbf{k},\lambda'' \lambda'}(t',t)-\Sigma^{<}_{\mathbf{k},\lambda \lambda''}(t,t')G^>_{\mathbf{k},\lambda'' \lambda'}(t',t)]  + h.c. \label{density_mat_rec2}
\end{equation}
\end{widetext}

This basis choice has the clear advantage that the equation of motion for the density matrix, Eq.~\ref{density_mat_rec2}, becomes formally identical to the spinless case. The price to pay is that the spin evolution is implicit, since each eigenstate identified by $\{\mathbf{k},\lambda\}$ is a spinor with components in both spins. I observe that if both $\hat{H}^{qp}$ and $\rho$ are diagonal in the $\lambda$ basis, no coherent spin precession is possible and the expectation value of the spin operator over any single particle electron state $\langle \mathbf{S}(t) \rangle_\lambda$:

\begin{equation}
    \langle \mathbf{S}(t) \rangle_{\lambda} =  \bra{\psi_{\mathbf{k}\lambda}} \hat{\mathbf{S}}(t)\ket{\psi_{\mathbf{k}\lambda}}
\end{equation}

is manifestly constant in the dynamics. A coherent spin precession can only be initiated by band off-diagonal $\lambda \neq \lambda'$ terms in the commutator, either in the single particle Hamiltonian (for example, through the explicitly time-dependent laser Hamiltonian or the time-dependent renormalization of the exchange splitting), or by electronic density matrix band off-diagonal terms, whose dynamics is initiated by the laser itself.
The complexity of a quantitative theory of ultrafast magnetic dynamics can thus be ascribed to the simultaneous presence of coherent and incoherent demagnetization channels in the non-equilibrium evolution triggered by the laser. 

Any solution of Eq.~\ref{density_mat_rec2} requires an approximate expression for the contour electronic Green’s function and self-energy. The objective is to develop physically motivated approximations that capture the dominant scattering mechanisms and quasiparticle renormalization effects, allowing to obtain a quantitative description of the time-dependent electron density matrix and, thus, of the time-dependent magnetization. The general idea of this work is to assume that the quasiparticle approximation for the retarded electronic Green's function holds and resort to the mirrored version of the generalized Kadanoff-Baym ansatz for the lesser and greater components~\cite{Schafer2002,Stefanucci2024}, properly extended to the spinor case. Approximations for the self-energy in the framework of many-body perturbation theory are instead discussed in detail in the next section. The approach proposed in this work is expected to be most naturally applicable to relatively weakly correlated itinerant systems far from the Mott limit~\cite{PhysRev.115.2}, where these assumptions are expected to be physically justified.

\section{Approximations for the electron self-energy}
\label{sec:selfen}

To build a physically grounded approximation for the self-energy, I adopt a many-body perturbation theory framework, and extend the considerations made in Ref.~\cite{Mocatti2026} to the magnetic case. In order to give a realistic description of the magnetization evolution, a scheme is necessary that is able to not only account for electron-electron and electron-phonon, but also for (emergent) electron-magnon effects, as the latter play an important role in ultrafast magnetic dynamics~\cite{PhysRevB.78.174422,7vr4-57mk,https://doi.org/10.1002/apxr.202300103}. In order to do so, additional diagrams must be included besides the ones present in the nonmagnetic case~\cite{Mocatti2026}, namely the ones constituting the transverse $GT$ self-energy in the particle-hole channel~\cite{PhysRevB.100.045130}. In the following section I will outline a strategy to consistently do so in an intuitive, yet consistent, way. I will start from the Fan-Migdal self-energy, originating from the electron-phonon interaction Hamiltonian, and then pass to $GW$ and $GT$ self energies coming from the electron-electron interaction Hamiltonian.

\vspace{2pt}\subsection{Fan-Migdal self-energy}

The electron-phonon self-energy contribution is included at the Fan-Migdal level, as discussed in Ref.~\cite{Mocatti2026}. The corresponding self-energy in the presence of spin-orbit coupling is expressed, in the spin-orbital basis $\{\mathbf{k},\lambda\}$  defined by Eq.~\ref{eq:basis_change}, as~\cite{Stefanucci2024,Mocatti2026}:

\begin{equation}
\begin{gathered}
    \Sigma^{FM,<}_{\mathbf{k},\lambda \lambda'}(t,t') =  i \sum_{\mathbf{Q}\lambda''\lambda'''\nu\nu'}   g^{\nu}_{\lambda \lambda'}(\mathbf{Q},\mathbf{k}-\mathbf{Q})~\times\\
 \mathcal{L}^{11,<}
_{\mathbf{k}-\mathbf{Q}\nu\nu'}(t,t')G^<_{\mathbf{Q}\lambda''\lambda'''}(t,t')g^{*\nu'}_{\lambda''\lambda'''}({\mathbf{Q},\mathbf{\mathbf{k}-\mathbf{Q}}})
\end{gathered}
\end{equation}

where $g_{\lambda \lambda'}^{\nu}({\mathbf{k},\mathbf{Q}})=\braket{\psi_{\mathbf{k+Q}{\lambda'}}| \Delta V^{el-ph}_{\mathbf{Q}\nu}|\psi_{\mathbf{k}\lambda}}$ represents the statically dressed electron-phonon coupling matrix element, see also the discussions in Refs.~\cite{Stefanucci2024,Mocatti2026}. It is related to the electron-phonon coupling in the $\{\mathbf{k},n,\alpha\}$ basis through the transformation:

 \begin{equation}
     g^{\nu}_{\alpha \alpha'nn'}(\mathbf{k},\mathbf{Q}) = U^{\mathbf{k+Q}\lambda}_{n \alpha} g_{\lambda \lambda'}^{\nu}({\mathbf{k},\mathbf{Q}})(U^\dagger)^{\mathbf{k}\lambda'}_{n'\alpha'}
 \end{equation}

I note that the screened electron-phonon coupling can in principle contain all the $\hat{\sigma}^I$ components in the form of Eq.~\ref{operator}, even if the bare electron-phonon interaction is purely charge-like in spinor space. This is the case, for example, at the self consistent density-functional perturbation theory level~\cite{urru2020lattice,Bistoni22}. The charge component proportional to the identity $\hat{\sigma}^0$ in spinor space is the one responsible for the Elliott-Yafet~\cite{PhysRev.96.266,YAFET19631} type scattering, while the three magnetic components are only present in the noncollinear magnetic case, contributing both to the the quasiparticle Hamiltonian, and to the collision integrals, as I will discuss in later sections.\\

\vspace{2pt}\subsection{$GW$ self-energy}

I now pass to the discussion of the  $GW$ self-energy, which  in the spin-dependent case takes the form~\cite{PhysRevLett.100.116402}:

\begin{equation}
\begin{gathered}
    \Sigma^{GW}_{\alpha \beta}(1,2)=i G_{\eta \gamma}(1,2) W_{\alpha \gamma \eta \beta}(1,2)= \\
    i ~\hat{\sigma}^I_{\alpha \eta} G_{\eta \gamma}(1,2) W_{IJ}(1,2)\hat{\sigma}^J_{\gamma \beta}
    \label{GW_selfen}
    \end{gathered}
\end{equation}

The corresponding lesser component is:

\begin{equation}
\begin{gathered}
    \Sigma^{GW,<}_{\alpha \beta}(1,2)=i G_{\eta \gamma}^<(1,2) W_{\alpha \gamma \eta \beta}^<(1,2)= \\
    i ~\hat{\sigma}^I_{\alpha \eta} G_{\eta \gamma}^<(1,2) W_{IJ}^<(1,2)\hat{\sigma}^J_{\gamma \beta}
    \label{GW_selfen}
    \end{gathered}
\end{equation}

Compared to the collinear nonmagnetic case, the $GW$ self-energy presents a much more complicated spin structure when the interaction $W$ is spin dependent, since both interaction vertexes can flip spin, thus mixing charge and spin components. However, I note that \textit{its time structure on the contour remains the one of a point-wise product}, as in the spinless case. Thus, one can conceptually operate in the same way on the lesser/greater self-energy, and express $W^\gtrless$ through the Dyson's equation for the screened potential, $W=v+vPW$, where $P$ is the polarization and $v$ is the bare Coulomb interaction. Employing the Langreth rules (see also Appendix B), one has~\cite{Marini2013,Stefanucci2024,Mocatti2026}

\begin{equation}
    \mathbf{W}^\gtrless= \mathbf{W}^R\mathbf{P}^\gtrless \mathbf{W}^A
\end{equation}

 $\mathbf{P}$ can then be approximated with the non-interacting polarizability $\mathbf{\chi^0}$: 

\begin{equation}
    P^<_{\eta \theta \mu \nu}(1,2) \simeq \chi^{0,<}_{\eta \theta \mu \nu}(1,2) =-i G^>_{\nu\theta}(2,1)G^<_{\eta \mu}(1,2)
\end{equation}

Furthermore, the retarded and advanced screened interactions $W$ are approximated as time-local~\cite{Mocatti2026}:

\begin{equation}
W^{R,A}_{IJ}(1,2) = W_{IJ}(\mathbf{r}_1,\mathbf{r}_2,z_1)\delta(z_1-z_2)   \label{TLWC}
\end{equation}

Within these approximation, the $GW$ contribution to the electron self-energy can be systematically framed as a perturbative  diagrammatic expansion in terms of a time-local interaction, see also Appendix C and Ref.~\cite{Mocatti2026}. I obtain the following expression for the  lesser $GW$ self-energy:

\begin{equation}
\begin{gathered}
    \Sigma^{GW,<}_{\alpha \beta}(1,2) = i G^<_{\eta \gamma}(1,2)\times\\
    W_{\alpha\delta\eta\theta}(1,2) \chi^{0,<}_{\delta\theta\zeta\kappa}(1,2) W_{\kappa\gamma\zeta\beta}(1,2)
    \end{gathered}
\end{equation}

In periodic systems, the screened Coulomb interaction is invariant under lattice translations, and its Fourier representation reads~\cite{PhysRevB.34.5390}

\begin{equation}
\begin{gathered}
W_{\alpha\beta\gamma\delta}(\mathbf{r}_1,\mathbf{r}_2,t) =\\
\frac{1}{V}\sum_{\mathbf{q}}\sum_{\mathbf{G},\mathbf{G}'}
  e^{i(\mathbf{q}+\mathbf{G})\cdot\mathbf{r}_1}\,
  W_{\alpha\beta\gamma\delta}^{\mathbf{G}\mathbf{G}'}(\mathbf{q},t)\,
  e^{-i(\mathbf{q}+\mathbf{G}')\cdot\mathbf{r}_2}
\label{eq:W_recip}
\end{gathered}
\end{equation}

where $\mathbf{q}$ belongs to the first Brillouin zone, $\mathbf{G},\mathbf{G}'$ are reciprocal-lattice vectors. Incidentally, I note that in metals it is sometimes a qualitatively good approximation to simplify this expression by only retaining the head of $W$ and neglecting local fields, i.e.
\begin{equation}
W^{\mathbf{G}\mathbf{G}'}(\mathbf{q},t)
\;\approx\;
W^{\mathbf{00}}(\mathbf{q},t)\,\delta_{\mathbf{G}\mathbf{0}}\,\delta_{\mathbf{G}'\mathbf{0}}.
\label{eq:head_approx}
\end{equation}

It is convenient to work in the Bloch space. I consider the $GW$ matrix element between Bloch states labeled by  $\{\mathbf{k},n,\alpha\}$, namely: 

\vspace{2.5pt}\begin{widetext}
\begin{equation}
  \Sigma_{\mathbf{k} nm,\alpha\beta}^{GW,<}(t, t') = \int d\mathbf{r}_1 d\mathbf{r}_2 \psi^{*,\alpha}_{\mathbf{k} n }(\mathbf{r}_1) G_{\eta\gamma}^<(1,2) W^<_{\alpha\gamma\eta\beta}(1,2)   \psi^{\beta }_{\mathbf{k} m}(\mathbf{r}_2) \label{GWselfenk}
\end{equation}

Expanding the Green's function through Eq.~\ref{bloch_fields} and employing crystalline momentum conservation one finds:

\begin{equation}
\begin{gathered}
\Sigma_{\mathbf{k}nm,\alpha\beta}^{GW,<}(t, t') = i  W_{\mathbf{k},\mathbf{k},\alpha\gamma\eta\beta}^{<,n m' n' m}(\mathbf{q},t, t') G_{\mathbf{q},\eta \gamma}^{<,n'm'}(t, t') \\
=  W^{nan'a'}_{\mathbf{k},\mathbf{q}\alpha\delta\eta\theta}(\mathbf{p},t) G^{>,b'a'}_{\mathbf{q}+\mathbf{p},\kappa\theta}(t', t) G_{\mathbf{k}+\mathbf{p},\delta\zeta}^{<,ab}(t, t') W^{b'm'bm}_{\mathbf{k}+\mathbf{p},\mathbf{q}+\mathbf{p},\kappa\gamma\zeta\beta}(-\mathbf{p},t) G_{\mathbf{q},\eta \gamma}^{<,n'm'}(t, t') \enspace,
\label{GWselfen}
\end{gathered}
\end{equation}
\end{widetext}
 
having defined the matrix element

\begin{equation}
\begin{gathered}
W^{n_1 n_2 n_3 n_4}_{\mathbf{k},\mathbf{k}',\alpha\beta\gamma\delta}(\mathbf{q},t)
=\\
\sum_{\mathbf{G}\mathbf{G}'}\rho^{n_1\alpha, n_3\gamma}_{\mathbf{k},\mathbf{k'}}(\mathbf{G})\,
W^{\mathbf{G}\mathbf{G}'}(\mathbf{k}-\mathbf{k}',t)\,
\rho^{*,n_2\beta, n_4\delta}_{\mathbf{k}+\mathbf{q},\mathbf{k'}+\mathbf{q}}(\mathbf{G'}),
\end{gathered}
\end{equation}
and the oscillator matrix element

\begin{equation}
    \rho^{n\alpha, m\beta}_{\mathbf{k},\mathbf{k}'}(\mathbf{G})
= \langle\psi_{\mathbf{k}n\alpha}|e^{i(\mathbf{k}-\mathbf{k}'+\mathbf{G})\cdot\mathbf{r}}|\psi_{\mathbf{k}'m\beta}\rangle
\end{equation} 

The complex spin structure of the lesser $GW$ self-energy can be  simplified considering a purely charge-like screened interaction:

\begin{equation}
    W_{\alpha \gamma \eta \beta}^{R,A}(1,2) = \hat{\sigma}^0_{\alpha \eta} W_{00}^{R,A}(1,2)\hat{\sigma}^0_{\gamma \beta}
\end{equation}

In this approximation, the real space expression of the $GW$ self-energy is:

\begin{equation}
        \Sigma^{GW,<}_{\alpha \beta}(1,2) = i G^<_{\alpha \beta}(1,2)W^R_{00}(1,2) P^<_{\eta\kappa\kappa\eta}(1,2) W^A_{00}(1,2)
\end{equation}

thus recovering a form similar to the one obtained in the spinless case. The self-energy Eq.~\ref{GWselfen}, expressed in the spin-orbital basis defined by Eq.~\ref{eq:basis_change}, reads: 

\vspace{2.5pt}\begin{widetext}
\begin{equation}
\begin{gathered}
\Sigma_{\mathbf{k},\lambda\lambda'}^{GW,<}(t, t') = i  W_{\mathbf{k},\mathbf{q}}^{<,\lambda \mu' \mu \lambda'}(t, t') G_{\mathbf{q},\mu\mu'}^{<}(t, t') \\
=  W^{\lambda \lambda'' \mu \lambda'''}_{\mathbf{k},\mathbf{q}}(\mathbf{p},t) G^{>}_{\mathbf{q}+\mathbf{p},\mu'''\lambda'''}(t', t) G_{\mathbf{k}+\mathbf{p},\lambda''\mu''}^{<}(t, t') W^{\mu'''\mu'\mu''\lambda'}_{\mathbf{k}+\mathbf{p},\mathbf{q}+\mathbf{p}}(-\mathbf{p},t) G_{\mathbf{q},\mu \mu'}^{<}(t, t') \enspace,
\end{gathered}
\end{equation}
\end{widetext}

with a structure that is formally identical to the collinear one~\cite{Stefanucci2024}, while the complex spin structure is hidden in the matrix element $W^{\lambda\lambda'\mu\mu'}_{\mathbf{k},\mathbf{k}'}(\mathbf{q},t)$, defined as:

\begin{equation}
\begin{gathered}
   W^{\lambda\lambda'\mu\mu'}_{\mathbf{k},\mathbf{k}'}(\mathbf{q},t) = \\
   (U^{\dagger})^{\mathbf{k}\lambda}_{n_1\alpha} (U^\dagger)^{\mathbf{k'+\mathbf{q}}\mu'}_{n_4\delta}W^{n_1 n_2 n_3 n_4}_{\mathbf{k},\mathbf{k}',\alpha\beta\gamma\delta}(\mathbf{q},t) U^{\mathbf{k}'\lambda'}_{n_2\beta} U^{\mathbf{k+\mathbf{q}}\mu}_{n_3\gamma} 
   \end{gathered}
\end{equation}

\vspace{2pt}\subsection{$GT$ self-energy and electron-magnon coupling}

As already mentioned, a quantitative description of ultrafast magnetic dynamics requires the inclusion of electron-magnon scattering. The strategy I follow is to derive an effective electron-magnon interaction from the particle-hole channel of the transverse $GT$ self-energy~\cite{PhysRevB.100.045130,muller_2024}, as electron-magnon scattering is known to emerge from this channel~\cite{Hertz_1973}. An important premise is due: care must be taken in the simultaneous inclusion of $GW$ together with particle-particle or particle-hole $T$ matrix, to prevent double counting of low-order terms, specifically the direct diagram in the second Born approximation~\cite{PhysRevB.85.155131}: a consistent inclusion of more than one correlation channel requires a suitable strategy~\cite{PhysRevB.43.8044,PhysRevLett.80.2389,PhysRevB.100.045130,muller_2024}. This can be done for example within the fluctuating exchange (FLEX) approximation~\cite{PhysRevLett.62.961,PhysRevB.43.8044,BICKERS1989206}, where one writes the exact self-energy up to the second Born and then sums each correlation channel's contribution starting from the third order. A fully self-consistent FLEX treatment  would remove
this overlap systematically at the price of solving four-point integral equations at every time step of the nonequilibrium evolution, currently out of reach for the \textit{ab initio}, real-time setting targeted in this work. Since here I am ultimately only interested in including the poles of the $GT$ self-energy and discard any finite-order term emerging from it, I adopt a different approach, where the pole contributions of the full transverse electron-hole $T$ matrix are included in combination with the $GW$ self-energy. A more detailed justification on this approach and its limitations is given in Appendix D. In the noncollinear case, the $GT$ self-energy on the contour reads

\begin{equation}
\Sigma_{\alpha \gamma}^{GT}(1,3) = -i G_{\beta \delta}(2,4)T_{\alpha \beta \gamma \delta} (1,2|3,4) \\ \label{GTselfen1}
\end{equation}

where the particle-hole $T$ matrix can be expanded in a spin basis as:

\begin{equation}
T_{\alpha \beta \gamma \delta}(1,2|3,4)= \hat{\sigma}^{I}_{\alpha \gamma} T_{I J}(1,2|3,4) \hat{\sigma}^{J}_{\beta \delta}    
\end{equation}

and obeys the following self-consistent Bethe-Salpeter equation (BSE) in the ladder approximation:
\vspace{2.5pt}\begin{widetext}
\begin{equation}
   T_{\alpha \beta \gamma \delta}(1,2|3,4) = \hat{\sigma}^I_{\alpha \gamma} W_{IJ}(1,2)\hat{\sigma}^J_{\beta \delta} ~\delta(1-3)\delta(2-4) + \hat{\sigma}^I_{\alpha \eta} W_{IJ}(1,2){\sigma_{\beta\theta}^J} K_{\eta \theta \mu \nu}(1,2|1',2')~T_{\mu \nu \gamma \delta}(1',2'|3,4) \label{$T$ matrix}
\end{equation}
\end{widetext}

where I defined the particle-hole propagator $K_{\eta \theta \mu \nu}(1,1'|2,2') = G_{\eta \mu}(1,1')G_{\nu\theta}(2',2)$. The $T$ matrix is a  four-time function on the contour, which makes the $GT$ self-energy a complicated object. A substantial simplification arises upon assuming a time-local screened interaction $W$, as in the $GW$ case:

\begin{equation}
W_{IJ}(1,2) = W_{IJ}(\mathbf{r}_1,\mathbf{r}_2,z_1)\delta(z_1-z_2)   \label{TLW}
\end{equation}

Within this approximation, the $T$ matrix will only depend on two times, $z_1=z_2$ and $z_3=z_4$. Explicitly:

\begin{equation}
\begin{gathered}
    T_{\alpha \beta \gamma \delta}(1,2|3,4) = \\
    T_{\alpha \beta \gamma \delta}(\mathbf{r}_1,\mathbf{r}_2,z_1|\mathbf{r}_3,\mathbf{r}_4,z_3)\delta(z_1-z_2)\delta(z_3-z_4)
    \end{gathered}
\end{equation}

This approximation has already been discussed in previous literature~\cite{PhysRevLett.80.2389,PhysRevB.100.045130,muller_2024} and is expected to correctly capture electron-magnon physics. Within this time-local approximation for $W$, I am explicitly neglecting potentially important dynamical effects like magnon-phonon coupling, which should emerge when a dynamically screened (time non-local) interaction is considered in the BSE equation for the $T$ matrix~\cite{PhysRevB.102.045136,PhysRevB.108.165101}, in analogy with the exciton case. Once the time-local expression  for $T$ is inserted in the self-energy $\Sigma^{GT}$, its time structure is much simpler, since the two internal time integrations over $t_2$ and $t_4$ become trivial, reducing to the simpler form:

\begin{equation}
\begin{gathered}
    \Sigma_{\alpha \gamma}^{GT}(1,3) = \\
    -i G_{\beta \delta}(\mathbf{r}_2,z_1,\mathbf{r}_4,z_3) T_{\alpha \beta\gamma\delta}(\mathbf{r}_1,\mathbf{r}_2,z_1|\mathbf{r}_3,\mathbf{r}_4,z_3)  
    \end{gathered}
\end{equation}

In this approximated form, the $GT$ self-energy is a simple a point-wise product on the contour, and its lesser component is:

\begin{equation}
\begin{gathered}
    \Sigma_{\alpha \gamma}^{GT,<}(1,3) = \\
    -i G^<_{\beta \delta}(\mathbf{r}_2,t_1,\mathbf{r}_4,t_3) T^<_{\alpha \beta\gamma\delta}(\mathbf{r}_1,\mathbf{r}_2,t_1|\mathbf{r}_3,\mathbf{r}_4,t_3) \label{GTless}
    \end{gathered}
\end{equation}

\vspace{2pt}\subsection{ Quasi-stationary limit and pole decomposition }

\begin{figure*}
    \centering
    \includegraphics[width=1\linewidth]{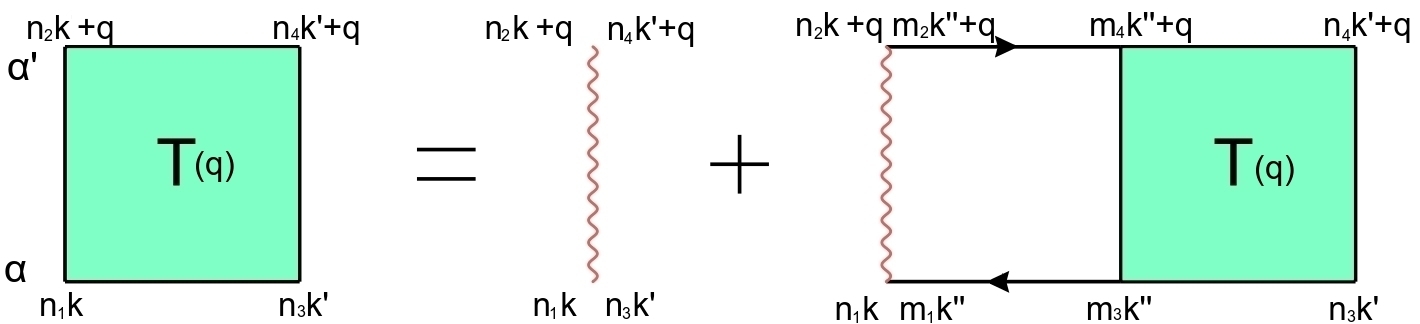}
    \caption{Diagrammatic representation~\cite{HARLANDER2020107465} of the transverse particle-hole $T$ matrix Bethe-Salpeter equation in the Bloch space.}
    \label{fig:Tmat}
\end{figure*}

 To extract the poles of the $T$ matrix and come to a simpler electron-boson form for the $GT$ self-energy, Eq.~\ref{GTless}, I first consider the more intuitive \textit{collinear} ferromagnetic case. Once the procedure has been clarified, I will discuss the effect of noncollinearity. Assuming that the screened interaction is a scalar and that the magnetic component can be neglected, transverse and longitudinal channels are separated, and it is possible to focus on the transverse channel only:

\begin{equation}
\begin{gathered}
    T_{\alpha \alpha'}(1,2|3,4) = W(\mathbf{r}_1,\mathbf{r}_2,z_1) \delta(1-3) \delta(2-4)+\\
    W(\mathbf{r}_1,\mathbf{r}_2,z_1)K_{\alpha \alpha'}(1,2|1',2') T_{\alpha,\alpha'}(1',2'|3,4)
    \end{gathered}
\end{equation}

where $\alpha$ and $\alpha'$ are opposite spins and I define ${T}_{\alpha \alpha'} = T_{\alpha \alpha' \alpha \alpha'}$ and ${K}_{\alpha \alpha'} = K_{\alpha \alpha' \alpha \alpha'}$. The $GT$ self-energy is:

\begin{equation}
\begin{gathered}
    \Sigma^{GT}_{\alpha}(1,3) = -i \int d\mathbf{r}_2 d\mathbf{r}_4 {G}_{\alpha'}(\mathbf{r}_2,z_1,\mathbf{r}_4,z_3) \\
 \times   {T}_{\alpha \alpha'}(\mathbf{r}_1,\mathbf{r}_2,z_1|\mathbf{r}_3,\mathbf{r}_4,z_3) \label{eq:self_energy}
    \end{gathered}
\end{equation}

where I define ${G}_{\alpha'} = G_{\alpha' \alpha'}$. I briefly anticipate the strategy here employed to build an approximated form for the lesser $T$ matrix, appearing in the lesser $GT$ self-energy collision integral: I first  employ the zero-temperature time-ordered formalism and build the $T$ matrix BSE equation using standard diagrammatic perturbation theory, and derive an approximated form for time-ordered $T$ by operating a pole decomposition. Having identified this simplified structure, I then extend it to the lesser $T$ matrix component, $T^<(t,t')$. In the time-ordered formalism, the transverse particle-hole $T$ matrix obeys the following BSE equation:

\begin{equation}
\begin{gathered}
    T_{\alpha \alpha'}(1,2|3,4) = W(\mathbf{r}_1,\mathbf{r}_2,t_1) \delta(1-3) \delta(2-4)+\\
    W(\mathbf{r}_1,\mathbf{r}_2,t_1)K_{\alpha \alpha'}(1,2|1',2') T_{\alpha,\alpha'}(1',2'|3,4)
    \end{gathered}
\end{equation}

Aiming at separating a fast ($\tau$) from a slow ($\bar{t}$) dynamics, I introduce Wigner coordinates:

\begin{equation}
    \bar{t} = (t_1+t_3)/2 ; ~~\tau = t_1-t_3
\end{equation}
and make a a quasi-stationary assumption on the $\bar{t}$ dependence in Eq.~\ref{TLW}. In this limit, it is possible to decouple the effect of the dependence on the time $\bar{t}$  and Fourier transform in the relative time $\tau$, obtaining the following expression for the Fourier component $T(\omega)$:
\vspace{2.5pt}\begin{widetext}
\begin{equation}
    T_{\alpha \alpha'}(\mathbf{r}_1,\mathbf{r}_2|\mathbf{r}_3,\mathbf{r}_4;\bar{t},\omega) = W (\mathbf{r}_1,\mathbf{r}_2,\bar{t})+W(\mathbf{r}_1,\mathbf{r}_2,\bar{t})K_{\alpha \alpha'}(\mathbf{r}_1,\mathbf{r}_2|\mathbf{r}_{1'}\mathbf{r}_{2'},\bar{t},\omega)T_{\alpha \alpha'}(\mathbf{r}_{1'},\mathbf{r}_{2'}|\mathbf{r}_3,\mathbf{r}_4;\bar{t},\omega)\label{Tfreq}
\end{equation}

For periodic systems, I work in the Bloch space as for the $GW$ self-energy. It is thus convenient to consider the matrix element of the $T$ matrix between Bloch states labeled by  $\{\mathbf{k},n,\alpha\}$, namely: 

\begin{equation}
   T^{n_1 n_2 n_3 n_4}_{\alpha \alpha'}(\mathbf{k}_1,\mathbf{k}_2,\mathbf{k}_3,\mathbf{k}_4,\bar{t},\omega) = \int d\mathbf{r}_1 d\mathbf{r}_2 d\mathbf{r}_3 d\mathbf{r}_4~\psi^{*,\alpha}_{\mathbf{k}_1n_1 }(\mathbf{r}_1) \psi^{*,\alpha'}_{\mathbf{k}_2n_2 }(\mathbf{r}_2) T_{\alpha \alpha'}(\mathbf{r}_1,\mathbf{r}_2|\mathbf{r}_3,\mathbf{r}_4;\bar{t},\omega)  \psi^{\alpha }_{\mathbf{k}_3n_3}(\mathbf{r}_3) \psi^{\alpha'}_{\mathbf{k}_4n_4 }(\mathbf{r}_4)\label{Tmatel}
\end{equation}

By expanding the Green's function through the Eq.~\ref{bloch_fields} and imposing crystalline momentum conservation at each vertex in the $T$ matrix, one of the four crystal momentum dependence in Eq.~\ref{Tmatel} is removed, obtaining:

\begin{equation}
        T^{n_1 n_2 n_3 n_4}_{\alpha \alpha'}(\mathbf{k},\mathbf{k}',\mathbf{q},\bar{t},\omega) = W^{n_1 n_2 n_3 n_4}_{\mathbf{k},\mathbf{k}'}(\mathbf{q},\bar{t})+W^{n_1 n_2 m_1m_2}_{\mathbf{k},\mathbf{k}''}(\mathbf{q},\bar{t})K^{m_1 m_2 m_3 m_4}_{\alpha \alpha'}(\mathbf{k}'',\mathbf{q},\bar{t},\omega)T^{m_3 m_4 n_3 n_4}_{\alpha \alpha'}(\mathbf{k}'',\mathbf{k}',\mathbf{q},\bar{t},\omega) \label{Tmateq}
\end{equation}

where I defined the electron-hole propagator in reciprocal space:

\begin{equation}
        K^{m_1 m_2 m_3 m_4}_{\alpha \alpha'}(\mathbf{k}',\mathbf{q},\bar{t},\omega)
    = 
    -i\int_{-\infty}^{+\infty} \frac{d\omega'}{2\pi}\,
    G^{m_1 m_3}_{\mathbf{k}'',\alpha}\!\left(\bar{t},\omega'\right)
    G^{m_2 m_4}_{\mathbf{k}''+\mathbf{q},\alpha'}\!\left(\bar{t},\omega'+\omega \right)
\end{equation}

and the spin-independent matrix element $W^{n_1 n_2 n_3 n_4}_{\mathbf{k},\mathbf{k}'}(\mathbf{q},\bar{t})=W^{n_1 n_2 n_3 n_4}_{\mathbf{k},\mathbf{k}'\alpha\alpha'\alpha\alpha'}(\mathbf{q},\bar{t})$. A pictorial depiction of Eq.~\ref{Tmateq} is given in Fig.~\ref{fig:Tmat}. For simplicity, the evaluation is now restricted to the band-diagonal case, $n_1=n_3$ and $n_2=n_4$:
\begin{equation}
        T^{n_1 n_2n_1n_2}_{\alpha \alpha'}(\mathbf{k},\mathbf{k}',\mathbf{q},\bar{t},\omega) = W^{n_1 n_2n_1n_2}_{\mathbf{k},\mathbf{k}'}(\mathbf{q},\bar{t})+W^{n_1 n_2m_1m_2}_{\mathbf{k},\mathbf{k}''}(\mathbf{q},\bar{t})K^{m_1m_2m_3m_4}_{\alpha \alpha'}(\mathbf{k}'',\mathbf{q},\bar{t},\omega)T^{m_3m_4n_1 n_2}_{\alpha \alpha'}(\mathbf{k}'',\mathbf{k}',\mathbf{q},\bar{t},\omega)
\end{equation}

 but the band-diagonal assumption can be readily lifted if necessary. The idea now is to retain the singular part of the $T$ matrix emerging from infinite ladder summation, related for the magnon emergence, and to neglect the Stoner continuum. Before doing so, I recast the $T$ matrix in a symmetric form: 
\begin{equation}
    T = W + WKW+WKWKW~+ ... = W+W \bar{K}~W   \Rightarrow
\end{equation}

\begin{equation}
  T^{n_1 n_2n_1n_2}_{\alpha \alpha'}(\mathbf{k},\mathbf{k}',\mathbf{q},\bar{t},\omega)  = W^{n_1 n_2n_1n_2}_{\mathbf{k},\mathbf{k}'}(\mathbf{q},\bar{t})+W^{n_1 n_2m_1m_2}_{\mathbf{k},\mathbf{k}''}(\mathbf{q},\bar{t}) \bar{K}^{m_1m_2m_3m_4}_{\alpha \alpha'}(\mathbf{k}'',\mathbf{k}''',\mathbf{q},\bar{t},\omega)W^{m_3m_4n_1 n_2}_{\mathbf{k'''},\mathbf{k'}}(\mathbf{q},\bar{t}) \label{T_w_bar}
 \end{equation}
\end{widetext}

The interacting kernel $\bar{K}$ contains the whole frequency dependence of the $T$ matrix. It is immediate to verify that $\bar{K}$ obeys the following BSE:

\begin{equation}
    \bar{K} = K + K W \bar{K} \label{collcore}
\end{equation}

which is the spin-flip analogue of the excitonic BSE in the longitudinal particle-hole channel~\cite{RevModPhys.74.601,Marini2009}. Practical solutions of this equation at equilibrium, in terms of magnon eigenvalues and eigenvectors have been discussed in various works~\cite{gbpw-zh1v,489b-n5pd,PhysRevB.81.054434,PhysRevB.60.7419,PhysRevB.62.3006,PhysRevLett.127.166402,Chen2026}, both for metallic and insulating magnetic systems. I rewrite Eq.~\ref{collcore} as:

\begin{equation}
     (K^{-1}-W) \bar{K}= I
\end{equation}

where I note that poles correspond to the zero-eigenvalues of the operator $(K^{-1}-W)$.  The pole decomposition is written:

\begin{equation}
    \bar{K}(\omega) = \sum_S \dfrac{\ket{S}\bra{\tilde{S}}}{\lambda^S(\omega)}
\end{equation}

with $\lambda^S$ being the eigenvalues and $\ket{S}$, $\bra{\tilde{S}}$  the left and right eigenvectors, respectively. I Taylor-expand the eigenvalue close to the pole, obtaining:

\begin{equation}
    \lambda_S(\omega) \simeq \lambda'_S(\omega)\Big|_{\omega=\Omega_S}(\omega-\Omega_S)
\end{equation}

The explicit expression for the interacting kernel  $\bar{K}$ is: 

\vspace{2.5pt}\begin{widetext}
\begin{equation}
          \bar{K}^{n_1n_2n_3n_4}_{\alpha \alpha'}(\mathbf{k},\mathbf{k}',\mathbf{q},\bar{t},\omega)  = \sum_{S} \dfrac{A^{n_1n_2}_{S,\alpha\alpha'}(\mathbf{k},\mathbf{q},\bar{t})\tilde{A}^{n_3n_4}_{S,\alpha \alpha'}(\mathbf{k}',\mathbf{q},\bar{t})~}{\dfrac{\partial{\lambda^{S,\alpha\alpha'}_{\mathbf{q}}}(\bar{t},\omega)}{\partial \omega}\Big|_{\omega=\Omega^{S\alpha\alpha'}_\mathbf{q}(\bar{t})}(\omega-\Omega^{S,\alpha\alpha'}_{\mathbf{q}}(\bar{t}))}~+ \bar{K}^{n_1n_2n_3n_4}_{0\alpha \alpha'}(\mathbf{k},\mathbf{k}',\mathbf{q},\bar{t},\omega) \label{kbar_poles}
\end{equation}
\end{widetext}

\begin{figure*}
    \centering
    \includegraphics[width=1\linewidth]{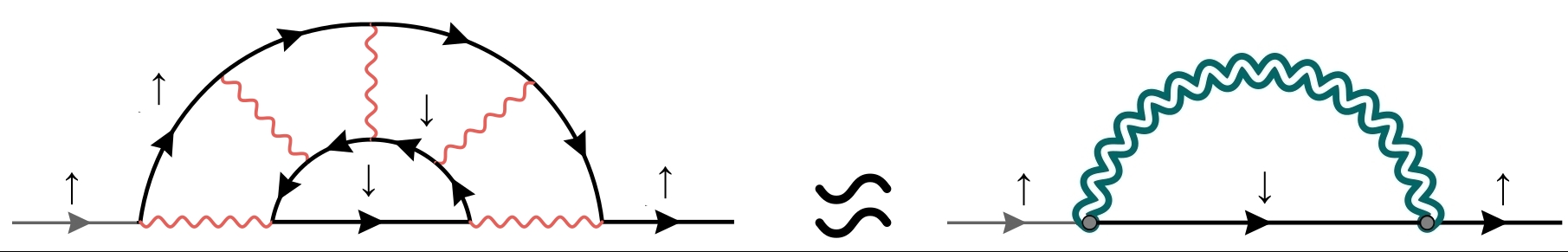}
    \caption{Pictorial representation~\cite{HARLANDER2020107465} of the $GT$ self-energy approximation as an emergent boson in proximity to a pole. Here, the wiggly red line represents the time-local screened interaction $W$, the black arrow represents the dressed electronic propagator, the grey arrow represents the bare electronic propagator and the wiggly double green line represents the magnon.}
    \label{fig:bosonic}
\end{figure*}

where I defined the magnon amplitude $A^{n_1n_2}_{S,\alpha\alpha'}(\mathbf{k},\mathbf{q},\bar{t})$ and the residue $R_{\mathbf{q}}^S  = \sqrt{(\dfrac{d\lambda^S_{\mathbf{q}}}{d \omega})^{-1}}\le 1$ due to the presence of Stoner particle-hole continuum. I explicitly isolated the magnon poles (first term), emerging from the infinite ladder resummation, from the continuum contribution, i.e.  the other zero eigenvalues, that I have put in $\bar{K}_0$ and that I will exclude from the evaluation of the $GT$ self-energy. This approximation is schematically represented in Fig.~\ref{fig:bosonic}, and is expected to work well close to the ladder poles. This expression for $\bar{K}$ allow to express the $T$ matrix in a simpler form of an effective electron-boson coupling. As a first step, I substitute the expression for $\bar{K}$, Eq.~\ref{kbar_poles}, inside Eq.~\ref{T_w_bar} (excluding the Stoner continuum $\bar{K}_0$):

\vspace{2.5pt}\begin{widetext}
\begin{equation}
\begin{gathered}
     T^{n_1 n_2n_1n_2}_{\alpha \alpha'}(\mathbf{k},\mathbf{k}',\mathbf{q},\bar{t},\omega)  = W^{n_1 n_2n_1n_2}_{\mathbf{k},\mathbf{k}'}(\mathbf{q},\bar{t})+ \\
     W^{n_1 n_2m_1m_2}_{\mathbf{k},\mathbf{k}''}(\mathbf{q},\bar{t})  \sum_{S} \dfrac{A^{m_1m_2}_{S,\alpha\alpha'}(\mathbf{k''},\mathbf{q},\bar{t})\tilde{A}^{m_3m_4}_{S,\alpha \alpha'}(\mathbf{k}''',\mathbf{q},\bar{t})~}{\dfrac{\partial{\lambda^{S,\alpha\alpha'}_{\mathbf{q}}}(\bar{t},\omega)}{\partial \omega}\Big|_{\omega=\Omega^{S\alpha\alpha'}_\mathbf{q}(\bar{t})}(\omega-\Omega^{S,\alpha\alpha'}_{\mathbf{q}}(\bar{t}))}~W^{m_3m_4n_1 n_2}_{\mathbf{k'''},\mathbf{k'}}(\mathbf{q},\bar{t})\label{Tbar_poles}
     \end{gathered}
\end{equation}
\end{widetext}

Since the aim is to evaluate the self-energy close to the $T$ matrix pole, I only retain the frequency-dependent second term in Eq.~\ref{Tbar_poles}, discarding the first term, and proceed to define the \textit{electron-magnon coupling}: 

\begin{equation}
    m^{S}_{n_1n_2,\alpha\alpha'}(\mathbf{k},\mathbf{q},\bar{t})=\dfrac{W_{\mathbf{k},\mathbf{k}''}^{n_1n_2m_1m_2}(\mathbf{q},\bar{t})A_{S,\alpha\alpha'}^{m_1m_2}(\mathbf{k''},\mathbf{q},\bar{t})}{\sqrt{\frac{\partial\lambda^{S,\alpha\alpha'}_{\mathbf{q}}(\bar{t},\omega)}{\partial \omega}}\Big|_{\omega=\Omega_\mathbf{q}^{S,\alpha\alpha'}(\bar{t})}} \label{electron-magnon}
\end{equation}

and recast Eq.~\ref{Tbar_poles} in the simpler form:

\begin{equation}
\begin{gathered}
       T^{n_1 n_2n_1n_2}_{\alpha \alpha'}(\mathbf{k},\mathbf{k}',\mathbf{q},\bar{t},\omega)  = \\
       \sum_{S} \dfrac{m^{S}_{n_1n_2,\alpha\alpha'}(\mathbf{k},\mathbf{q},\bar{t})  m^{S*}_{n_1n_2,\alpha\alpha'}(\mathbf{k'},\mathbf{q},\bar{t})}{\omega-\Omega^{S,\alpha\alpha'}_{\mathbf{q}}(\bar{t})+i\eta}\label{Tbar_poles2}
       \end{gathered}
\end{equation}

where I assume real isolated poles and add an infinitesimal positive broadening $\eta=0^+$. In the assumption of Hermitianity, left and right eigenvectors are related though $\tilde{A}=A^*$. I make another simplifying assumption, namely that the magnon amplitude $A^{n_1n_2}_{S,\alpha\alpha'}(\mathbf{k},\mathbf{q})$ is time independent. These approximations to the $T$ matrix allow one to reformulate the $GT$ self-energy as an effective electron-boson coupling, whose effect in the one-body electron density matrix evolution can be understood in analogy to the electron-phonon Fan-Migdal term~\cite{Stefanucci2024,Mocatti2026}: indeed, I note that the $T$ matrix pole has the form of a boson propagator for an infinitely long-lived quasiparticle: 

\begin{equation}
    D_{\mathbf{q},S}^{m,\alpha\alpha'}(\bar{t},\omega) = \dfrac{1}{\omega-\Omega_\mathbf{q}^{S,\alpha\alpha'}(\bar{t})+i \eta} \label{eq:Dmag}
\end{equation}

 With this definition for the bosonic propagator, the $GT$ self-energy in Bloch space now takes the form:

\vspace{2.5pt}\begin{widetext}
\begin{equation}
\begin{gathered}
\Sigma^{GT,n_1 n_1}_{\mathbf{k},\alpha}(\bar{t},\tau) = -\dfrac{i}{\Omega_{BZ}}\int_{\Omega_{BZ}} d \mathbf{q} ~ G^{n_2 n_2}_{\mathbf{k}-\mathbf{q},\alpha'}(\bar{t},\tau)~T^{n_1 n_2 n_1 n_2}_{\alpha \alpha'}(\mathbf{k},\mathbf{k},\mathbf{q},\bar{t},\tau)= \\
-\dfrac{i}{\Omega_{BZ}}\int_{\Omega_{BZ}} d \mathbf{q}~ G^{n_2 n_2}_{\mathbf{k}-\mathbf{q},\alpha'}(\bar{t},\tau)~\sum_{S} |m^{S}_{n_1n_2,\alpha\alpha'}(\mathbf{k},\mathbf{q},\bar{t})|^2 D^{m,\alpha \alpha'}_{\mathbf{q},S}(\bar{t},\tau)
\end{gathered}
\end{equation}
\end{widetext}

where $D^{m,\alpha \alpha'}_{\mathbf{q},S}(\bar{t},\tau)$ is the Fourier antitransform in $\omega$ of Eq.~\ref{eq:Dmag}, and I introduced the Brillouin zone volume $\Omega_{BZ}$.\\

\vspace{2pt}\subsection{Effect of noncollinearity in the $GT$ self-energy}

Now that an insight about the $GT$ self-energy in the electron-boson approximation has been developed, I discuss noncollinear effects in spinor space. The purely transverse character of magnon excitations is generally lost in the noncollinear case, where spin can flip during propagation and at any interaction vertex. I start from Eq.~\ref{$T$ matrix} and consider once again a simplified ``charge-only'' form for $W$, as already done for the $GW$ case, $W_{\alpha \beta \gamma \delta}^<(1,2) = \hat{\sigma}^0_{\alpha \gamma} W_{00}^<(1,2)\hat{\sigma}^0_{\beta \delta}$. In this approximation, the $T$ matrix equation reads:

\vspace{2.5pt}\begin{widetext}
\begin{equation}
    T_{\alpha\beta\gamma\delta}(1,2|3,4) = \sigma_{\alpha\gamma}^0 W_{00}(1,2)\sigma_{\beta\delta}^0 + \hat{\sigma}^0_{\alpha \eta}W_{00}(1,2)\hat{\sigma}^0_{\beta\theta}K_{\eta\theta\mu\nu}(1,2|1',2')T_{\mu\nu\gamma\delta}(1',2'|3,4)
\end{equation}
\end{widetext}

Even if the screened Coulomb lines does not flip spins, spin can still flip during propagation (off-diagonal terms in the electron propagator in spinor space) at any point between interaction lines in the  $T$ matrix ladder. The consequence is made evident if I explicitly rewrite the $GT$ self-energy Eq.~\ref{GTselfen1} in the form of Eq.~\ref{operator}, namely:

\begin{equation}
    \Sigma^{GT}_{\alpha \gamma} = \Sigma^{GT}_0 \hat{\sigma}^0_{\alpha \gamma} + \Sigma^{GT}_x\hat{\sigma}^x_{\alpha \gamma}+\Sigma^{GT}_y\hat{\sigma}^y_{\alpha \gamma}+\Sigma^{GT}_z\hat{\sigma}^z_{\alpha \gamma}
\end{equation}

It is possible to divide the contributing processes in two classes, namely the ones with an odd number of spin flips between 1 and 3, which will contribute to the off-diagonal $\alpha\neq\gamma$ part of the self-energy, and the ones with an even number of spin flips, contributing to the diagonal $\alpha=\gamma$ term. In this classification, the collinear case considered in the previous discussion belongs to the diagonal part, since no spin flip was possible during propagation. The off-diagonal contributions are the ones that break $SU(2)$ symmetry, gapping the Goldstone mode in the $\mathbf{q}\to0$ limit. Differently from the collinear case, it is not possible to strictly isolate a transverse channel and a longitudinal channel, see also the discussions in Refs.~\cite{PhysRevB.103.155152,PhysRevLett.127.166402,gbpw-zh1v}. All poles of the $T$ matrix can in principle mix, and are obtained from the zeros of the now $4\times4$ matrix determinant (in spinor space)

\begin{equation}
    \det[K^{-1}-W]_{\alpha \beta \gamma \delta}=0 \label{eqdet}
\end{equation}

which in principle can describe collective modes in any type of noncollinear magnetic configuration. In a ferromagnet,  assuming that one has a way to find solutions of Eq.~\ref{eqdet}, it would still be possible to identify ``quasi-transverse'' modes by calculating their spin expectation value and select the ones with largest spin component, see also the discussion in Sec.~\ref{secmag}.

\vspace{2pt}\subsection{The transverse limit}

Purely transverse excitations can be recovered under suitable approximations for the $GT$ self-energy, by forcing the spin of the electron Green's function to be transverse, $i.e.$: 

\begin{equation}
\Sigma_{\alpha \gamma}^{GT}(1,3) = -i G_{\bar{\alpha}\bar{\gamma}}(2,4)T_{\alpha \bar{\alpha}, \gamma \bar{\gamma}} (1,2|3,4)  
\end{equation}

where $\bar{\alpha}$ is constrained to be the opposite of $\alpha$, $i.e.$ 
\begin{equation}
    \begin{cases}
    \alpha=\downarrow \enspace \Rightarrow  \enspace \bar{\alpha}=\uparrow\\ 
    \alpha=\uparrow \enspace \Rightarrow \enspace \bar{\alpha}=\downarrow
\end{cases}
\end{equation}

and same for $\gamma$. Since the screened interaction does not allow for spin flips in its charge-only form, the self-energy in this case is 

\vspace{2.5pt}\begin{widetext}
\begin{equation}
    T_{\alpha\bar{\alpha}\gamma\bar{\gamma}}(1,2|3,4) = \sigma_{\alpha\gamma}^0 W_{00}(1,2)\sigma_{\bar{\alpha}\bar{\gamma}}^0 + \hat{\sigma}^0_{\alpha \eta}W_{00}(1,2)\hat{\sigma}^0_{\bar{\alpha}\bar{\eta}}K_{\eta\bar{\eta}\mu\bar{\mu}}(1,2|1',2')T_{\mu\bar{\mu}\gamma\bar{\gamma}}(1',2'|3,4) \label{eq:negT}
\end{equation}
\end{widetext}

 i.e., while spin can flip during propagation, the two spins $\alpha$ and $\alpha'$ are constrained to be opposite at any time in the $T$ matrix. This self-energy automatically reduces to the usual transverse particle-hole $T$ matrix of the collinear case if spin-orbit coupling is not present. An even more drastic approximation, justified if SOC is weak, would be to force no spin flips in the $GT$ self-energy, i.e. only consider contributions of the form $\alpha = \eta = \mu = \gamma$ in Eq.~\ref{eq:negT}, recovering the same self-energy of the collinear case, $\Sigma_{\alpha}^{GT}$. Physically, this amounts to assume that SOC is weak enough that no spin-flip is possible during the ladder propagation, effectively assuming that the electron-boson interaction is much larger than SOC. Within this spin-diagonal approximation, the $GT$ self-energy in the noncollinear case becomes:

\vspace{2.5pt}\begin{widetext}
\begin{equation}
\begin{gathered}
\Sigma^{GT,n_1 n_1}_{\mathbf{k},\alpha \alpha}(\bar{t},\tau) = -\dfrac{i}{\Omega_{BZ}}\int_{\Omega_{BZ}} d \mathbf{q} ~ G^{n_2 n_2}_{\mathbf{k}-\mathbf{q},\bar{\alpha}\bar{\alpha}}(\bar{t},\tau)~T^{n_1 n_2 n_1 n_2}_{\alpha \bar{\alpha}\alpha\bar{\alpha}}(\mathbf{k},\mathbf{k},\mathbf{q},\bar{t},\tau)= \\
-\dfrac{i}{\Omega_{BZ}}\int_{\Omega_{BZ}} d \mathbf{q}~ G^{n_2 n_2}_{\mathbf{k}-\mathbf{q},\bar{\alpha}\bar{\alpha}}(\bar{t},\tau)~\sum_{S} |m^{S}_{n_1n_2,\alpha\bar{\alpha}}(\mathbf{k},\mathbf{q},\bar{t})|^2 D^{m,\alpha\alpha'}_{\mathbf{q},S}(\bar{t},\tau) \label{GTselfen2}
\end{gathered}
\end{equation}
\end{widetext}

The $GT$ self-energy in the form of Eq.~\ref{GTselfen2} in the purely transverse limit is the starting point to treat the emerging electron-boson coupling. In analogy to the $GW$ and Fan-Migdal self-energies~\cite{Mocatti2026,Stefanucci2024}, its contour representation can be decomposed into a singular contribution, which determines the quasiparticle properties, and a lesser/greater component that will contribute a new term to the scattering integrals in the right-hand side of Eq.~\ref{density_mat_rec}, which I later identify in the electron-magnon contribution. The effect of the singular part can be understood looking at the
spin-antisymmetric part:
\begin{equation}
\Sigma^{\delta,GT,z}_\mathbf{k}(\bar{t}) =\tfrac{1}{2}\bigl[
\Sigma_{\mathbf{k},
\uparrow\uparrow}^{\delta,GT}(\bar{t})
-\Sigma_{\mathbf{k},\downarrow\downarrow}^{\delta,GT}(\bar{t})
\bigr],
\label{eq:Sigmaz}
\end{equation}
and similarly for the the spin-symmetric part $\Sigma^{\delta,GT,0}_\mathbf{k} = \frac{1}{2}(
\Sigma^{\delta,GT}_{\mathbf{k},\uparrow\uparrow} + \Sigma^{\delta,GT}_{\mathbf{k},\downarrow\downarrow})$, which simply renormalizes the
quasiparticle dispersion. In matrix form:
\begin{equation}
\Sigma^{\delta,GT}_\mathbf{k}(\bar{t})
= \Sigma^{\delta,GT,0}_\mathbf{k}(\bar{t})\,\hat{\sigma}^0
+ \Sigma^{\delta,GT,z}_\mathbf{k}(\bar{t})\,\hat{\sigma}^{z}.
\label{eq:matrix_form}
\end{equation}

Thus, the singular part of the $GT$ self-energy will contribute to an effective magnetic field term in the quasiparticle Hamiltonian, which will be discussed later in Sec.~\ref{singular}, together with the other contributions to the singular part of the self-energy. It is important to stress that, even in the spin-diagonal approximation, an effective boson-induced spin flip is still possible due to the Elliott-Yafet mechanism in the presence of the spin-orbit coupling ($i.e.$, the diagonal self-energy still contributes to off-diagonal components of the Green's function). Once an approximation for the $GT$ self-energy has been chosen, it is possible to express it in the basis of spin orbitals through the basis change $U$, defined in Eq.~\ref{eq:basis_change}:

 \begin{equation}
     \Sigma_{\mathbf{k},\lambda \lambda'}^{GT}(t,t') = U^{\mathbf{k}\lambda*}_{n \alpha}\Sigma^{GT,nn'}_{\mathbf{k},\alpha\gamma}(t,t')U^{\mathbf{k}\lambda'}_{n'\gamma}
 \end{equation}

In the spin-orbital basis, the self-energy maintains the same structure, as can be explicitly shown by the insertion of the $UU^\dagger$ identity between the Green's function and the vertex and between the two vertices, obtaining:

\vspace{2.5pt}\begin{widetext}
\begin{equation}
\begin{gathered}
    \Sigma_\mathbf{k}^{GT,\lambda\lambda}(\bar{t},\tau) = -\dfrac{i}{\Omega_{BZ}}\int_{\Omega_{BZ}} d \mathbf{q}~ G_{\mathbf{k}-\mathbf{q},\lambda'\lambda'}(\bar{t},\tau)~ T^{\lambda\lambda'\lambda\lambda'}(\mathbf{k},\mathbf{k},\mathbf{q},\bar{t},\tau)=\\
-\dfrac{i}{\Omega_{BZ}}\int_{\Omega_{BZ}} d \mathbf{q}~ G_{\mathbf{k}-\mathbf{q},\lambda'\lambda'}(\bar{t},\tau)~\sum_{S} |m^{S}_{\lambda \lambda'}(\mathbf{k},\mathbf{q},\bar{t})|^2 D^{m}_{\mathbf{q},S}(\bar{t},\tau) \label{GTselfen3}
\end{gathered}
\end{equation}
\end{widetext}

I suppressed the explicit $\alpha, \alpha'$ (or $\lambda, \lambda')$ dependence from the $D$ propagator: the reason will become evident in Sec.~\ref{magKB}, where I discuss the magnon propagator in detail. For simplicity, in the rest of the work I will employ the transverse approximation to the $GT$ self-energy, Eq.~\ref{GTselfen3}. The connection between the obtained $GT$ self-energy and formulations of electron-magnon scattering in itinerant ferromagnets in the context of the Hubbard model is discussed in Appendix E. 

\section{Magnon propagator and magnon Kadanoff-Baym equation}
\label{magKB}

Since the only two-time function in the $T$ matrix in Eq.~\ref{GTselfen3} is the bosonic propagator, the lesser $GT$ self-energy in this approximation is simply:

\begin{equation}
\begin{gathered}
    \Sigma^{<,GT}_{\mathbf{k},\lambda\lambda}(\bar{t},\tau) =  -\dfrac{i}{\Omega_{BZ}} \times\\\int_{\Omega_{BZ}} d \mathbf{q}~ G^{<}_{\mathbf{k}-\mathbf{q},\lambda'\lambda'}(\bar{t},\tau)~\sum_{S} |m^{S}_{\lambda \lambda'}(\mathbf{k},\mathbf{q},\bar{t})|^2 D^{m,<}_{\mathbf{q},S}(\bar{t},\tau)
    \end{gathered}
\end{equation}

The one-body electronic density matrix evolution is thus coupled to the new boson appearing in the lesser $GT$ self-energy, and an additional equation of motion for the new boson is necessary to close the system of equations. As I will now discuss, this new boson is the magnon. In what follows I show how this additional equation can be practically written for the ferromagnetic case with purely transverse magnons, while the  general mixed longitudinal-transverse case is left for future work. The boson propagator appearing in the $GT$ self-energy in the purely transverse limit inherits its Dyson equation from the BSE defining the $T$ matrix in the pole approximation, Eq.~\ref{Tbar_poles}. Namely: 

\begin{equation}
D^m_{\mathbf{q}}(\bar{t},\omega) = D^m_{0,\mathbf{q}}(\bar{t})+D^m_{0,\mathbf{q}}(\bar{t})\Pi_{\mathbf{q}}(\bar{t},\omega) D^m_{\mathbf{q}}(\bar{t},\omega)  \label{dyson_D}  
\end{equation}

where I identify the bare propagator $D_0 = m^{-1}W\tilde{m}^{-1}$, the magnon-electron self-energy $\Pi= mK\tilde{m}$ and $D = m^{-1}T\tilde{m}^{-1}$ is the dressed magnon propagator. Incidentally, I note that it would be possible to systematically go beyond this level of theory by considering other diagrams beyond the ladder series in the $T$ matrix. A ``dressed'' $T$ matrix could include phonon insertions, resulting in an effective magnon-phonon coupling, to be derived from first principles similarly to what it has been done for exciton-phonon coupling~\cite{PhysRevResearch.2.012032,PhysRevB.102.045136,PhysRevB.108.165101}, as well as magnon-magnon processes, with the understanding that care must be taken to avoid double counting. Before proceeding to the formal definition of a magnon propagator, it is necessary to make some additional considerations on the $GT$ self-energy. I assume for definiteness that the system under investigation has spin up majority. The majority spin self-energy $\Sigma_{\uparrow}^{GT}$ will describe a majority electron absorbing a boson, flipping to minority down spin, then re-emitting the boson and flipping back up. Explicitly: 

\vspace{2.5pt}\begin{widetext}
\begin{equation}
\begin{gathered}
\Sigma^{n_1n_1}_{\mathbf{k},\uparrow \uparrow}(\bar{t},\tau) = -\dfrac{i}{\Omega_{BZ}}\int_{\Omega_{BZ}} d \mathbf{q} ~ G^{n_2n_2}_{\mathbf{k}-\mathbf{q},\downarrow \downarrow}(\bar{t},\tau)~T^{n_1 n_2 n_1 n_2}_{\uparrow \downarrow \uparrow \downarrow}(\mathbf{k},\mathbf{k},\mathbf{q},\bar{t},\tau)= \\
-\dfrac{i}{\Omega_{BZ}}\int_{\Omega_{BZ}} d \mathbf{q}~ G^{n_2 n_2}_{\mathbf{k}-\mathbf{q}\downarrow \downarrow}(\bar{t},\tau)~\sum_{S} |m^{S}_{n_1n_2,\uparrow\downarrow}(\mathbf{k},\mathbf{q},\bar{t})|^2 D^{m,\uparrow \downarrow}_{\mathbf{q},S}(\bar{t},\tau)
\end{gathered}
\end{equation}

Conversely, the minority spin self-energy describes a minority electron emitting a boson, flipping to majority up spin, then the re-absorbing the boson and flipping back down:

\begin{equation}
\begin{gathered}
\Sigma^{n_1 n_1}_{\mathbf{k},\downarrow \downarrow}(\bar{t},\tau) = -\dfrac{i}{\Omega_{BZ}}\int_{\Omega_{BZ}} d \mathbf{q} ~ G^{n_2 n_2}_{\mathbf{k}-\mathbf{q},\uparrow \uparrow}(\bar{t},\tau)~T^{n_1 n_2 n_1 n_2}_{\downarrow \uparrow \downarrow \uparrow}(\mathbf{k},\mathbf{k},\mathbf{q},\bar{t},\tau)= \\
-\dfrac{i}{\Omega_{BZ}}\int_{\Omega_{BZ}} d \mathbf{q}~ G^{n_2 n_2}_{\mathbf{k}-\mathbf{q},\uparrow \uparrow}(\bar{t},\tau)~\sum_{S} |m^{S}_{n_1n_2,\downarrow\uparrow}(\mathbf{k},\mathbf{q},\bar{t})|^2 D^{m,\downarrow \uparrow}_{\mathbf{q},S}(\bar{t},\tau)
\end{gathered}
\end{equation}
\end{widetext}

 The above discussed magnon absorption and emission processes are the so-called Stoner processes. The anti-Stoner processes, consisting of magnon emission from spin-majority electrons and magnon absorption from spin-minority electrons, are also possible in the presence of SOC. A fundamental symmetry of the retarded boson propagators entering the majority and minority $GT$ self-energies is (see Appendix F):

\begin{equation}
     D^{R,m,\alpha \alpha'}_{\mathbf{q},S}(\bar{t},\omega) = [D^{R,m,\alpha' \alpha}_{-\mathbf{q},S}(\bar{t},-\omega)]^* \label{propD}
\end{equation}

Thus, both electron-boson self energies couple the electron to the same emergent bosonic quasiparticle, the magnon, defined in this context as the boson that carries an up-to-down electron spin flip.  Correspondingly, I define magnon annihilation and creation operator $\hat{b}_{\mathbf{q}S}(t)$ and $\hat{b}^\dagger_{\mathbf{q}S}(t)$:

\begin{equation}
    \begin{pmatrix}
        \hat{b}_{\mathbf{q}S}(t)\\
        \hat{b}^\dagger_{\mathbf{q}S}(t)
    \end{pmatrix}
\end{equation}
 
Note that, in principle, these operators may have non-zero expectation values in the presence of coherent bosons. For this reason, it is desirable to define the fluctuation operators in analogy with phonons, as:

\begin{equation}
    \begin{pmatrix}
        \delta \hat{b}_{\mathbf{q}S}(t)  \\
      \delta \hat{b}^\dagger_{\mathbf{q}S}(t)
    \end{pmatrix} =\begin{pmatrix}
          \hat{b}_{\mathbf{q}S}(t) - \langle \hat{b}_{\mathbf{q}S}(t) \rangle  \\\hat{b}^\dagger_{\mathbf{q}S}(t)- \langle \hat{b}^\dagger_{\mathbf{q}S}(t) \rangle
    \end{pmatrix}
\end{equation}

Given the electron-boson coupling structure, a natural choice for the bosonic propagator is the following matrix form:

\begin{equation}
\begin{gathered}
    \mathbf{D}_{\mathbf{q},S}(t,t') = \begin{pmatrix}
        D^{11}_{\mathbf{q},S}(t,t') ~D^{12}_{\mathbf{q},S}(t,t')\\[1ex]
        D^{21}_{\mathbf{q},S}(t,t') ~D^{22}_{\mathbf{q},S}(t,t')
    \end{pmatrix} = \\
    \begin{pmatrix}
        \langle T \delta \hat{b}_{-\mathbf{q}S}^\dagger(t) ~\delta \hat{b}_{-\mathbf{q}S}(t')\rangle  \enspace\langle T \delta \hat{b}^\dagger_{-\mathbf{q}S}(t) ~\delta \hat{b}^\dagger_{\mathbf{q}S}(t') \rangle \\[1ex]
        \langle T \delta \hat{b}_{\mathbf{q}S}(t) ~\delta \hat{b}_{-\mathbf{q}S}(t')\rangle  \enspace \langle T \delta \hat{b}_{\mathbf{q}S}(t)~\delta \hat{b}^\dagger_{\mathbf{q}S}(t') \rangle \label{Dmat}
 \end{pmatrix}
    \end{gathered}
\end{equation}

This bosonic propagator form is different from the one presented in Ref.~\cite{Karlsson2021}, whose approach is natural in the case of phonons~\cite{Stefanucci2024}. For the magnon propagator, I find the present definition to be more convenient due to the form of the effective electron-magnon interaction, which is not proportional to the displacement operator. One recognizes in the $D_{11}$ component the $D^{\uparrow \downarrow}$ propagator and in the $D_{22}$ the $D^{\downarrow \uparrow }$ propagator.  Instead, the off-diagonal ``coherent'' components do not couple with electrons in these approximations and are assumed to be zero. Starting from Eq.~\ref{dyson_D}, the aim is to now obtain Kadanoff-Baym equations for magnons. The first step is to define an auxiliary bare quasiparticle propagator, since $D_0$ has no poles. I note that the pole  $\Omega_\mathbf{q}^{S,\alpha\alpha'}(\bar{t})$ in Eq.~\ref{eq:Dmag} is generally complex,  and is solution to the following transcendental equation:

\begin{equation}
  \omega  = \Omega_{0\mathbf{q}}^{S,\alpha\alpha'}(\bar{t})+ \Pi^{S,\alpha\alpha'}_{\mathbf{q}}(\bar{t},\omega)
\end{equation}

where $\Omega_{0\mathbf{q}}^{S,\alpha\alpha'}(\bar{t})=0$ since no pole is present in the absence of the interaction. I adopt an analogous approach to the one employed for phonons and electrons, namely I divide the magnon self-energy as follows:

\begin{equation}
      \Pi^{S,\alpha\alpha'}_{\mathbf{q}}(\bar{t},\omega)  =  \Pi^{\delta,S,\alpha\alpha'}_{\mathbf{q}}(\bar{t})+ \Delta \Pi^{S,\alpha\alpha'}_{\mathbf{q}}(\bar{t},\omega)
\end{equation}

where $\Pi^{\delta}$ is a purely real time-local part of the magnon self-energy, while $\Delta \Pi$ is the time non-local part, which is neglected in the $T$ matrix solution. This defines an infinitely-lived quasiparticle propagator: 

\begin{equation}
    \bar{D}_{\mathbf{q},0S}^{m,\alpha\alpha'}(\bar{t},\omega) = \dfrac{1}{\omega-\Pi^{\delta,S,\alpha\alpha'}_{\mathbf{q}}(\bar{t})+i\eta} \label{eq:Dauxb}
\end{equation}

The notation can be simplified by specializing Eq.~\ref{eq:Dauxb} in the two spin components and employing Eq.~\ref{propD}:

\begin{equation}
\begin{gathered}
\bar{D}_{\mathbf{q},0S}^{m,\downarrow\uparrow}(\bar{t},\omega) = \dfrac{1}{\omega-\Omega^{S}_{\mathbf{q}}(\bar{t})+i\eta} \\
\bar{D}_{-\mathbf{q},0S}^{m,\uparrow\downarrow}(\bar{t},\omega) = -\dfrac{1}{\omega+\Omega^{S}_{\mathbf{q}}(\bar{t})+i\eta}\\\label{eq:Dauxb2}
\end{gathered}
\end{equation}

where I defined $\Omega^{S}_{\mathbf{q}}(\bar{t}) =\Pi^{\delta,S,\alpha\alpha'}_{\mathbf{q}}(\bar{t}) = \Pi^{\delta,S,\alpha'\alpha}_{\mathbf{q}}(\bar{t})$. The quasiparticle will then be dressed by the time non-local self-energy $\Delta\Pi$. On the Keldysh contour, the magnon propagator preserves the same pole structure as the time-ordered propagator because the contour formalism only modifies  the time-ordering prescription, while the poles are still determined by the quasiparticle spectrum, which follow from the equation of motion and therefore remain unchanged on the contour. The contour magnon Green’s function is thus obtained by extending the real-time propagator to a contour-ordered object. The Dyson's equation for the magnon Eq.~\ref{dyson_D} can be expressed in terms of $\bar{D}_{0}$, since

\begin{equation}
    D^{-1} = \bar{D}_0^{-1}+\Delta \Pi
\end{equation}

and thus, in the Dyson equation form 

\begin{equation}
    D =  \bar{D}_0 + \bar{D}_0\Delta\Pi D \label{dyson_DD}
\end{equation}

The Dyson equation for the magnon propagator, Eq.~\ref{dyson_DD}, is the starting point for the magnon Kadanoff-Baym equations.  From the Fourier antitransform in $\omega$ of Eq.~\ref{eq:Dauxb2} extended to the contour, $\bar{D}_{0S}$ satisfies:

\begin{equation}
\begin{gathered}
    [i \partial_z - \Omega^{S}_{\mathbf{q}}(z)]\bar{D}^{m,\downarrow\uparrow}_{\mathbf{q},0S}(z,z') = \delta(z-z')\\
      [-i \partial_z - \Omega^{S}_{\mathbf{q}}(z)]\bar{D}^{m,\uparrow\downarrow}_{\mathbf{q},0S}(z,z') = \delta(z-z')\label{barD0}
      \end{gathered}
\end{equation}

which defines the auxiliary inverse bare magnon propagator. Through the application of the inverse propagator, Eq.~\ref{barD0}, on the Dyson's equation for the magnon, Eq.~\ref{dyson_DD}, one obtains the (left) Kadanoff-Baym equation for the magnon propagator, in matrix notation over magnon mode space:

\vspace{2.5pt}\begin{widetext}
\begin{equation}
\begin{gathered}
  [ -i \hat{\sigma}^z \partial_t - \mathbf{\Omega}_{\mathbf{q}}(t)]\mathbf{D}^{<,m}_{\mathbf{q}}(t,t')
   =\int_0^t dt'[\Delta\mathbf{\Pi}_\mathbf{q}^>(t,t')\mathbf{D}_\mathbf{q}^{m,<}(t',t)-\Delta\mathbf{\Pi}_{\mathbf{q}}^<(t,t')\mathbf{D}_{\mathbf{q}}^{m,>}(t',t)]+h.c.
   \label{KB_mag}
   \end{gathered}
\end{equation}
\end{widetext}

where, due to the canonical commutation relations inherent to magnon propagator, the coherent dynamics of the density matrix exhibit pseudo-Hermitian evolution~\cite{doi:10.1142/S0219887810004816}. Furthermore, I defined the diagonal matrix:

\begin{equation}
    \mathbf{\Omega}_{\mathbf{q}}(t) = \begin{pmatrix}
       \Omega_{\mathbf{q}}(t)  & 0 \\
         0 & \Omega_\mathbf{q}(t)
    \end{pmatrix}
\end{equation}

and all quantities should be intended as matrices in the magnon modes space. By an analogous procedure already employed for the electron Green's function and in Ref.~\cite{Stefanucci2024} for the phonon Green's function, an equation formally analogous to the one for the phonons can be formulated for the magnon density matrix (cfr. Eq.~\ref{density_mat_ph}):

\vspace{2.5pt}\begin{widetext}
\begin{equation}
\begin{gathered}
       -\partial_{t} \beta_\mathbf{q}^<(t)+i(\hat{\sigma}^z \,\mathbf{\Omega}_{\mathbf{q}}(t)\beta_\mathbf{q}^<(t)-\beta_{\mathbf{q}}^<(t)\mathbf{\Omega}_{\mathbf{q}}(t)\hat{\sigma}^z)\\
   =\hat{\sigma}^z \int_0^t dt'[\Delta\mathbf{\Pi}_\mathbf{q}^>(t,t')\mathbf{D}_\mathbf{q}^{m,<}(t',t)-\Delta\mathbf{\Pi}_{\mathbf{q}}^<(t,t')\mathbf{D}_{\mathbf{q}}^{m,>}(t',t)]+h.c.
   \label{density_mat_mag}
   \end{gathered}
\end{equation}
\end{widetext}

where I defined  $\beta^<_{\mathbf{q}}(t) = i \mathbf{D}_{\mathbf{q}}^<(t,t)$. This is the equation of motion for the one-body magnon density matrix, which closes the set of equations of motion for electrons, phonons and magnons.

\section{Mirrored generalized Kadanoff-Baym ansatz}

In order to reduce the complexity of Kadanoff-Baym equations and obtain practically solvable equations I resort to the mirrored generalized Kadanoff-Baym ansatz (MGKBA)~\cite{Schafer2002,Stefanucci2024}, expressing the lesser and greater Green's functions using a relation that is exact in the non-interacting case. I now discuss how the MGKBA can be extended for the case of electron, magnons and phonons in the noncollinear magnetic case. 

\vspace{2pt}\subsection{Spinor}

The MKGBA in the spinor case is taken to be of the same form as the one for the spinless case~\cite{Stefanucci2024}, and it is written, in matrix form:

\begin{equation}
    G_{\mathbf{k}}^\gtrless(t,t') \simeq -\rho^{\gtrless}_{\mathbf{k}}(t)G^R_{\mathbf{k}}(t,t')+G_{\mathbf{k}}^A(t,t')\rho^{\gtrless}_{\mathbf{k}}(t')
\end{equation}

The difference with he spinless case is in the fact that Green's function in spinor space is non-diagonal. The retarded Green's function is written in the quasiparticle approximation as:

\begin{equation}
    G_{\mathbf{k}\alpha\beta}^R (t,t') = -i\theta(t-t')T\Big\{\exp(-i\int_{t'}^t d\bar{t}~h^{qp}_{\alpha \beta}(\mathbf{k},\bar{t}))\Big\}
\end{equation}

where I introduced the time-ordered exponential operator. Additional care must be taken with respect the spinless case, since the quasiparticle Hamiltonian $\hat{H}^{qp}$ evaluated at $t$ and $t' \neq t$ don't generally commute.

It is convenient to now define the unitary transformation that diagonalizes $h^{qp,nn'}_{\alpha \beta}$, namely I consider a short time interval $\tau =t-t'$ where it is meaningful to adiabatically approximate the time evolution as follows:

\begin{equation}
\begin{gathered}
    U_T(t,t-\tau) = \Big\{\exp(-i\int_{t'}^t d\bar{t}~{h}^{qp}_{\alpha \beta}(\mathbf{k},\bar{t}))\Big\} \simeq\\ U^{\mathbf{k}\lambda}_{n\alpha }(t) [\exp(-i\Lambda_{\mathbf{k}}(t) \tau)]_{\lambda \lambda} (U^\dagger)^{\mathbf{k}\lambda}_{n \beta}(t)
    \end{gathered}
\end{equation}

where 
$\exp(-i\Lambda_{\mathbf{k}}(t)~\tau)$ is the diagonal matrix containing the exponential of the eigenvalues $e^{-i\varepsilon_{\lambda\mathbf{k}}(t)}$ while $U(t)$ is the unitary basis change matrix that instantly diagonalizes the quasiparticle Hamiltonian ( at $t=0$, this is the basis change of Eq.~\ref{eq:basis_change}). In this basis of Eq.~\ref{eq:basis_change}, the Hamiltonian at time $t$ will be generally non-diagonal, as I discuss in detail in Sec.~\ref{singular}.

In analogy with the spinless case, a  simplification consists then in assuming that quasiparticles remain band-diagonal along propagation,i.e. 
$G_{\mathbf{k},\lambda \lambda'}^{\gtrless}(t,t') = \delta_{\lambda \lambda'}G_{\mathbf{k},,\lambda \lambda}^{\gtrless}(t,t')$. In this case it is possible to write the lesser/greater propagators in the  form:

\begin{equation}
\begin{gathered}
    G^{\gtrless}_{\mathbf{k},\lambda\lambda}(t,t') =  i \exp(-i\varepsilon_{\mathbf{k}\lambda}({t})(t-t')) \times\\
   [\theta(t-t')f_{\mathbf{k}\lambda}(t)+\theta(t'-t)f_{\mathbf{k}\lambda}(t')]
\end{gathered}
\end{equation}

or, in the $\{\mathbf{k},n,\alpha\}$ basis:

\begin{equation}
      G^{\gtrless,n n'}_{\alpha \beta,\mathbf{k}}(t,t') = \\
   \sum_{\lambda \lambda'}U^{\mathbf{k}\lambda}_{\alpha n} G^{\gtrless}_{\mathbf{k},\lambda \lambda'}(t,t')U^{*,\mathbf{k}\lambda}_{\beta n'} 
\end{equation}

\vspace{2pt}\subsection{Magnon}

The spinor equation for the electron is coupling to the bosonic Kadanoff-Baym equations for the magnon through the electron-magnon self-energy, Eq.~\ref{density_mat_mag}. In order to solve the coupled equations, a Kadanoff-Baym ansatz for the magnon is needed. 
The specific algebraic structure of the ansatz is fundamentally dictated by the requirement to preserve the equal-time boundary conditions. The proposed MGKBA for the magnon propagator is:

\begin{equation}
\begin{gathered}
    \mathbf{D}^{m,\gtrless}_{\mathbf{q}}(t,t') \simeq \\
    -\beta_{\mathbf{q}}^\gtrless(t) \hat{\sigma}^z \mathbf{D}^{m,R}_{\mathbf{q}}(t,t')+\mathbf{D}^{m,A}_{\mathbf{q}}(t,t')\hat{\sigma}^z \beta^{\gtrless}_{\mathbf{q}}(t')
    \end{gathered}
\end{equation}

together with the following ansatz for the retarded magnon propagator:

\begin{equation}
   \mathbf{D}^R_{\mathbf{q}} (t,t') = -i\theta(t-t')(-\hat{\sigma}^z)\mathcal{W}^m_{\mathbf{q}}(t)(\mathcal{W}^{m}_{\mathbf{q}}(t))^{-1}
\end{equation}

where I defined

\begin{equation}
\mathcal{W}^m_{\mathbf{q}}(t) = T ~\mathrm{exp}(-i\int_0^t d\bar{t} ~\mathbf{\Omega}_\mathbf{q}(\bar{t})(-\hat{\sigma}^z~))\quad.
\end{equation}

The simple diagonal form of the commutator metric, combined with the adiabatic time dependent of $\mathbf{\Omega}_{\mathbf{q}}(t)$ immediately gives the diagonal expression for the retarded magnon propagator:

\begin{widetext}
\begin{equation}
 \mathbf{D}^R_\mathbf{q}(t,t') = i \theta(t-t')~\hat{\sigma}^z \begin{pmatrix}\exp( ~{i \Omega}_\mathbf{q}(t)(t-t')) & 0\\ 0 & \exp( -i{\Omega}_\mathbf{q}(t)(t-t')) \end{pmatrix}
\end{equation}
\end{widetext}

The lesser/greater magnon propagators in the MGKBA correctly reduce to the equal-time density matrix, $\mathbf{D}_{\mathbf{q}}^{m,\gtrless}(t, t) = -i\beta_{\mathbf{q}}^{\gtrless}(t)$. To verify it, I consider the equal-time limit of the retarded and advanced propagators:

\begin{equation}
    \mathbf{D}_{\mathbf{q}}^{m,R}(t, t) = i\hat{\sigma}^z, \quad ; \quad \mathbf{D}_{\mathbf{q}}^{m,A}(t, t) = -i\hat{\sigma}^z.
\end{equation}

In forward-time propagation ($t > t'$), the advanced component vanishes, reducing the ansatz to:
\begin{equation}
    \mathbf{D}_{\mathbf{q}}^{m, \gtrless}(t, t') \simeq -\beta_{\mathbf{q}}^{\gtrless}(t) \hat{\sigma}^z \mathbf{D}_{\mathbf{q}}^{m,R}(t, t').
\end{equation}

and substituting the value of $\mathbf{D}^{m,R}_{\mathbf{q}}(t,t)$ in the equal-time limit correctly gives 

\begin{equation}
    \mathbf{D}_{\mathbf{q}}^{m, \gtrless}(t, t) = -\beta_{\mathbf{q}}^{\gtrless}(t) \hat{\sigma}^z (i\hat{\sigma}^z) = -i\beta_{\mathbf{q}}^{\gtrless}(t)
\end{equation}

\vspace{2pt}\subsection{Phonon}

For the phonon propagator, there is no formal difference in the noncollinear magnetic case with respect to the collinear nonmagnetic case, and I can adopt the same procedure explained in Ref.~\cite{Stefanucci2024}, and employ the MGKBA in the form of Ref.~\cite{Stefanucci2024}:

\begin{equation}
    \mathcal{L}^{\gtrless}_{\mathbf{Q}}(t,t')\simeq \gamma^\gtrless_{\mathbf{Q}}(t)\mathcal{J}\mathcal{L}^R_{\mathbf{Q}}(t,t')-\mathcal{L}^A_{\mathbf{Q}}(t,t')\mathcal{J}\gamma^{\gtrless}_{\mathbf{Q}}(t)
\end{equation}

accompanied by the following quasiparticle ansatz for the retarded propagator:

\begin{equation}
   \mathcal{L}^R_{\mathbf{Q}} (t,t') = -i\theta(t-t')\mathcal{J}\mathcal{W}_{\mathbf{Q}}(t)\mathcal{W}^{-1}_{\mathbf{Q}}(t)
\end{equation}

where 

\begin{equation}
\mathcal{W}_{\mathbf{Q}}(t) = T ~\mathrm{exp}(-i\int_0^t d\bar{t} ~h^{ph}_{qp}(\mathbf{Q},\bar{t})\mathcal{J}~)\quad.
\end{equation}

The singular part of the self-energy appearing in $h^{ph}_{qp}(\mathbf{Q},\bar{t})$, Eq.~\ref{hqp}, is chosen as the adiabatic phonon self-energy~\cite{Calandra2010,Caldarelli2025,PhysRevB.111.024307,PhysRevX.13.041009,Mocatti2026}, to recover a phonon quasiparticle description consistent with the equilibrium BO normal modes at $t=0$. In the no-mixing approximation employed here, $h^{ph}_{qp}(\mathbf{Q},\bar{t})$ remains diagonal at any time in the BO basis, and the retarded phonon propagator is~\cite{Stefanucci2024,Stefanucci_van_Leeuwen_2025}:

\begin{widetext}
\begin{equation}
\begin{gathered}
   \mathcal{L}_{\mathbf{Q},\nu\nu'}^R(t,t')= i \delta_{\nu\nu'}\dfrac{\theta(t-t')}{2\omega_{\mathbf{Q}\nu}}\Big[ \exp(i\omega_{\mathbf{Q}\nu}(t) (t-t'))\begin{pmatrix}1 & -i\omega_{\mathbf{Q}\nu}(t)\\ i\omega_{\mathbf{Q}\nu}(t) & \omega^2_{\mathbf{Q}\nu}(t)
   \end{pmatrix} \\
   - \exp(-i\omega_{\mathbf{Q}\nu}(t) (t-t'))\begin{pmatrix}1 & i\omega_{\mathbf{Q}\nu}(t)\\ -i\omega_{\mathbf{Q}\nu}(t) & \omega^2_{\mathbf{Q}\nu}(t)
   \end{pmatrix}\Big]
\end{gathered}
\end{equation}
\end{widetext}

\section{Markov approximation and collision integrals}

Once the MGKBA has been performed, the Markov approximation consists in approximating the time integrals in the right-hand side of the density matrix evolution for electron, magnons and phonons in  Eqs.~\ref{density_mat_rec2},\ref{density_mat_mag} and \ref{density_mat_ph}. Having defined $\tau = t-t'$ and $\bar{t}=\frac{t+t'}{2}$, one assumes that the only $\tau$ dependence in the Green's function and the self-energy are of the oscillating exponential form $e^{iE\tau}$ and employs the identity

\begin{equation}
   \int_{-\infty}^\infty d\tau  ~e^{i E~ \tau} =   2\pi~\delta(E)
\end{equation}

to approximate the time integrals as 

\begin{equation}
   \int_0^t d\tau  ~e^{i \Omega_{qp}(\bar{t})~ \tau} \simeq   \pi~\delta(\Omega_{qp}(\bar{t}))
\end{equation}

where $\Omega_{qp}$ represents an adiabatically time-dependent quasiparticle energy. In this approximation, one neglects memory effects in the kernel, and formulates the problem in terms of a time-local real-time dynamics. This is a reasonably good approximation as long as time $t$ is long enough with respect to $1/\Omega_{qp}(\bar{t})$; the earliest
stage of the dynamics is correspondingly the least controlled. Nevertheless, the coherent evolution
generated by $h^{qp}$ is treated exactly in time and is unaffected: the approximation is on the scattering part only. Thus, the present 
framework should be expected to describe well the very early dynamics at the level of the
coherent terms, with the incoherent channels becoming quantitative only a bit later.  I note
that this approximation is consistent with those introduced earlier, since the
quasiparticle form assumed for $G^R$ in the MGKBA, the time-local approximation for
$W$, and the pole approximation for the $T$ matrix all rest on the same assumption
that the relevant excitations are well-defined quasiparticles, of which the Markov
approximation is the counterpart at the level of the time integration.

\begin{figure*}
    \centering
    \includegraphics[width=1\linewidth]{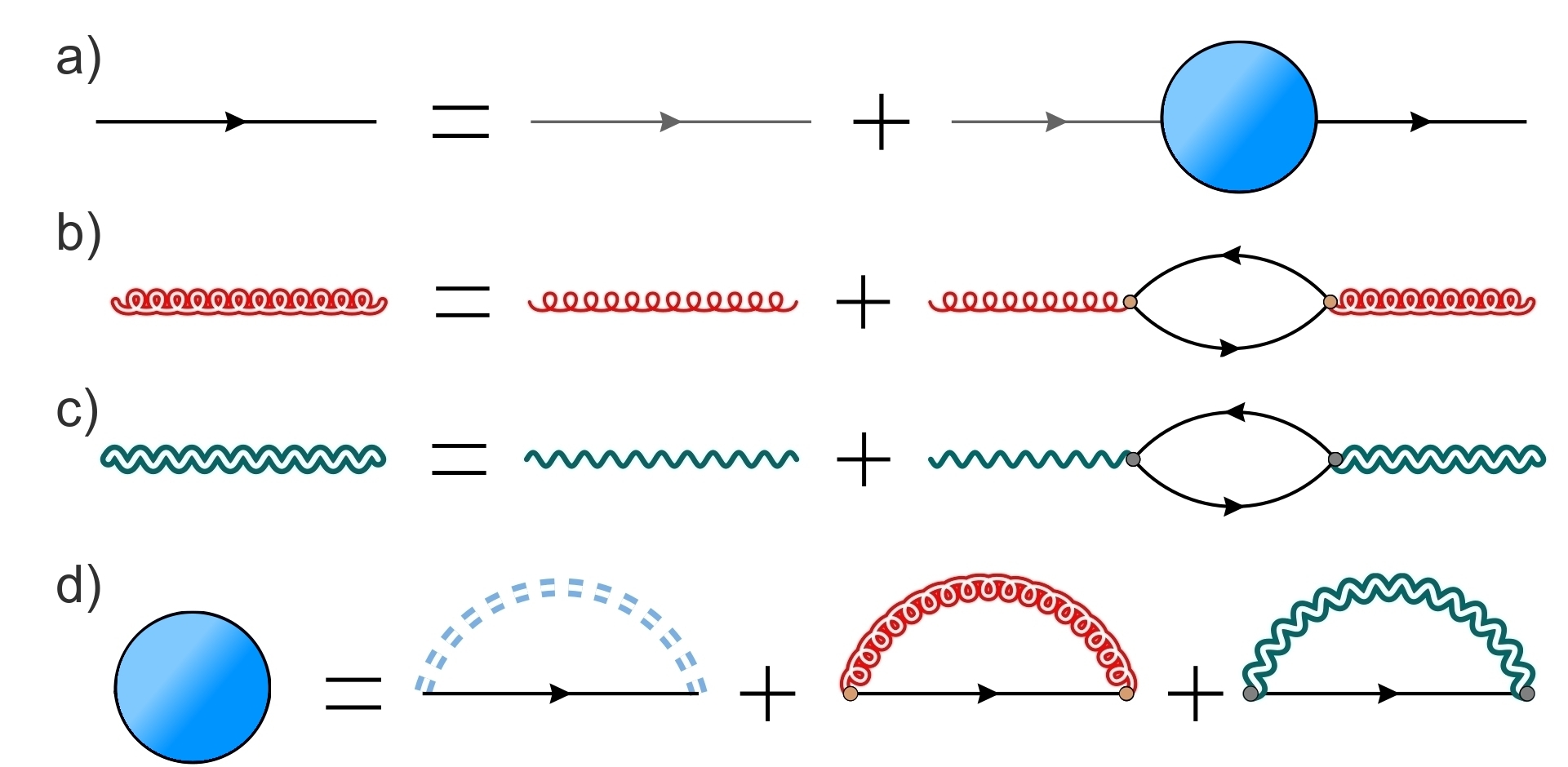}
    \caption{Diagrammatic representation~\cite{HARLANDER2020107465} of the electron (a), phonon (b) and magnon (c) Dyson equation. Panel (d): electron self-energy. The black arrow represents the dressed electron propagator, the grey arrow represents the bare electron propagator, the single wiggly red line represents the bare phonon propagator, the double wiggly red line represents the dressed phonon propagator, the orange vertex represents the screened electron-phonon coupling, the single wiggly green line represents the bare magnon propagator, the double wiggly green line represents the dressed magnon propagator, the grey vertex represents the electron-magnon interaction, the double light-blue dashed line represents the screened Coulomb interaction. }
    \label{figdyson}
\end{figure*}

 The logic up to this point can be summed up as follows: after choosing a self-energy for electrons, magnons and phonons, one proceeds to obtain Kadanoff-Baym equations from the Dyson equations show in Fig.~\ref{figdyson}. Then, through the use of the Markov approximation in combination with the Kadanoff-Baym ansatz, collision integrals terms in the electron, magnon and phonon density matrices, Eqs.~\ref{density_mat_rec2},\ref{density_mat_mag} and \ref{density_mat_ph}, are simplified. The obtained equations for the density matrices form a system of coupled differential equations, which I discuss below, starting from the collision integrals appearing in the right-hand side of the three equations. For the electron, three contributions are present, namely electron-electron, electron-phonon and electron-magnon:

\begin{equation}
    I^{el}(t) = I^{el-el}(t)+I^{el-ph}(t)+I^{el-mag}(t)
\end{equation}

For the magnon, I only consider the magnon-electron contribution:

\begin{equation}
        I^{mag}(t) = I^{mag-el}(t)
\end{equation}

while for the phonons one has the phonon-electron contribution:

\begin{equation}
    I^{ph}(t) = I^{ph-el}(t) 
\end{equation}

with the understanding that phonon-phonon scattering can also be included within this formalism if necessary, as discussed in Ref.~\cite{Mocatti2026}. Band off-diagonal scattering terms are instead approximated through a relaxation-time approximation. I define the time-dependent electronic occupations $f_{\mathbf{k}\lambda}(t) = \rho^<_{\mathbf{k}\lambda \lambda}(t)$, magnon occupations $N_{\mathbf{q}S}(t) =\beta^<_{\mathbf{q}SS}(t)$,  phonon occupations $n_{\mathbf{Q}\nu}(t) = \gamma^<_{\mathbf{Q}\nu\nu}(t)$ and electronic polarization $p_{\mathbf{k}\lambda\lambda'}(t) = \rho_{\mathbf{k}\lambda\lambda'}(t)$ from their respective density matrices. The diagonal electron-electron collision integral takes the form~\cite{Marini2013, Steinhoff2016, Stefanucci2024}:
\vspace{2.5pt}\begin{widetext}
\begin{equation}
\label{eq:elel_collision_int}
    \begin{gathered}
    I_{\mathbf{k}\lambda}^{el-el}(t)  = \dfrac{\pi}{N^2} \enspace \sum_{\mathclap{\substack{\mathbf{k'},\mathbf{q}  \lambda',\mu,\mu'}}} \big|W_{\mathbf{k},\mathbf{k}-\mathbf{q}}^{\lambda \mu \lambda'\mu'}(\mathbf{k'-k+q},t) - W_{\mathbf{k,k'+q}}^{\lambda \mu\mu'\lambda'}(-\mathbf{q},t)\big|^2 \times \\
     \big[f_{\mathbf{k-q}\mu}(t)f_{\mathbf{k'+q}\mu'}(t)(1-f_{\mathbf{k}\lambda}(t))(1-f_{\mathbf{k'}\lambda'}(t)) - 
     f_{\mathbf{k}\lambda}(t)f_{\mathbf{k'}\lambda'}(t)(1-f_{\mathbf{k-q}\mu}(t))(1-f_{\mathbf{k'+q}\mu'}(t))\big] \times \\
    \delta(\varepsilon_{\mathbf{k}\lambda}(t) + \varepsilon_{\mathbf{k'}\lambda'}(t) - \varepsilon_{\mathbf{k'+q}\mu'}(t) - \varepsilon_{\mathbf{k-q}\mu}(t)),
    \end{gathered}
  \end{equation}
  \end{widetext}

and the form of the vertex $\big|W_{\mathbf{k},\mathbf{k}-\mathbf{q}}^{\lambda \mu \lambda'\mu'}(\mathbf{k'-k+q},t) - W_{\mathbf{k,k'+q}}^{\lambda \mu\mu'\lambda'}(-\mathbf{q},t)\big|^2$ is due to the inclusion of the second-Born exchange diagram, see discussions in Refs.~\cite{Stefanucci2024,Mocatti2026}. The screened interaction $W$ generally has all magnetic components. It is important to remark that spin can be flipped due to electron-electron interaction even if $W$ is approximated as charge-only, see also the discussion about Elliott-Yafet electron-electron scattering in Ref.~\cite{7vr4-57mk}. The band off-diagonal collision integral is approximated using the relaxation-time ansatz:
\begin{equation}
\label{eq:rta_polarization}
    I_{\mathbf{k}\lambda\mu}(t) \simeq - \Gamma_{\mathbf{k}\lambda\mu}(t)p_{\mathbf{k}\lambda\mu}(t),
\end{equation}
where $\Gamma_{\mathbf{k}nm}(t)$ denotes the time-dependent dephasing rate. These rates are often treated as phenomenological quantities in the nonmagnetic case~\cite{Marini2008, Selig2016}. Here, they are written in the quasiparticle approximation~\cite{Stefanucci2024}:
\begin{equation}
\label{eq:RTA_quasiparticle_approximation}
    \Gamma_{\mathbf{k}\lambda\mu}(t) \simeq \Gamma_{\mathbf{k}\lambda}(t) + \Gamma_{\mathbf{k}\mu}(t),
\end{equation}
with $\Gamma_{\mathbf{k}\lambda}(t)$ denoting the relaxation rate of an electron in the $\lambda$th spin-orbital with momentum $\mathbf{k}$. The electron-electron relaxation rate  is then recast as:
\vspace{2.5pt}\begin{widetext}
\begin{equation}
\begin{gathered}
\label{eq:elel_dephasing_rate}
        \Gamma_{\mathbf{k}\lambda}^{el-el}(t) = \dfrac{\pi}{2N^2} \sum_{\mathclap{\substack{\mathbf{k'},\mathbf{q} \mu,\lambda',\mu'}}} \big|W_{\mathbf{k,}\mathbf{k}-\mathbf{q}}^{\lambda \mu \lambda'\mu'}(\mathbf{k'-k+q},t) - W_{\mathbf{k,k'+q}}^{\lambda \mu\mu'\lambda'}(-\mathbf{q},t)\big|^2 \\
         \big[f_{\mathbf{k'}\lambda'}(t)(1-f_{\mathbf{k-q}\mu}(t))(1-f_{\mathbf{k'+q}\mu'}(t)) + 
      f_{\mathbf{k-q}\mu}(t)f_{\mathbf{k'+q}\mu'}(t)(1-f_{\mathbf{k'}\lambda'}(t))\big] \times\\
     \delta(\varepsilon_{\mathbf{k}\lambda}(t) + \varepsilon_{\mathbf{k'}\lambda'}(t) - \varepsilon_{\mathbf{k'+q}\mu'}(t) - \varepsilon_{\mathbf{k-q}\mu}(t)).
     \end{gathered}
\end{equation}  
\end{widetext}

The explicit form of the electron-phonon collision term is derived from the full FM self-energy discussed earlier and in Ref.~\cite{Mocatti2026}. Within this approximation, the diagonal contribution to the collision integral reads~\cite{Marini2013, OMahony2019, Stefanucci2024}:

\begin{equation}
\label{eq:elph_collision_int}
\begin{aligned}
& I^{el-ph}_{\mathbf{k}\lambda}(t) = \dfrac{2\pi}{N} \sum_{\mathclap{\substack{\mathbf{Q} \\ \lambda',\nu}}} |g_{\lambda'\lambda}^\nu(\mathbf{k,Q})|^2 \Big\{  \\
& \big[f_{\mathbf{k+Q}\lambda'}(t)(1 - f_{\mathbf{k}\lambda}(t)) - n_{\mathbf{Q}\nu}(t)(f_{\mathbf{k}\lambda}(t) - f_{\mathbf{k+Q}\lambda'}(t))\big] \times \\ 
&\delta({\varepsilon}_{\mathbf{k+Q}\lambda'}(t) - {\varepsilon}_{\mathbf{k}\lambda}(t) - \omega_{\mathbf{Q}\nu}(t)) + \\
&\big[n_{\mathbf{q}\nu}(t)(f_{\mathbf{k+Q}\lambda'}(t) - f_{\mathbf{k}\lambda}(t)) - f_{\mathbf{k}\lambda}(t)(1-f_{\mathbf{k+Q}\lambda'}(t))\big] \times
\\
&\delta({\varepsilon}_{\mathbf{k+Q}\lambda'}(t) - {\varepsilon}_{\mathbf{k}\lambda}(t) + \omega_{\mathbf{Q}\nu}(t)) \Big\}.
\end{aligned}
\end{equation}

 In Eq.~\ref{eq:elph_collision_int}, $n_{\mathbf{Q}\nu}(t)$ and $\omega_{\mathbf{Q}\nu}(t)$ denote the time-dependent phonon occupation and frequency of mode $\nu$ with momentum $\mathbf{Q}$. Similarly, for the electron-magnon scattering one has:

\begin{equation}
\label{eq:elph_collision_int}
\begin{aligned}
& I^{el-mag}_{\mathbf{k}\lambda}(t) = \dfrac{2\pi}{N} \sum_{\mathclap{\substack{\mathbf{q} \\ \lambda',S}}} |m_{\lambda\lambda'}^S(\mathbf{k,q},t)|^2 \Big\{  \\
& \big[f_{\mathbf{k+q}\lambda'}(t)(1 - f_{\mathbf{k}\lambda}(t)) - N_{\mathbf{q}S}(t)(f_{\mathbf{k}\lambda}(t) - f_{\mathbf{k+q}\lambda'}(t))\big] \times \\ 
&\delta({\varepsilon}_{\mathbf{k+q}\lambda'}(t) - {\varepsilon}_{\mathbf{k}\lambda}(t) - \Omega_{\mathbf{q}S}(t)) + \\
&\big[N_{\mathbf{q}S}(t)(f_{\mathbf{k+q}\lambda'}(t) - f_{\mathbf{k}\lambda}(t)) - f_{\mathbf{k}\lambda}(t)(1-f_{\mathbf{k+q}\lambda'}(t))\big] \times
\\
&\delta({\varepsilon}_{\mathbf{k+q}\lambda'}(t) - {\varepsilon}_{\mathbf{k}\lambda}(t) + \Omega_{\mathbf{q}S}(t)) \Big\}.
\end{aligned}
\end{equation}

The off-diagonal component of the electron-phonon collision integral is obtained analogously to the electron-electron case and the corresponding quasiparticle decay rate is:
\begin{equation}
\begin{aligned}            
&{\Gamma}^{el-ph}_{\mathbf{k}\lambda}(t) = \dfrac{\pi}{N}\sum_{\mathclap{\substack{\mathbf{Q} \\ \lambda',\nu}}} |g_{\lambda'\lambda}^\nu(\mathbf{k},\mathbf{Q})|^2 \Big\{ 
\\
&\big[n_{\mathbf{Q}\nu}(t) + 1-f_{\mathbf{k+Q}\lambda'}(t)\big]\delta({\varepsilon}_{\mathbf{k+Q}\lambda'}(t) - {\varepsilon}_{\mathbf{k}\lambda}(t) + {\omega}_{\mathbf{Q}\nu}(t)) \\
& + \big[n_{\mathbf{Q}\nu}(t) + f_{\mathbf{k+Q}\lambda'}(t)\big]\delta({\varepsilon}_{\mathbf{k+Q}\lambda'}(t) - {\varepsilon}_{\mathbf{k}\lambda}(t) - {\omega}_{\mathbf{Q}\nu}(t))\Big\}.
\end{aligned}
    \label{eq:elph_dephasing_rate}
\end{equation}

A formally equivalent contribution appears due to electron-magnon scattering:

\begin{equation}
\begin{aligned}            
&{\Gamma}^{el-mag}_{\mathbf{k}\lambda}(t) = \dfrac{\pi}{N}\sum_{\mathclap{\substack{\mathbf{q} \\ \lambda',S}}} |m_{\lambda\lambda'}^S(\mathbf{k},\mathbf{q},t)|^2 \Big\{ 
\\
&\big[N_{\mathbf{q}S}(t) + 1-f_{\mathbf{k+q}\lambda'}(t)\big]\delta({\varepsilon}_{\mathbf{k+q}\lambda'}(t) - {\varepsilon}_{\mathbf{k}\lambda}(t) + {\Omega}_{\mathbf{q}S}(t)) \\
& + \big[N_{\mathbf{q}S}(t) + f_{\mathbf{k+q}\lambda'}(t)\big]\delta({\varepsilon}_{\mathbf{k+q}\lambda'}(t) - {\varepsilon}_{\mathbf{k}\lambda}(t) - {\Omega}_{\mathbf{q}S}(t))\Big\}.
\end{aligned}
    \label{eq:elph_dephasing_rate}
\end{equation}

For the phonon self-energy, I employ the approximated electron-hole bubble structure with two statically screened vertices~\cite{Calandra2010, Caldarelli2025, PhysRevB.111.024307,Mocatti2026,Ziman2001}:
\begin{align}
\label{eq:phel_collision_int}
&I^{ph-el}_{\mathbf{Q}\nu}(t) = -\dfrac{2\pi}{N}\sum_{\mathbf{k},\lambda',\lambda} |g_{\lambda'\lambda}^\nu(\mathbf{k},\mathbf{Q})|^2 \times \nonumber \\
&
\big[ n_{\mathbf{q}\nu}(t)(f_{\mathbf{k}\lambda}(t) - f_{\mathbf{k+Q}\lambda'}(t)) - f_{\mathbf{k+Q}\lambda'}(t)(1 - f_{\mathbf{k}\lambda}(t)) \big] \times \nonumber
\\
&\delta(\varepsilon_{\mathbf{k+Q}\lambda'}(t) - \varepsilon_{\mathbf{k}\lambda}(t) - \omega_{\mathbf{Q}\nu}(t))
\end{align}

\begin{figure*}
    \centering
    \includegraphics[width=1\linewidth]{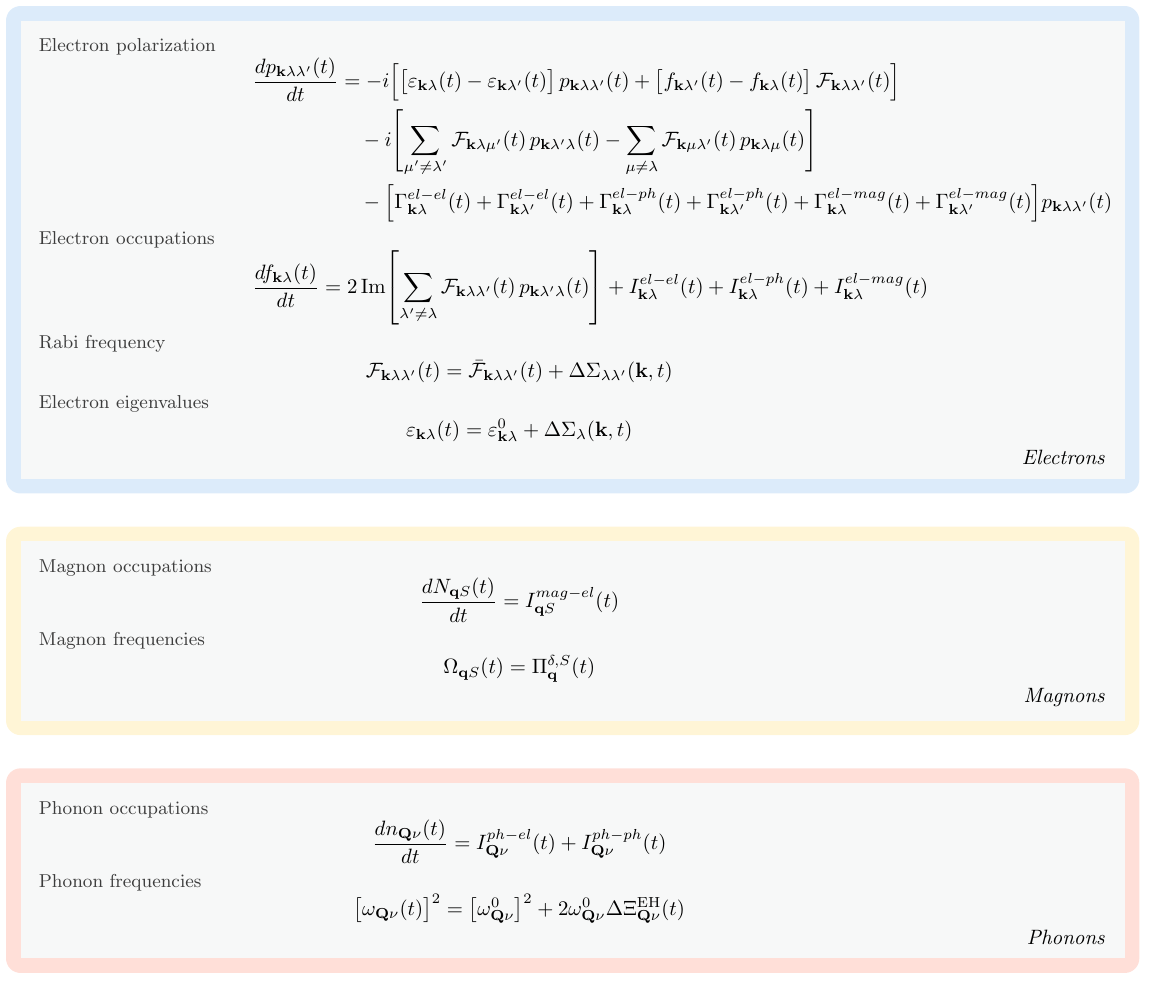}
    \caption{Equations of motion for coupled electrons (light-blue rectangle), magnons (yellow rectangle) and phonons (red rectangle).}
    \label{fig:EOM}
\end{figure*}

A formally equivalent expression is again obtained for the magnon-electron scattering integral, coming from the electron-hole bubble self-energy: 

\begin{align}
\label{eq:magel_collision_int}
&I^{mag-el}_{\mathbf{q}S}(t) = -\dfrac{2\pi}{N}\sum_{\mathbf{k},\lambda,\lambda'} |m_{\lambda \lambda'}^S(\mathbf{k},\mathbf{q},t)|^2 \times \nonumber \\
&
\big[ N_{\mathbf{q}S}(t)(f_{\mathbf{k}\lambda}(t) - f_{\mathbf{k+q}\lambda'}(t)) - f_{\mathbf{k+q}\lambda'}(t)(1 - f_{\mathbf{k}\lambda}(t)) \big] \times \nonumber
\\
&\delta(\varepsilon_{\mathbf{k+q}\lambda'}(t) - \varepsilon_{\mathbf{k}\lambda}(t) - \Omega_{\mathbf{q}S}(t))
\end{align}

I note that the spin selectivity in both electron-magnon  and magnon-electron terms is contained in the electron-magnon matrix elements:

\begin{equation}
    m^{S}_{\lambda\lambda'}(\mathbf{k},\mathbf{q},t) = (U^\dagger)^{\mathbf{k+q}\lambda}_{n \alpha} m_{nn'\alpha\alpha'}^{S}({\mathbf{k},\mathbf{q}},t)(U)^{\mathbf{k}\lambda'}_{n'\alpha'}
\end{equation}

\section{Quasiparticle renormalization terms and coherent dynamics}
\label{singular}

The quasiparticle electron Hamiltonian $h^{qp}_{\alpha \beta}(\mathbf{k},t)$ plays a central role for coherent magnetization dynamics. In order to shed light on its role, I perform the usual decomposition in terms of Pauli matrices: 

\begin{equation}
\begin{gathered}
    h^{qp,nn'}_{\alpha \beta}(\mathbf{k},t) = h^{qp,nn'}_{0}(\mathbf{k},t) \hat{\sigma}^0_{\alpha \beta} + h_x^{qp,nn'}(\mathbf{k},t)\hat{\sigma}^x_{\alpha \beta}+\\
    h_y^{qp,nn'}(\mathbf{k},t)\hat{\sigma}^y_{\alpha \beta}+h_{z}^{qp,nn'}(\mathbf{k},t)\hat{\sigma}^z_{\alpha \beta}\label{spinh}
\end{gathered}
\end{equation}

The idea, once again, is that the $\hat{\sigma}^0$ component of $\hat{H}^{qp}$ represents the scalar component also present in the non-magnetic case, while the other three components $h^{qp,nn'}_{x,y,z}$ represent a (generally time dependent) effective magnetic field. Thus, even at $t=0$, all components are generally present. I now discuss all the time-dependent contributions to the quasiparticle electron Hamiltonian one by one and establish to what channel they contribute, while keeping in mind that I perform the time evolution in the basis $\{\lambda ,\mathbf{k}\}$ that diagonalizes ${h}^{qp}(\mathbf{k},0)$, with the usual basis transformation defined by Eq.~\ref{eq:basis_change}, as previously discussed. For this reason, it is instructive to rewrite the quasiparticle Hamiltonian in this basis, in analogy to what we did in Ref.~\cite{Mocatti2026}, as follows:

\begin{equation}
    {h}^{qp}_{\lambda\lambda'}(\mathbf{k},t) = \varepsilon_{\mathbf{k}\lambda}^0+ \mathcal{\bar{F}}_{\mathbf{k}\lambda\lambda'}(t)+\Delta \Sigma_{\lambda\lambda'}(\mathbf{k},t)
\end{equation}

where I introduced $\Delta \Sigma_{\lambda\lambda'}(\mathbf{k},t) = \Sigma_{\mathbf{k},\lambda\lambda'}(t) - \Sigma_{\mathbf{k},\lambda\lambda}(0)$, with the understanding that $\Sigma_{\mathbf{k},\lambda\lambda}(0)$ enters in the definition of ${h}^{qp}(\mathbf{k},0)$. I will start by discussing the contributions that are diagonal in the $\{\lambda,\mathbf{k}\}$ basis, $\Delta \Sigma_{\lambda\lambda}(\mathbf{k},t) = \Delta \Sigma_{\lambda}(\mathbf{k},t)$. The idea is that both $\mathcal{\bar{F}}_{\mathbf{k}\lambda\lambda'}(t)$ and $\Delta \Sigma_{\lambda\lambda'}(\mathbf{k},t)$ will contribute both a scalar term, as in the non-magnetic case, and an effective time-dependent magnetic field, responsible for a coherent magnetic dynamics (see the decomposition in Eq.~\ref{spinh}). Namely, the effective band- and momentum-dependent magnetic field $\mathbf{B}^{\mathrm{eff}}(\mathbf{k},t)$ is defined as: 

\begin{equation}
     B^{\mathrm{eff}}_{i}(\mathbf{k},t) =\dfrac{{h}^{qp}_{i}(\mathbf{k},t)}{\mu_B}=  B^{\mathrm{eff}}_i(\mathbf{k},0)+\dfrac{\mathcal{\bar{F}}_{i\mathbf{k}}(t)+\Delta \Sigma_{i}(\mathbf{k},t)}{\mu_B}
\end{equation}

where $i=x,y,z$ is the component in spinor space, according to the decomposition Eq.~\ref{spinh}, $\mu_B = 1/2$ in Rydberg atomic units and $B^{\mathrm{eff}}_i(\mathbf{k},0)=\dfrac{h^{qp}_{i}(\mathbf{k},0)}{\mu_B}$. I first discuss the bare Rabi frequency $\bar{\mathcal{F}}_{\mathbf{k}\lambda\lambda'}(t)$. While this term only contributes to the scalar charge channel, it connects different spin-orbitals in a Elliott-Yafet fashion. For this reason, it activates a polarization and occupation dynamics that is nontrivial in the magnetic sense. The self-energy variation $\Delta\Sigma_{\lambda\lambda'}(\mathbf{k},t)$ is formed by a band-diagonal part, contributing to the renormalization of electronic eigenvalues, and an off-diagonal part, which renormalizes the bare Rabi frequency,  $\bar{\mathcal{F}}_{\mathbf{k}\lambda\lambda'}(t)$, and to which I assume only the $GW$ and $GT$ self-energies contribute. I first discuss the band-diagonal part of $\Delta \Sigma$. The considered time-dependent contributions consists of four terms: Hartree (H), static GW, Fan-Migdal (FM), and coherent atomic motion (AM) (I note that the inclusion of the AM contribution requires to also propagate equations of motion for ions, see Ref.~\cite{Mocatti2026}):

\begin{equation}
\begin{gathered}
\label{eq:Re_el_selfenergy}
    \Delta\Sigma_{\lambda}(\mathbf{k},t) = \Delta \Sigma_{\lambda}^{\text{H}}(\mathbf{k},t) + \Delta \Sigma_{\lambda}^{\text{AM}}(\mathbf{k},t)\\
    + \Delta \Sigma_{\lambda}^{\text{GW}}(\mathbf{k},t) + \Delta \Sigma_{\lambda}^{\text{FM}}(\mathbf{k},t)+\Delta\Sigma_{\lambda}^{GT}(\mathbf{k},t)
    \end{gathered}
\end{equation}

The explicit expressions for the H, FM and AM terms can be found in the SM of Ref.~\cite{Mocatti2026} and remain conceptually the same also in the magnetic case, with the understanding that, for the magnetic case, both the Fan-Migdal and the AM self-energies can in principle also contribute to the magnetic component ${h}^{qp}_{x,y,z}(\mathbf{k},t)$, since the electron phonon coupling is dressed and thus not purely in the scalar charge channel, rather it can carry magnetic components. In any case, it should be reasonable to assume that, in most cases, the magnetic contribution due to the FM and AM self-energies is relatively small. The idea is that most of the time-dependent renormalization of $\hat{H}^{qp}_{x,y,z}$ comes from the the $GW$ self-energy's singular part, $\Sigma_{\lambda,\lambda'}^{\text{GW}}(\mathbf{k},t)$, which can in principle contribute to all magnetic components, and potentially the $GT$ self-energy. I first discuss the $GW$ contribution. When treated at the COHSEX level~\cite{Hedin1965, Steinhoff2016, Steinhoff2017, Erben2018,Mocatti2026}, its diagonal part can be split into screened exchange (SEX) and Coulomb hole (COH) components:
\begin{equation}
    \Delta \Sigma_{\lambda}^{\text{GW}}(\mathbf{k},t) = \Delta \Sigma_{\lambda}^{\text{SEX}}(\mathbf{k},t)+ \Delta \Sigma_{\lambda}^{\text{COH}}(\mathbf{k},t)
\end{equation}

Both terms are computed as the difference between their nonequilibrium and equilibrium values:

\vspace{2.5pt}\begin{widetext}
\begin{equation}
\begin{gathered}
\label{eq:Delta_sigma_GW_diag}
\Delta \Sigma^{\text{GW}}_{\lambda}(\mathbf{k},t) = - \dfrac{1}{N} \sum_{\mathbf{k}',\lambda'} \bigg[
W_{\mathbf{k},\mathbf{k'}}^{\lambda\lambda'\lambda'\lambda}(\mathbf{q}=0,t) \, \Delta f_{\mathbf{k'}\lambda'}(t) - \left(\dfrac{1}{2} - f_{\mathbf{k'}\lambda'}(0)\right)
\left( W_{\mathbf{kk'}}^{\lambda\lambda'\lambda'\lambda}(\mathbf{q}=0,t) - W_{\mathbf{kk'}}^{\lambda\lambda'\lambda'\lambda}(\mathbf{q}=0,0) \right)
\bigg]
\end{gathered}
\end{equation}
\end{widetext}

The COH term can only indirectly contribute to a change in the effective magnetic field 
${h}^{qp}_{x,y,z}(\mathbf{k},t)$ through a the time-dependent screening change, involving a change in the spin-orbital occupations $\lambda$ (see the Supplementary Material of Ref.~\cite{Mocatti2026} for further details on the time-dependent screening calculation).
Instead, I expect the SEX term to play the most fundamental role for the coherent magnetic dynamics. In a magnetic system, its $\hat{\sigma}^z$ component is the exchange splitting responsible for the splitted band structure. For the particularly simple case of a collinear ferromagnetic system, this term is proportional to $\hat{\sigma}^z$. Its renormalization, when expressed in the $\{n\alpha\}$ basis, reads:

\begin{equation}
\begin{gathered}
    \Delta \Sigma^{\text{SEX}}_{n\alpha}(\mathbf{k},t) =\\ - \dfrac{1}{N} \sum_{\mathbf{k}',n'} \bigg[
W_{\mathbf{kk'}}^{nn'n'n}(\mathbf{q}=0,t) \, \Delta f_{\mathbf{k'}n'\alpha}(t) \bigg]
\end{gathered}
\end{equation}

Thus, even assuming a purely charge-like interaction $W$ and weak screening renormalization (in a metal this seems reasonable), the out-of-equilibrium occupations directly enter $\Delta \Sigma^{\text{SEX}}_{n\alpha}(\mathbf{k},t)$  and differently for up and down spins. Finally, an additional contribution to the eigenvalue renormalization comes from the diagonal $GT$ self-energy, namely $\Delta \Sigma_{\lambda}^{\text{GT}}(\mathbf{k},t)$. The off-diagonal $GW$ and $GT$ terms, $\lambda\neq\lambda'$, will contribute to the renormalization of the bare Rabi frequency, see also Fig.~\ref{fig:EOM}: 

\begin{equation}
    \mathcal{F}_{\mathbf{k}\lambda\lambda'}(t) = \bar{\mathcal{F}}_{\mathbf{k}\lambda\lambda'}+\Delta\Sigma^{GW}_{\lambda\lambda'}(\mathbf{k},t)+\Delta\Sigma_{\lambda\lambda'}^{GT}(\mathbf{k},t)
\end{equation}

with $\lambda\neq\lambda'$. A fundamental difference with respect to the non-magnetic case is that in this case the renormalizing term are not necessarily ``charge-like'', and generally contribute to the magnetic components of $h^{qp}$. Thus, the renormalized Rabi frequency will not be purely proportional to $\hat{\sigma}^0$ in spinor space. The band off-diagonal renormalization is dominated by the SEX contribution to the $GW$ part:

\begin{equation}
\begin{gathered}
    \Delta \Sigma^{\text{SEX}}_{nm\alpha}(\mathbf{k},t) =\\ - \dfrac{1}{N} \sum_{\mathbf{k}',n'} \bigg[
W_{\mathbf{kk'}}^{nn'm'm}(\mathbf{q}=0,t) \, p _{\mathbf{k'}m'n'\alpha}(t) \bigg]
\end{gathered}
\end{equation}

Both the diagonal and off-diagonal SEX contributions are thus expected to contribute to a time-dependent effective magnetic field renormalization, analogous to the one observed in TDDFT~\cite{Krieger2015,PhysRevB.95.024412,Krieger_2017}.\\

For the magnon, it is important to stress the nature of the magnon lifetime described at this level of
theory. Once the pole decomposition has been performed, the magnon amplitude is a static input and the only
dynamical variable retained is the occupation $N_{\mathbf{q}S}(t)$, so that the
magnon acquires a finite lifetime solely through the time non-local part $\Delta\Pi$
of its self-energy, i.e.\ through the magnon-electron collision integral. This is a {population} lifetime, counting the events in which a
magnon is created or destroyed by an electron. What is not described is instead the
loss of coherence of the magnon as a collective object: being a phase-coherent
superposition of particle-hole pairs, the magnon is degraded whenever one
of its constituents is scattered {without} the magnon being absorbed or emitted,
since such an event randomizes the relative phase of the pairs at fixed magnon
number. This channel is governed by the quasiparticle rates
$\Gamma_{\mathbf{k}\lambda}(t)$, which the present
framework does compute, but it cannot be captured here, since the coherent components of the magnon propagator are set to zero.  This is the magnon
counterpart of excitation-induced dephasing in the exciton problem, and there is no
argument to suggest that it should be negligible after photoexcitation, when the
constituent scattering rates are largest. Its inclusion would require propagating the
magnon amplitude alongside its occupation, and is left to future work.

Finally, I do not include a detailed discussion of phonon renormalization terms for the phonon frequency in Fig.~\ref{fig:EOM}and instead direct the reader to the treatment provided in Ref.~\cite{Mocatti2026}, both for the definition of electron-hole bubble frequency renormalization $\Delta\Xi^{\mathrm{EH}}(t)$, and for the anharmonic phonon bubble renormalization, $\Delta\Xi^{\mathrm{PB}}(t)$, where needed. 

\section{Equations of motion and magnetization dynamics}
\label{secmag}
The final equations of motion are depicted in Fig.~\ref{fig:EOM}. Here, the dynamics of phonon coherences is neglected (see also Refs.~\cite{Stefanucci2024,Mocatti2026,Furci2026-un}). I did not explicitly discuss the coupling with coherent ionic motion and photoinduced atomic forces~\cite{Mocatti2026}, however they can be readily integrated in the present scheme if coherent phonon are present. The equations in Fig.~\ref{fig:EOM} represent a coupled set of differential equations that can be solved in real time, through numerical techniques~\cite{Mocatti2026}. I note that magnon-phonon coupling, which in principle can play an important role, is explicitly neglected here. While its formal inclusion is left to future work, its effect can be understood in close analogy to exciton-phonon coupling~\cite{PhysRevB.108.165101,PhysRevB.105.085111,PhysRevB.108.165101,489b-n5pd}. Model calculations~\cite{7vr4-57mk} suggest that its inclusion in the magnon scattering integral could be decisive to obtain realistic results for magnon population dynamics, especially for their relaxation towards equilibrium.

The real-time evolution described by the equations in Fig.~\ref{fig:EOM} allow to also describe how the macroscopically observed magnetization $\mathbf{M}(t)$ evolves in time. In a non-periodic system, the total magnetic moment operator of the electron is given by the expectation value of the magnetization operator
\begin{equation}
    \hat{\mathbf{M}} = -\frac{\mu_B}{\hbar}\left(\hat{\mathbf{L}} + 2\hat{\mathbf{S}}\right),
\end{equation}
where $\hat{\mathbf{L}}$ is the orbital angular momentum
operator, and the factor of 2 in front of $\hat{\mathbf{S}}$ accounts for the electron
spin $g$-factor ($g_s \approx 2$). The macroscopic electronic magnetization is then obtained performing the appropriate
expectation value of $\hat{\mathbf{M}}$, resulting in a spin plus a orbital contribution: 

\begin{equation}
    \mathbf{M}_{el}(t) = \mathbf{M}^{\mathrm{spin}}(t) + \mathbf{M}^{{orb}}(t),
\end{equation}

In periodic systems, the situation is more nuanced. First, total magnetization is not a conserved quantity because the presence of the crystal lattice always breaks the continuous rotational symmetry, which is the fundamental prerequisite for the conservation of total angular momentum $\hat{\mathbf{J}} = \hat{\mathbf{L}} + \hat{\mathbf{S}}$. Because electrons effectively scatter off the discrete periodic potential of the solid, the lattice exerts an effective torque that couples the orbital angular momentum to the rigid structure of the crystal~\cite{Elliott2018}. The total angular momentum is never conserved, even if spin is conserved and even in the absence of SOC, due to the non-conservation of $\hat{\mathbf{L}}$. In the presence of SOC, neither $\hat{\mathbf{S}}$, nor $\hat{\mathbf{L}}$ or $\hat{\mathbf{J}}$ are generally conserved in the evolution. The spin contribution to the magnetization can be evaluated as the spin trace over the electron density matrix:

\begin{equation}
\begin{gathered}
    \mathbf{M}^{spin}(t) = -\frac{\mu_B}{V}\sum_{\mathbf{k}}
      \mathrm{Tr}\big[\rho^{<}_{\mathbf{k}}(t)\,2\hat{\mathbf{S}}_{\mathbf{k}}\big]\\
    = -\frac{\mu_B}{V}
    \sum_{\mathbf{k},\lambda\lambda'} \rho^{<}_{\mathbf{k}\lambda'\lambda}(t)\,
    \braket{\psi_{\mathbf{k}\lambda} | 2\hat{\mathbf{S}} | \psi_{\mathbf{k}\lambda'}}
      \end{gathered}
\end{equation}

where $V$ is the system volume. In the absence of coherences, this expression reduces sum of individual quasiparticle contributions: 

\begin{equation}
    \mathbf{M}^{spin}(t) = -\frac{\mu_B}{V}
    \sum_{\lambda, \mathbf{k}} f(\varepsilon_{\mathbf{k}\lambda}(t),t)
    \braket{\psi_{\mathbf{k}\lambda} |  2\hat{\mathbf{S}} | \psi_{\mathbf{k}\lambda}},
    \label{eq:mag_soc}
\end{equation}

The situation is more complicated for orbital magnetization, as the naive bulk expectation value of $\hat{\mathbf{L}}=\hat{\mathbf{r}}\times\hat{\mathbf{p}}$ cannot be calculated due to the ill-defined position operator under periodic boundary conditions. The expression obtained in modern theory of magnetization reads, at thermodynamical equilibrium~\cite{PhysRevLett.95.137205,PhysRevLett.95.137204}: 

\begin{equation}
\begin{gathered}
  \mathbf{M}^{{orb}}
  = -\frac{\sqrt2}{2}
    \,\mathrm{Im}
    \sum_{n} \int_{\mathrm{BZ}} \frac{d\mathbf{k}}{(2\pi)^3}\,
    f_{\mathbf{k}\lambda}
    \left\langle \partial_{\mathbf{k}} u_{\mathbf{k}\lambda} \right|
    \times\\
    \bigl(H_{\mathbf{k}} + \varepsilon_{\mathbf{k}\lambda} - 2\mu\bigr)
    \left| \partial_{\mathbf{k}} u_{\mathbf{k}\lambda} \right\rangle
  \label{eq:RCSV}
  \end{gathered}
\end{equation}
 
where $\mu$ is the chemical potential and $\ket{u_{\mathbf{k}\lambda}}$ represent the periodic part of Bloch eigenstates.  The generalization of orbital magnetization, Eq.~\ref{eq:RCSV}, to the out-of equilibrium case is non-trivial and represents an interesting avenue for future work. For now, I neglect its contribution. The obtained expression for the electronic magnetization $\mathbf{M}_{el}$ correctly reduces to the usual $\mathbf{M}_{el}(t) \propto \sum_{\mathbf{k}n} [f_{\mathbf{k}n}^{\uparrow\uparrow}-f^{\downarrow\downarrow}_{\mathbf{k}n}]~\hat{u}_z$ in the collinear case, and if the orbital contribution is neglected. The total magnetization per unit cell should also include the magnon contribution:

\begin{equation}
    \mathbf{M}(t) = \mathbf{M}_{\text{el}}(t) + \mathbf{M}_{\text{mag}}(t)
\end{equation}
where $\mathbf{M}_{\text{mag}}(t)$ represents the reduction in magnetization due to the excitation of  magnon modes. In the most general case, the magnon creation operator $\hat{b}^\dagger_{\mathbf{q}S}$ for a branch $S$ at momentum $\mathbf{q}$ is defined as a coherent linear combination of multi-band particle-hole pairs across all spin configurations ($\alpha,\alpha' \in \{\uparrow, \downarrow\}$):
\begin{equation}
    \hat{b}^\dagger_{\mathbf{q}S} = \sum_{\mathbf{k}, n_1, n_2} \sum_{\alpha \alpha'} A_{S,\alpha\alpha'}^{n_1, n_2}(\mathbf{k}, \mathbf{q}) \, \hat{c}^\dagger_{\mathbf{k}+\mathbf{q}n_1\alpha} \, \hat{c}_{\mathbf{k}n_2\alpha'}
\end{equation}
where $A_{S,\alpha \alpha'}^{n_1, n_2}(\mathbf{k}, \mathbf{q})$ is the normalized magnon wavefunction discussed earlier. To determine the net spin projection carried by a single quantum of this collective mode, one considers the expectation value of the total spin operator $\hat{\mathbf{S}}$:
\begin{equation}
    \mathbf{M}_S(\mathbf{q}) = \braket{S,\mathbf{q}|\hat{\mathbf{S}}|S,\mathbf{q}}= \braket{0|\hat{b}_{\mathbf{q}S}\hat{\mathbf{S}}~\hat{b}^\dagger_{\mathbf{q}S}|0}
\end{equation}

Under this definition, transverse spin-flip transitions ($\uparrow \to \downarrow$) contributes a factor of $-1$, while any longitudinal mixing does not contribute to the expectation value, reducing the effectively carried spin. The magnon contribution to magnetization is thus expressed as:
\begin{equation}
    \mathbf{M}_{\text{mag}}(t) = \frac{2 \mu_B}{N_{\mathbf{q}}} \sum_{\mathbf{q}, S} \mathbf{M}_S(\mathbf{q}) \, N_{\mathbf{q}S}(t) \label{magnon_mag}
\end{equation}

If magnon excitations are purely transverse, each magnon carries a $\Delta S^z = -1$ and Eq.~\ref{magnon_mag} reduces to:
\begin{equation}
    \mathbf{M}_{\text{mag}}(t) = - N_{\text{mag}}(t) \hat{u}_z = -\frac{1}{N_{\mathbf{q}}} \sum_{\mathbf{q}, S} N_{\mathbf{q}S}(t) \hat{u}_z
\end{equation}

In this limit, the model dynamics of Ref.~\cite{7vr4-57mk} is recovered: following an external laser pump, minority electrons can scatter to unoccupied majority states by emitting magnons through an electron-magnon interaction~\cite{PhysRevB.78.174422,https://doi.org/10.1002/apxr.202300103,7vr4-57mk}. Consequently, $\mathbf{M}_{\text{el}}(t)$ increases as electrons flip from $\downarrow$ to $\uparrow$, and $N_{\mathbf{q}S}(t)$ simultaneously increases, too. To conclude this discussion, I note that I am neglecting possible contributions to magnetization coming from chiral phonons~\cite{PhysRevLett.115.115502}. 

\subsection{Conservation properties of the equations of motion}
\label{sec:conservation}

I now briefly discuss the conservation properties of the derived equations of motion. The Kadanoff-Baym equations are exact as long as the self-energy is treated as an
exact functional of the Green's function, and the Baym-Kadanoff
theorem~\cite{Baym1961,PhysRev.127.1391} guarantees that any $\Phi$-derivable approximation,
$\Sigma=\delta\Phi/\delta G$, preserves the macroscopic conservation laws. Here, it is not possible to formally inherit the conservation laws from the Baym-Kadanoff argument directly due to the MGKBA form of the propagator and the Markov approximation. Instead, conservation properties are more easily discussed by directly looking at the structure of the final equations of Fig.~\ref{fig:EOM}. 

The number of electrons is
conserved: the coherent Rabi term redistributes occupations but does not change the total electron number, while each scattering process removes one electron from
$(\mathbf{k}+\mathbf{q},\lambda')$ and adds one to $(\mathbf{k},\lambda)$ at the same
rate, so that
\begin{equation}
\sum_{\mathbf{k}\lambda}I^{el-el}_{\mathbf{k}\lambda}(t)=
\sum_{\mathbf{k}\lambda}I^{el-ph}_{\mathbf{k}\lambda}(t)=
\sum_{\mathbf{k}\lambda}I^{el-mag}_{\mathbf{k}\lambda}(t)=0
\label{eq:numbercons}
\end{equation}
identically. In the absence of spin-orbit coupling, spin is a good quantum number, the electron
Green's function is diagonal in spinor space, and $\langle \hat S^z\rangle_{\mathbf{k}
\lambda}=\pm\hbar/2$ exactly. The coherent evolution then conserves the total
electronic spin, since $h^{qp}$ possesses only $\hat\sigma^0$ and $\hat\sigma^z$
components and therefore commutes with $\hat S^z$. The electron-phonon and
electron-electron collision integrals conserve spin separately. The electron-magnon term does not: each process flips one electron from minority to majority while creating one magnon (and vice-versa for magnon absorption). The
two contributions cancel exactly, and the conserved quantity is
\begin{equation}
S^z_{tot}(t)=S^z_{el}(t)-\frac{\hbar}{N_q}\sum_{\mathbf{q}S}N_{\mathbf{q}S}(t),
\qquad \frac{dS^z_{tot}}{dt}=0 ,
\label{eq:spincons}
\end{equation}

When spin-orbit coupling is present, the coherent term
no longer commutes with $\hat S^z$, since $h^{qp}$ acquires $\hat\sigma^{x,y}$
components; the electron-phonon and electron-electron vertices acquire spin-flip
matrix elements through the spinor structure of the eigenstates, giving the
Elliott-Yafet channels; and the electron-magnon vertex is no longer purely
transverse.  \\
Within the Markov approximation, energy conservation can only be studied in the quasiparticle sense. For the electron-magnon scattering, each process transfers an energy exactly equal to
$\Omega_{\mathbf{q}S}$ from the electronic to the magnonic subsystem, as enforced by
the $\delta$ function. Assuming for now time-independent quasiparticle energies and summing the corresponding contributions
to the electron and boson equations of motion one obtains
\begin{equation}
\sum_{\mathbf{k}\lambda}\varepsilon_{\mathbf{k}\lambda}
I^{el-mag}_{\mathbf{k}\lambda}
+\sum_{\mathbf{q}S}\Omega_{\mathbf{q}S}I^{mag-el}_{\mathbf{q}S}=0 ,
\label{eq:encons}
\end{equation}
and identically for the electron-phonon pair
$(I^{el-ph},I^{ph-el})$; the electron-electron integral conserves the electronic
energy on its own, due to the $\delta$ function in the electron-electron collision integral. The collision integrals therefore conserve the total
quasiparticle energy
$E_{qp}=\sum_{\mathbf{k}\lambda}\varepsilon_{\mathbf{k}\lambda}f_{\mathbf{k}\lambda}
+\sum_{\mathbf{Q}\nu}\omega_{\mathbf{Q}\nu}n_{\mathbf{Q}\nu}
+\sum_{\mathbf{q}S}\Omega_{\mathbf{q}S}N_{\mathbf{q}S}$ at frozen quasiparticle
energies. It is important to note that the quasiparticle energies are themselves time dependent through
the renormalizations discussed previously. Quasiparticle energy conservation is then exact only in the
adiabatic limit in which the renormalizations vary slowly compared with the
collisional dynamics, which is the regime in which the quasiparticle picture
underlying the whole construction is meaningful.

Finally, a property closely related to the above is that
the equations of motion of Fig.~\ref{fig:EOM} possess the correct stationary state. Each rate vanishes identically when the electronic occupations are
Fermi-Dirac and the bosonic ones Bose-Einstein at a common temperature, guaranteeing that a thermalized final state is stable.

\section{Spatial inhomogeneity and spin diffusion}
Following ultrafast laser excitation, spatial inhomogeneity is typically created in several physical quantities in the magnetic system. This aspect is particularly relevant in the case of ultrafast demagnetization, since it can in principle lead to spin (super)diffusion~\cite{Battiato2010}. Furthermore, assuming the generation of a hot magnon population~\cite{7vr4-57mk,https://doi.org/10.1002/apxr.202300103}, magnon diffusion could also play a role during magnetization recovery. For this reason, it could be important to account for laser-induced spatial inhomogeneity in electron, phonon and magnon dynamics. To this aim, the derivation outlined in this paper could be straightforwardly extended to a spatially inhomogeneous case through a gradient expansion of the density matrix in the slow spatial variable $\mathbf{R}=\frac{\mathbf{r}_1+\mathbf{r}_2}{2}$ in the spatial Wigner representation, as outlined in Ref.~\cite{Maciejko2007}, and by deriving an inhomogeneous counterpart to the equations of Fig.~\ref{fig:EOM}. Besides diffusion of out-of-equilibrium electrons, phonon and magnons, the effective magnetic field discussed in Sec.~\ref{singular} would also acquire an explicit spatial dependence, $B_{i}^{\mathrm{eff}}(\mathbf{k},\mathbf{R},t)$, causing the appearance of an effective force acting on both magnons and electrons.

\section{Towards a practical implementation of \textit{ab initio} electron-magnon-phonon coupled dynamics}
I now discuss a practical approach towards the first principles implementation of the equations of motion presented in Fig.~\ref{fig:EOM}. Working in the Wannier basis, we showed in Ref.~\cite{Mocatti2026} how equations of motion for coupled electrons and phonons (and ions) can be can be practically solved. In order to extend such framework to the magnetic noncollinear case, some extra steps need to be taken, illustrated below: 

\begin{enumerate}
     \item ~Ground state electron quasiparticle energies are needed for the magnetic equilibrium state. In metals, Kohn-Sham eigenvalues may already represent a good starting point~\cite{PhysRevB.100.045130}. These can be calculated through any (noncollinear) magnetic density-functional theory (DFT) implementation, for example employing Quantum ESPRESSO~\cite{QE,Giannozzi2017,PhysRevB.100.045115}, or within many-body perturbation theory in a static approximation ($e.g.$ COHSEX) employing any many-body perturbation theory code, for example yambo~\cite{Sangalli2019}, in cases where this is needed (for example, semiconductors). 
    \item Phonon frequencies for the magnetic ground state must be calculated, for example employing the linear response algorithm implemented within Quantum ESPRESSO~\cite{urru2020lattice} or through finite differences methods. 
    \item ~The Wannier interpolation procedure for the electron-phonon coupling~\cite{Marini2024} must be extended to the noncollinear magnetic case. This can be done for example starting from the electron-phonon coupling calculation for the noncollinear magnetic case, as implemented in Quantum ESPRESSO~\cite{QE,Giannozzi2017,PhysRevB.100.045115}.
    \item Optical dipole matrix elements for the magnetic noncollinear case are needed, Eq.~\ref{dipole}. These can be obtained for example in the Kohn-Sham spin orbital basis at DFT level.
    \item Electron-electron interaction matrix elements $W$ in the noncollinear magnetic case for metallic systems, through any many body perturbation theory code, for example yambo or Berkeley GW~\cite{Sangalli2019,DESLIPPE20121269,PhysRevB.103.155152}.
    \item ~ Magnon dispersions. Magnon frequencies in Fig.~\ref{fig:EOM} are in principle time-dependent, however the time-dependence should be weak in metallic systems because it is ultimately related to the time-dependent change in screening. Thus, employing equilibrium (or linear response) magnon frequencies should be a good approximation in metallic systems. In practice, magnon frequencies can be calculated either within the TDDFT framework~\cite{PhysRevB.84.174418,PhysRevB.85.054305,PhysRevB.103.245110,GORNI2022108500} or as a solution of spin-flip BSE~\cite{gbpw-zh1v,489b-n5pd,PhysRevB.81.054434,PhysRevB.60.7419,PhysRevB.62.3006,PhysRevLett.127.166402,Chen2026}. I note that the solution of the BSE returns both the collective modes and the Stoner particle-hole continuum; in practice the former can be identified by the character of their eigenvectors, which are delocalized over many particle-hole transitions and carry large spectral weight. For a TDDFT calculation, the same separation must instead be performed on $\operatorname{Im}\chi^{+-}(\mathbf{q},\omega)$ by fitting a Lorentzian on a smooth background, which also provides the magnon linewidth $\Gamma_\mathbf{q}$. 
    \item ~ A practical way to calculate the electron-magnon coupling, Eq.~\ref{electron-magnon}, from first principles. The coupling
\begin{equation}
\begin{gathered}
m^S_{n_1n_2,\alpha\alpha'}({\bf k},{\bf q})\propto\\\sum_{m_1m_2}
W^{n_1n_2m_1m_2}_{{\bf k},{\bf k}''}({\bf q})\,
A^{m_1m_2}_{S,\alpha\alpha'}({\bf k}'',{\bf q})
\end{gathered}
\end{equation}
requires the knowledge of the statically screened interaction $W$, which can be calculated at the RPA  level as routinely done in \textit{ab initio} many-body perturbation theory, and the  magnon wavefunction, which is obtained through the solution of the BSE (see e.g. Refs.~\cite{gbpw-zh1v,489b-n5pd,PhysRevB.81.054434,PhysRevB.60.7419,PhysRevB.62.3006,PhysRevLett.127.166402}),  neglecting the residue $R_{\mathbf{q}}$ and renormalizing magnon eigenvectors so that the magnetization sum rule Eq.~\ref{sumrule} is satisfied. Finally, the electron-magnon coupling time dependence should be weak in metals, since screening is not expected to change drastically after photoexcitation, and could be neglected.

\end{enumerate}
Since the magnon enters the final equations only through its dispersion
and its coupling to electrons, $\Omega_{\mathbf{q}S}$ may equivalently be obtained
from linear spin-wave theory applied to a DFT-derived spin Hamiltonian, in close
analogy with the treatment of phonons, provided the electron-magnon coupling, which involves the itinerant states and is not contained in a
localized-spin model, is supplied independently, for instance as the matrix element of the exchange–correlation magnetic field variation induced by the mode $(\mathbf{q},S)$:
\begin{equation}
m^S_{\lambda\lambda'}(\mathbf{k},\mathbf{q})\simeq
\langle\psi_{\mathbf{k}+\mathbf{q}\lambda}|\,
\Delta_{\mathbf{q}S}B_{\rm xc}\cdot\hat{\boldsymbol\sigma}\,
|\psi_{\mathbf{k}\lambda'}\rangle
\end{equation}

   A possible strategy to obtain an approximated time-dependent renormalization of magnon frequencies without having to solve the BSE many times during the dynamics, would be to operate in analogy to what discussed in Ref.~\cite{Mocatti2026} for phonon frequencies, practically useful is the electron-magnon coupling can be approximated as time-independent. In the adiabatic approximation, the electronic screening contribution to the magnon mode $S$ at a specific time $t$ is governed by the time-local electron-hole polarization bubble $\Pi_{\mathbf{q}}^{\delta,S}(t)$. At Kohn-Sham level:

\vspace{2.5pt}\begin{widetext}
\begin{equation}
\Pi_{\mathbf{q}}^{\delta,S}(t) = \frac{1}{\Omega_{BZ}} \int_{\Omega_{BZ}} d\mathbf{k} \sum_{\lambda, \lambda'} |m_{\lambda\lambda'}^S(\mathbf{k}, \mathbf{q}, t)|^2 \mathcal{P} \left[ \frac{f_{\mathbf{k}\lambda}(t) - f_{\mathbf{k}-\mathbf{q}\lambda'}(t)}{\varepsilon_{\mathbf{k}\lambda}^{\text{KS}} - \varepsilon_{\mathbf{k}-\mathbf{q}\lambda'}^{\text{KS}}} \right]
\end{equation}
\end{widetext}

where $\mathcal{P}$ denotes the Cauchy principal value, and $\varepsilon_{\mathbf{k}\lambda}^{\text{KS}}$ represents the ground-state Kohn-Sham electronic eigenvalues. The transient self-energy correction $\delta\Pi_{\mathbf{q}}^S({t})$ is defined by subtracting the equilibrium contribution:

\begin{equation}
\delta\Pi_{\mathbf{q}}^{\delta,S}(t) = \Pi_{\mathbf{q}}^{\delta,S}(t) - \Pi_{\mathbf{q}}^{\delta,S}(0) \label{deltaP}
\end{equation}

thus isolating the time-dependent part. The time-dependent renormalized magnon frequency $\Omega_{\mathbf{q}}^S({t})$ is evolved by adding this correction directly to equilibrium magnon dispersion $\Omega_{\mathbf{q}}^S(0)$, calculated just once:

\begin{equation}
\Omega_{\mathbf{q}}^S(t) = \Omega_{\mathbf{q}}^S(0) + \text{Re}\left[ \delta\Pi_{\mathbf{q}}^S(t) \right] \label{eq:adiabmag}
\end{equation}

If the electron-magnon coupling  can be approximated as time independent, |$m_{\lambda\lambda'}^S(\mathbf{k}, \mathbf{q}, t)|^2 \approx |m_{\lambda\lambda'}^S(\mathbf{k}, \mathbf{q})|^2$, only states close to the Fermi level, where occupations vary, contribute to the correction in Eq.~\ref{deltaP}, making this expression practically useful.

More elaborate simulation schemes could include magnon-phonon scattering. Matrix elements could be calculated either in the scheme proposed in Ref.~\cite{489b-n5pd} or in the one presented in Ref.~\cite{PhysRevB.111.104431}. Finally, the implementation could be coupled to a Boltzmann equation solver~\cite{Cepellotti_2022,PONCE2016116,ZHOU2021107970} to treat spatially inhomogeneous systems and describe an inhomogeneous spin dynamics.\\

\section{Solution of the equations of motion for a simple model system}
\label{sec:toy}

\begin{figure*}[t]
\includegraphics[width=0.9\textwidth]{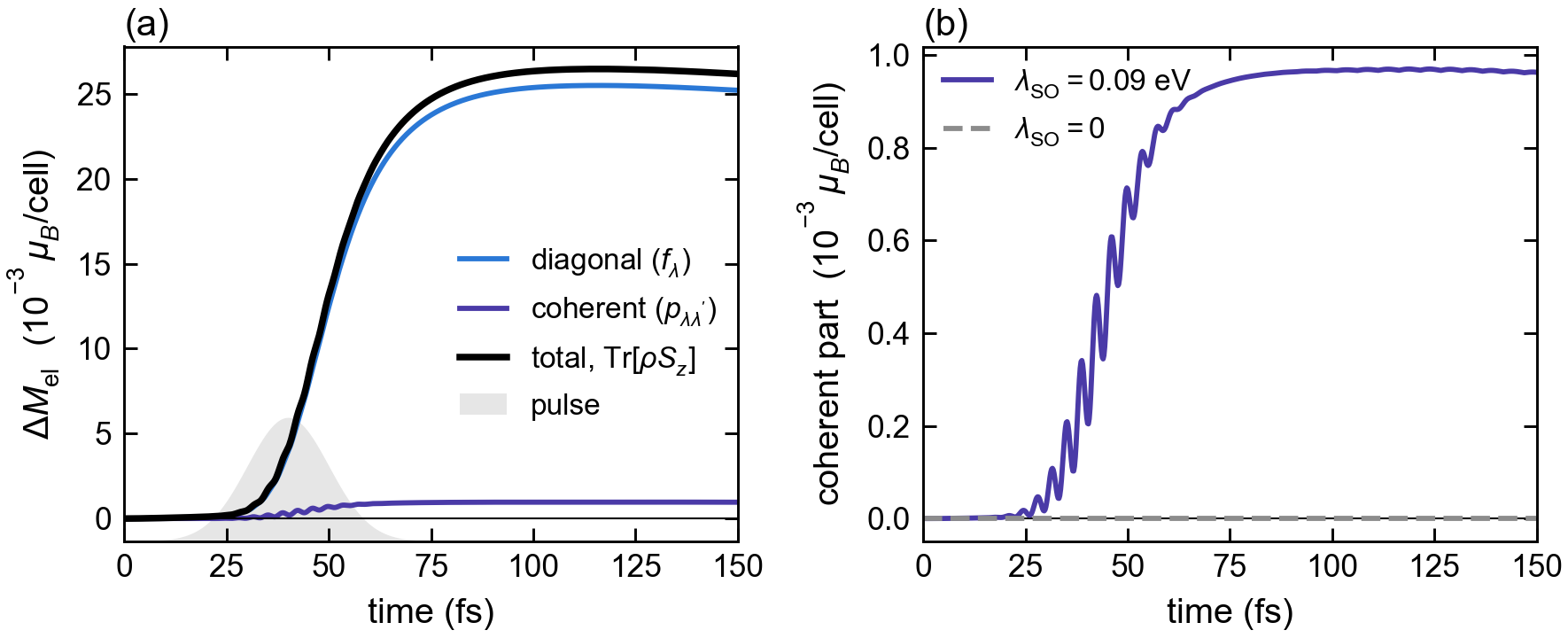}
\caption{Spin magnetization dynamics in the first 150 fs for the considered model system: coherent and incoherent contributions (panel (a)) and role of SOC (panel (b)). The grey area in (a) shows the pump envelope.}
\label{fig:coherent}
\end{figure*}

The derivation presented so far leads to the closed set of equations of motion
summarized in Fig.~\ref{fig:EOM} for the electron density matrix, the magnon
occupations and the phonon occupations. To illustrate their physics, I present the solution of the equations in Fig.~\ref{fig:EOM} in
real time for a minimal model. The considered model system is a one-dimensional chain with two orbitals ($n=1,2$) and two
spin projections, i.e. four spin-orbitals $\{\mathbf{k},\lambda\}$ per crystal
momentum, sampled on $N_k=256$ points. The
single-particle Hamiltonian is
\begin{equation}
\begin{gathered}
h_{\alpha\beta}^{nn'}(k)=
\big[\epsilon_n-2t_n\cos k\big]\,\hat\sigma^0_{\alpha\beta}\,\delta_{nn'} \\
-\frac{\Delta_{\rm ex}}{2}\,\hat\sigma^z_{\alpha\beta}\,\delta_{nn'}
+\lambda_{\rm SO}\,\hat\tau^x_{nn'}\hat\sigma^x_{\alpha\beta},
\label{eq:toyham}
\end{gathered}
\end{equation}
where $\hat\tau$ are Pauli matrices acting in the orbital space. The second term is a collinear exchange
field producing the equilibrium spin splitting, while the third term plays the role
of the spin-orbit Hamiltonian: it does not commute with
$\hat\sigma^z$, so that no spin projection is a good quantum number. Orbital 1 is partially filled and carries the magnetization;
orbital 2 is empty at $t=0$ and acts as the target of the optical excitation. With
the parameters of Table~\ref{tab:toypar} the ground state is ferromagnetic, with
$n_\uparrow=0.659$ and $n_\downarrow=0.372$ electrons per unit cell, a spin
polarization $\langle \hat S^z\rangle=0.143$, and a band splitting of $0.241$~eV.

A single magnon branch is assumed, with equilibrium dispersion
$\Omega_q=\Omega_0+D(1-\cos q)$, where $\Omega_0$ is a gap opened by the
spin-orbit interaction. Two phonon branches are included, one dispersionless and one
acoustic-like.

All couplings
are taken to be constant and purely charge-like, i.e. proportional to $\hat\sigma^0$,
in the $\{n,\alpha\}$ basis. Their entire spin structure in the spin-orbital basis is generated
by the unitary transformation $U^{k\lambda}_{n\alpha}$. The electron-magnon
vertex is taken in the purely transverse form, in which the emission of a magnon is accompanied by a minority-to-majority electron
spin flip.

The system is driven out of equilibrium by a laser term, with a Gaussian envelope of
standard deviation $10$~fs (FWHM $23.5$~fs) centred at $t=40$~fs, and a photon
energy of $0.6$~eV, slightly below the $1\to2$ interorbital transitions, which are reached through the spectral width of the pulse. All parameters are collected in Table~\ref{tab:toypar}.

\begin{table}[t]
\caption{Model parameters, with energies in eV, time in fs. The contact interaction $W_{\rm ee}$ enters the SEX renormalization and, in Appendix G, the electron-electron collisions.}
\label{tab:toypar}
\begin{ruledtabular}
\begin{tabular}{lll}
 & symbol & value \\
\hline
orbital energies      & $\epsilon_1,\epsilon_2$        & $0.00$, $0.70$ \\
hoppings              & $t_1,t_2$                      & $0.15$, $0.12$ \\
exchange field        & $\Delta_{\rm ex}$              & $0.25$ \\
spin-orbit coupling  & $\lambda_{\rm SO}$             & $0.09$ \\
initial temperature   & $k_BT$                         & $0.035$ \\
magnon gap, dispersion& $\Omega_0$, $D$                & $0.08$, $0.16$ \\
electron-magnon      & $m_0$                          & $0.10$ \\
phonon energies       & $\omega_{\rm opt}$, $\omega_{\rm ac}$ & $0.055$, $\le 0.040$ \\
electron-phonon      & $g_{\rm opt}$, $g_{\rm ac}$    & $0.075$, $0.060$ \\
electron-electron        & $W_{\rm ee}$    &  $0.50$ \\
pump                  & $\hbar\omega_L$, $\sigma_t$, peak   & $0.60$, $10$, $0.030$ \\
grid, broadening      & $N_k$, $\eta$                  & $256$, $0.008$ \\
\end{tabular}
\end{ruledtabular}
\end{table}

\subsection{Numerical solution}

The full set of Fig.~\ref{fig:EOM} is propagated with a fourth-order Runge-Kutta
scheme: the electron density matrix including its band off-diagonal components, with
the dephasing rates, the magnon
occupations $N_{qS}(t)$ and the phonon occupations $n_{{Q}\nu}(t)$. The
quasiparticle Hamiltonian is updated at every step,
including the screened-exchange renormalization in its band-diagonal and band off-diagonal
parts, and the singular contributions of the $GT$ and Fan-Migdal self-energies. The magnon frequencies are
renormalized adiabatically according to Eq.~\ref{eq:adiabmag}. The electron-phonon and electron-magnon collisional terms are included, while the electron-electron collisions are here neglected; their inclusion is studied (out of convergence) in Appendix G.

Two numerical remarks are in order. First, the energy-conserving $\delta$ functions
appearing in the collision integrals are represented by Gaussians of width $\eta$.
On a finite grid $\eta$ must exceed the level spacing, while detailed balance
requires $\eta\ll k_BT$ and $\eta\ll\Omega_{qS},\omega_{{Q}\nu}$; both
conditions are met by the values in Table~\ref{tab:toypar}. Second, the collision
integrals are re-evaluated every few femtoseconds as the quasiparticle energies drift.\\
Finally, the results are converged with respect to the sampling: at $t=300$~fs the
demagnetization is $M/M(0)=0.877$ with $N_k=256$ ($\eta=8$~meV) and $0.871$ with $N_k=384$ ($\eta=6$~meV), a change of less than $1\%$.

\subsection{Results}
\label{sec:toyresults}

\begin{figure*}[t]
\includegraphics[width=\textwidth]{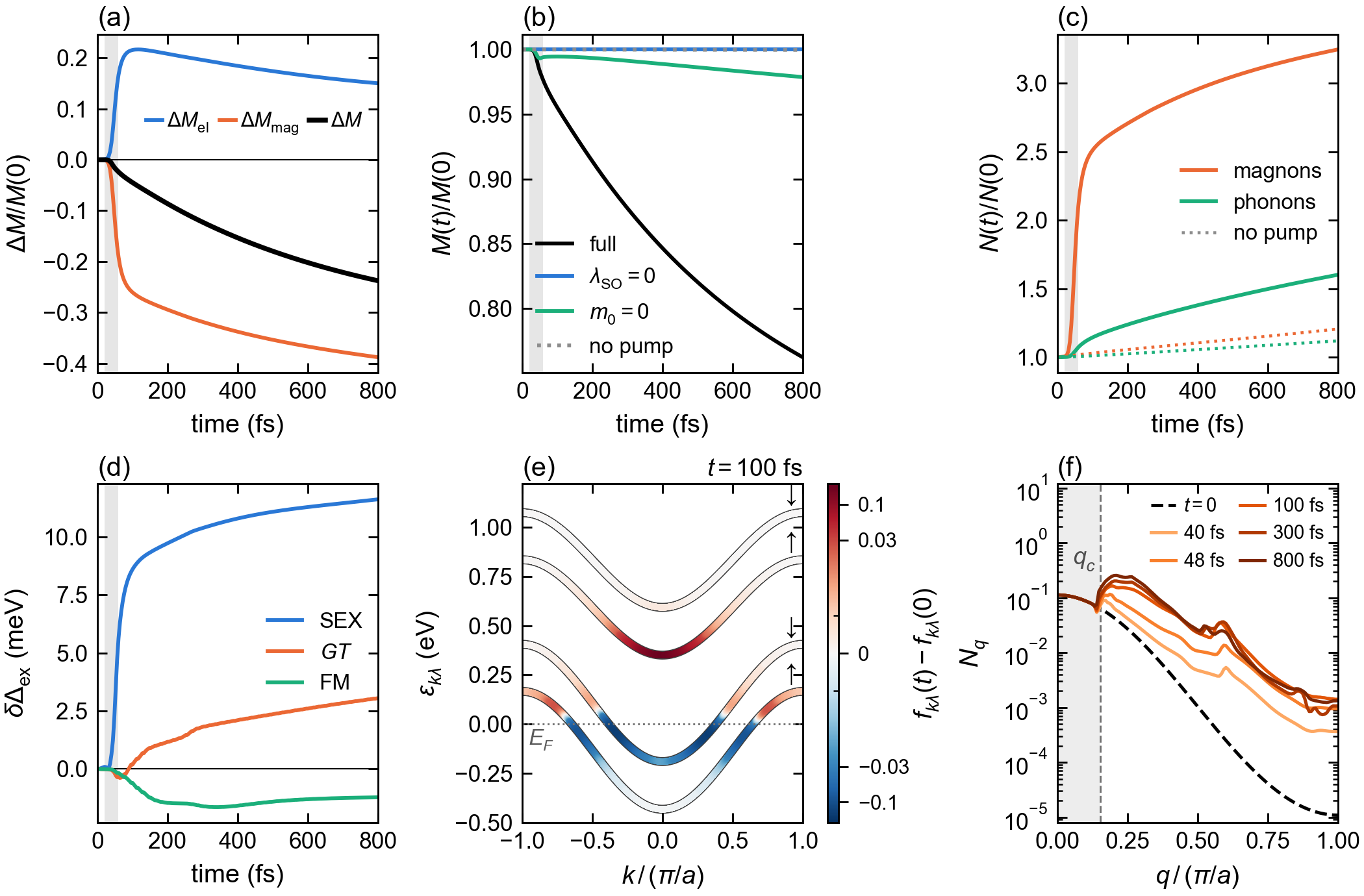}
\caption{Solution of the coupled equations of Fig.~\ref{fig:EOM} for the model of
Sec.~\ref{sec:toy}. (a) Change of the electronic and magnonic contributions to the
magnetization, and of the total magnetization, normalized to the equilibrium value.
(b) Total magnetization for the full calculation, for $\lambda_{\rm SO}=0$, for
$m_0=0$, and for the unpumped system (dotted). (c) Total magnon and phonon
populations, $N(t)=\sum_q N_q(t)$, normalized to their equilibrium values; dotted
lines are the corresponding unpumped baselines. (d) Change of the exchange
splitting, $\delta\Delta_{\rm ex}$, i.e.\ renormalization of the $\hat\sigma^z$
component of the quasiparticle Hamiltonian, from the SEX, $GT$ and Fan-Migdal (FM)
terms. (e) Band structure of the model, coloured by the pump-induced change of the
occupations, $f_{k\lambda}(t)-f_{k\lambda}(0)$, at $t=100$~fs (red: electrons,
blue: holes); arrows mark the majority and minority character of the bands. (f) Magnon occupations
across the Brillouin zone at selected times; the dashed line marks the threshold
$q_c$ at which the magnon branch enters the spin-flip Stoner continuum. In panels (a)-(d)
the grey band marks the pump pulse.}
\label{fig:toy}
\end{figure*}
Figs.~\ref{fig:coherent} and~\ref{fig:toy} summarize the main results of the model calculation.

In Fig.~\ref{fig:coherent} the electronic magnetization  is decomposed into its band-diagonal and band
off-diagonal contributions (panel (a)). The off-diagonal term, carried by the electronic
polarizations $p_{\mathbf{k}\lambda\lambda'}$, is the coherent magnetization channel:
it builds up while the pulse is on, oscillates on the timescale set by the driving
field, and saturates once the polarizations have dephased. This is the same
contribution that is responsible for the coherent magnetization dynamics obtained in
real-time TDDFT simulations~\cite{Krieger2015,PhysRevB.95.024412,Krieger_2017}. In the present model its magnitude is small, of the
order of $1\%$ of the equilibrium moment, against a total demagnetization exceeding
$20\%$ at $800$~fs; the incoherent channel therefore dominates the long-time evolution,
although this balance is model and parameter dependent and no general conclusion should be drawn
from it. Fig.~\ref{fig:coherent}(b) shows that this contribution vanishes
when the spin-orbit coupling is switched off. This is expected: for
$\lambda_{\rm SO}=0$, when
$\hat{S}^z$ is diagonal in the $\{\mathbf{k},\lambda\}$ basis, and there is no
off-diagonal matrix element for the polarizations to couple to. In Fig.~\ref{fig:toy}, additional information is reported:

\emph{(i) Carrier dynamics.} The pump creates a polarization between
the two orbitals which decays on the timescale set by the dephasing rates, leaving a
hot carrier distribution (Fig.~\ref{fig:toy}(e)) that subsequently relaxes by
emitting magnons and phonons. After the pulse, the photoexcited electrons are found
mainly in the majority antibonding band, while the holes below the Fermi level are
mostly in the minority band. 

\emph{(ii) The two demagnetization channels operate in combination.} This is shown in
Fig.~\ref{fig:toy}(b). Setting $\lambda_{\rm SO}=0$, the magnon population still
roughly doubles under the pump, but the total moment $M=M_{\rm el}+M_{\rm mag}$
remains substantially constant: within the electron-magnon channel
alone, angular momentum is transferred from the itinerant electrons to the
magnon bath, and no demagnetization occurs, as expected. Conversely, retaining the spin-orbit coupling but switching off the electron-magnon vertex ($m_0=0$) produces a change of the moment of $\approx 2\%$ over the whole simulated window. Only when both channels
are active does the moment decrease appreciably, by $20\%$ within about $600$~fs. The
mechanism, visible in Fig.~\ref{fig:toy}(a), is that magnon emission
overpolarizes the itinerant electrons, $\Delta M_{\rm el}>0$, and the excess
spin polarization is subsequently removed by the spin-orbit-mediated Elliott-Yafet
channels, which transfer angular momentum to the lattice. 

\emph{(iii) Time-dependent effective magnetic field.} The renormalization of the magnetic components of
$h^{qp}$ is dominated by the screened-exchange term, which is
considerably larger than the contribution of the singular $GT$ and
Fan-Migdal self-energies with this choice of parameters (Fig.~\ref{fig:toy}(d)). In the present model the exchange
splitting {increases} during the dynamics, following the increase of the
itinerant spin polarization.

\begin{figure}[t]
\includegraphics[width=0.75\columnwidth]{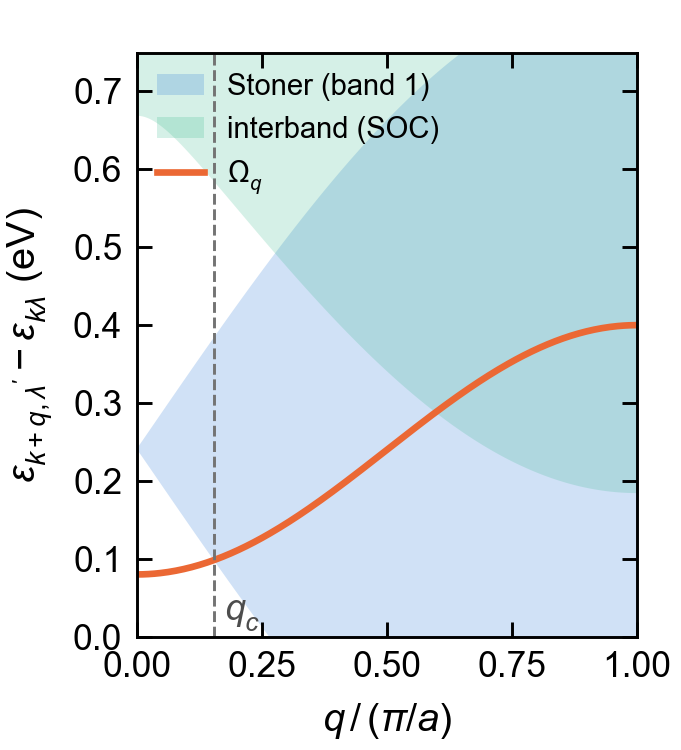}
\caption{Magnon dispersion $\Omega_q$ (orange) and spin-flip transition energies
$\varepsilon_{k+q,\lambda'}-\varepsilon_{k\lambda}$ allowed by the electron-magnon
vertex: Stoner continuum of the partially filled band (blue) and SOC-mediated
interband transitions (green).}
\label{fig:stability}
\end{figure}

\emph{(iv) Nonthermal magnon distribution.} The magnon occupations
(Fig.~\ref{fig:toy}(f)) develop a nonthermal profile within the first $\sim 50$~fs, reflecting the phase space available for
electron-magnon scattering at the relevant carrier energies, and subsequently tend to relax
towards lower energy magnons. Below $q_c\simeq0.15\,\pi/a$ the magnon occupations remain at their equilibrium
value at all times, and a sharp onset is visible at $q_c$. This threshold is of
kinematic origin: the electron-magnon vertex connects minority states at $k+q$ with
majority states at $k$, so that energy conservation requires
$\Omega_q=\varepsilon_{k+q\downarrow}-\varepsilon_{k\uparrow}$, and these spin-flip
energies are bounded from below by $\Delta_{\rm ex}-4t_1\sin(q/2)$ for $\lambda_{\rm SO}=0$.
Since the exchange splitting exceeds the magnon gap, at long wavelength the magnon
branch lies below this Stoner continuum and enters it only at $q_c$, as is the case
for long-wavelength magnons in itinerant ferromagnets. For $q<q_c$ no
energy-conserving electron-magnon process exists, and, in the absence of
magnon-magnon and magnon-phonon scattering, these modes do not take part in the
dynamics. Fig.~\ref{fig:stability}  better illustrates the argument: the magnon branch lies below the Stoner continuum of the
partially filled band at long wavelength and enters it at $q_c\simeq0.15\,\pi/a$. A
second weaker channel opens at $q\simeq0.6\,\pi/a$, where $\Omega_q$
reaches the interband spin-flip transitions allowed by the spin-orbit coupling; this
is the origin of the small structure visible around this value in
Fig.~\ref{fig:toy}(f).

\emph{(v) Absence of recovery.} No recovery of the magnetization is observed on the
simulated timescale. This is the expected combined consequence of both the relatively short simulation time and the neglect of magnon-phonon
coupling: without it, the angular momentum stored in the magnon bath has no channel
to return to the electronic system, and the magnon population can only relax by
re-absorption. 
\subsection{Conservation laws and convergence}
\label{sec:toyconv}

The calculation also provides a numerical verification of the conservation properties discussed above for Eqs. in Fig.~\ref{fig:EOM}. Indeed: \\
\textit{(i)} The total electron number is numerically conserved over the entire trajectory.\\
\textit{(ii)} In the absence of spin-orbit
coupling the total spin
is numerically conserved over the trajectory. \\
\textit{(iii)} Energy is conserved by the collision integrals at fixed quasiparticle energies, with the caveat that the Gaussian representation of the $\delta$ functions
violates detailed balance, producing a slow spurious
heating, which tends to become zero at infinitesimal broadening; the baselines, realized without laser and shown as dotted lines in
Fig.~\ref{fig:toy}(b) and \ref{fig:toy}(c), allow it to be assessed directly. \\

\section{Conclusions}
In this work, I have presented an \textit{ab initio} framework based on the Kadanoff-Baym equations in spinor space, where coherent and incoherent ultrafast magnetic dynamics can be treated on equal footing. Within well-defined approximations, I demonstrate the formal emergence of collective magnons and detail the structural modifications required in the scattering integrals relative to the standard nonmagnetic case. Furthermore, I outline how this formalism can be practically implemented in a first principles computational workflow to yield a tractable set of equations. Importantly, within this framework it is possible to describe both the incoherent demagnetization regime, typically discussed in literature employing electron-phonon-spin scattering models, and the TDDFT approach employed for coherent dynamics. I discuss these features through the explicit numerical solution of a minimal model, which captures the fundamental physics presented in this work. Ultimately, this work provides a robust foundation for fully predictive calculations, which will be instrumental in identifying the dominant microscopic mechanisms driving femtomagnetic dynamics in a wide range of experimental scenarios.

\section*{Acknowledgments}
I acknowledge useful discussions with Dr. Stefano Mocatti, Prof. Matteo Calandra and Prof. Pierluigi Cudazzo. This work was funded by the European Union (ERC, DELIGHT, 101052708). Views and opinions expressed are however those of the author(s) only and do not necessarily reflect those of the European Union or the European Research Council. Neither the European Union nor the granting authority can be held responsible for them.

\section*{Data availability statement}

The software employed to obtain the results presented in this work will be made available in a zenodo repository (link to be added) upon publication.

\vspace{2.5pt}\begin{widetext}
\section*{Appendix A: relativistic single particle terms in the Hamiltonian}

Here I briefly introduce the relativistic single particle terms that I include in the Hamiltonian. Formally, the additional relativistic single particle terms which I include in this work emerge from a second-order expansion in powers of $v/c$ of the Dirac equation for the free electron~\cite{urru2020lattice,PhysRevB.100.045115,Bistoni22}: 

\begin{equation}
    \hat{H}^{0,e} = \hat{\mathbf{p}}^2 + \hat{V}(\hat{\mathbf{r}})+\hat{H}_{mv}+\hat{H}_{D}+\hat{H}_{SO}
\end{equation}

The scalar relativistic mass velocity term $\hat{H}_{mv}$ and the Darwin term $\hat{H}_D$ represent trivial additions to the non-relativistic Hamiltonian, being proportional to the identity in spinor space. Conversely, the SOC Hamiltonian $\hat{H}_{SO}$ possesses a more complex spin structure,  $\hat{H}_{SO}\propto \hat{\mathbf{L}}\cdot \hat{\mathbf{S}}$~\cite{urru2020lattice,Bistoni22}. To understand why, it is convenient to represent any single particle operator as  a $2\times2$ matrix acting in the spin-$1/2$ Hilbert space, $span~\{\ket{\uparrow},\ket{\downarrow}\}$, which can be expressed in the complete basis constituted by the identity plus Pauli matrices: 

\begin{equation}
    \hat{\mathbf{A}} = \sum_{I=0,x,y,z} A_I~\hat{\sigma}^I \label{operator}
\end{equation}

here, $A_I$ are scalars (or operators acting in other spaces), $\hat{\sigma}^0$ is the $\hat{I}_2$ identity and $\hat{\sigma}^{x,y,z}$ are Pauli matrices. In the absence of SOC, all the $ab~initio$ Hamiltonian terms only possess the ``charge'' component proportional to the identity $\hat{\sigma}^0$. Thus, in the absence of spontaneous symmetry breaking, both the electron Green's function and self-energy are proportional to the identity, and the usual scalar spinless treatment is possible. Clearly, both the electron Green's function and self-energy can still acquire nonzero $\hat{\sigma}^{x,y,z}$ components in broken-symmetry magnetic states or in the presence of an external magnetic field, even if SOC is absent. 
\section*{Appendix B: Kadanoff-Baym Equations in Spinor Space, Spin matrices on the contour and Langreth rules}

 The electronic Green's function satisfies the contour Dyson equation in left or right form~\cite{PhysRevLett.100.116402,Maciejko2007}:

\begin{equation}
\begin{gathered}
    {G}_{\alpha \beta}(1,2) = G_{0,\alpha\beta}(1,2)+G_{0,\alpha\gamma}(1,3) \Sigma_{\gamma\eta}(3,4)G_{\eta\beta}(4,2) \\
    {G}_{\alpha \beta}(1,2) = G_{0,\alpha\beta}(1,2)+G_{\alpha\gamma}(1,3) \Sigma_{\gamma\eta}(3,4)G_{0,\eta\beta}(4,2) \label{eq:Dyson}
\end{gathered}
\end{equation}

where Greek letters are spin indices, repeated spin indices are summed over, indicating a matrix multiplication in spinor space, and repeated coordinates are integrated over when they do not appear on both sides of the equation. From the contour Dyson's equation in the left form it is possible to employ the Langreth rules in spinor space (see Appendix B) to formulate the Keldysh equation for the lesser/greater parts of the electronic propagator:

\vspace{2.5pt}\
\begin{align}
\begin{gathered}
        {G}^{\lessgtr}_{\alpha \beta}(1,2) = G^\lessgtr_{0,\alpha\beta}(1,2)+G^R_{0,\alpha\gamma}(1,3) \Sigma^R_{\gamma\eta}(3,4)G^\lessgtr_{\eta\beta}(4,2) + G^R_{0,\alpha\gamma}(1,3) \Sigma^\lessgtr_{\gamma\eta}(3,4)G^A_{\eta\beta}(4,2) \\
        + G^\lessgtr_{0,\alpha\gamma}(1,3) \Sigma^A_{\gamma\eta}(3,4)G^A_{\eta\beta}(4,2) \label{Keldysh}
        \end{gathered}
\end{align}

where the time integration is understood on the real axis.
In order to obtain the Kadanoff-Baym equations for the electronic propagator, I first define the inverse single particle propagator $\mathbf{G}_0^{-1}$ through:

\begin{equation}
\begin{gathered}
    [i\partial_{z_1}-\hat{H}^{qp}(\mathbf{r}_1,-i \nabla_1,z_1)]_{\alpha \gamma}{G}_{0,\gamma \beta}(1,2)=\\
    \delta(\mathbf{r}_1-\mathbf{r}_2)\delta(z_1-z_2)\delta_{\alpha \beta}
    \end{gathered}
\end{equation}

where I introduced a quasiparticle electron Hamiltonian that absorbs the singular (time-local) part of the electron self-energy, $\hat{H}^{qp}_{\alpha \beta} = \hat{H}^{0,e}_{\alpha \beta}+\Sigma^\delta_{\alpha \beta}$, see also the discussions in Refs.~\cite{Stefanucci2024,Mocatti2026}, $\nabla_1$ is the gradient with respect to coordinate $\mathbf{r}_1$. The explicit form for the inverse single particle bare propagator $\mathbf{G}_0^{-1}$ is thus: 

\begin{equation}
\begin{gathered}
G^{-1}_{0,\alpha\beta}=[i\partial_{z_1}-\hat{H}^{qp}(\mathbf{r}_1,-i \nabla_1,z_1)]_{\alpha \gamma} =\\
i\partial_{z_1}\delta_{\alpha \gamma}+\sum_I H^{qp}_{I}(z_1)\hat{\sigma}^I_{\alpha\gamma}
\end{gathered}
\end{equation}

 where the $I$ index runs over $0,x,y,z$. I apply it from the left to the Keldysh equation, Eq.~\ref{Keldysh}, obtaining the Kadanoff-Baym equations for the electron Green's function in the spinor case:

\begin{equation}
\begin{gathered}
        [i\partial_{t_1}-\hat{H}^{qp}(\mathbf{r}_1,-i \nabla_1,t_1)]_{\alpha \gamma}{G}^{\lessgtr}_{\gamma \beta}(1,2) = \\
        \Sigma^R_{\alpha\eta}(1,3)G^\lessgtr_{\eta\beta}(3,2) + \Sigma^\lessgtr_{\alpha\eta}(1,3)G^A_{\eta\beta}(3,2) \enspace ;\\
        {G}^{\lessgtr}_{\alpha \gamma}(1,2)[-i\overleftarrow{\partial}_{t_2}-\hat{H}^{qp}(\mathbf{r}_2,-i \overleftarrow{\nabla}_2,t_2)]_{\gamma \beta}= \\
        G^R_{\alpha\eta}(1,3)\Sigma^\lessgtr_{\eta\beta}(3,2) + G^\lessgtr_{\alpha\eta}(1,3)\Sigma^A_{\eta\beta}(3,2) \label{Kadanoff-Baym}
\end{gathered}
\end{equation}

where the second equation has been obtained by the same procedure but operating on the right form of the Dyson equation, and the arrow over the time derivative and gradient implies that they act from the right. Subtracting the second equation from the first one in Eqs.~\ref{Kadanoff-Baym}, I obtain:

\vspace{2.5pt}
\begin{equation}
\begin{gathered}
        [i\partial_{t_1}-\hat{H}^{qp}(\mathbf{r}_1,-i \nabla_1,t_1)]_{\alpha \gamma}{G}^{\lessgtr}_{\gamma \beta}(1,2) -  {G}^{\lessgtr}_{\alpha \gamma}(1,2)[-i\overleftarrow{\partial}_{t_2}-\hat{H}^{qp}(\mathbf{r}_2,-i \overleftarrow{\nabla}_2,t_2)]_{\gamma \beta} = \\
        \Sigma^R_{\alpha\eta}(1,3)G^\lessgtr_{\eta\beta}(3,2) + \Sigma^\lessgtr_{\alpha\eta}(1,3)G^A_{\eta\beta}(3,2) + G^R_{\alpha\eta}(1,3)\Sigma^\lessgtr_{\eta\beta}(3,2) + G^\lessgtr_{\alpha\eta}(1,3)\Sigma^A_{\eta\beta}(3,2) \label{Kadanoff-Baym2}
        \end{gathered}
\end{equation}

The evolution of the single particle density matrix $\rho$ is then obtained by operating  the $t_2 \to~ t_1$ limit on the equation for the $G^<$ (or $G^>$) component in Eq.~\ref{Kadanoff-Baym2},  and employing the explicit expression of retarded and advanced components in terms of the lesser and greater ones: 

\begin{equation}
\begin{gathered}
    \mathbf{G}^R(1,2) = \theta(t_1-t_2)[\mathbf{G}^>(1,2)-\mathbf{G}^<(1,2)]\\
    \mathbf{G}^A(1,2) = -\theta(t_2-t_1)[\mathbf{G}^>(1,2)-\mathbf{G}^<(1,2)]
\end{gathered}
\end{equation}

finally arriving to the result: 

\begin{equation}
\begin{gathered}
   i\partial_{t} {G}^{<}_{\alpha \beta}(1,2)+[\hat{H}^{qp}(\mathbf{r},-i\nabla,t),\mathbf{G}^{<}(1,2)]_{\alpha \beta} = \\
   -~[\Sigma^>_{\alpha \gamma}(1,3)G^<_{\gamma \beta}(3,2)-\Sigma^<_{\alpha \gamma}(1,3)G^>_{\gamma \beta}(3,2)]  + h.c.\label{density_mat}
   \end{gathered}
\end{equation}

with $t=t_1=t_2$ and the understanding that $\hat{H}^{qp}$ acts on coordinate 1 from the right and on coordinate 2 from the left, while $h.c.$ represents the Hermitian conjugate. 

A generic scalar function on the contour takes the form~\cite{Danielewicz1984}:

\begin{align}
    {F}(1,2) &= \theta(z_1,z_2) {F}^>(1,2) + \theta(z_2,z_1) {F}^<(1,2) + {F}^{\delta}(\mathbf{r}_1,\mathbf{r}_2,z_1) \delta(z_1-z_2) \label{scalar_decomposition}
    \end{align}

The same decomposition can be applied to a spin matrix propagator living in the contour, since the decomposition holds component-wise: 

\begin{align}
    \mathbf{F}(1,2) &= \theta(z_1,z_2) \mathbf{F}^>(1,2) + \theta(z_2,z_1) \mathbf{F}^<(1,2) + \mathbf{F}^{\delta}(\mathbf{r}_1,\mathbf{r_2},z_1) \delta(z_1-z_2)
    \end{align}

The retarded and advanced components are given by

\begin{align}
    F^{R/A}(1,2) &= \pm \theta(\pm t \mp t') \left[F^{<}(1,2) - F^{>}(1,2)\right] + F^{\delta}(\mathbf{r}_1,\mathbf{r}_2,t_1) \delta(t_1 - t_2).
\end{align}

and equivalently for the spin matrices: 
\begin{align}
    \mathbf{F}^{R/A}(1,2) &= \pm \theta(\pm t \mp t') \left[\mathbf{F}^{<}(1,2) - \mathbf{F}^{>}(1,2)\right] + \mathbf{F}^{\delta}(\mathbf{r}_1,\mathbf{r}_2,t_1) \delta(t_1 - t_2).
\end{align}

The Langreth theorem allows one to relate products of two contour-ordered functions of the form~\cite{Maciejko2007,Stefanucci_van_Leeuwen_2025,Langreth1976}
\begin{align}
    C(1,1')  = \int d2 A(1,2) B(2,1')
\end{align}

to real-time integrals, involving the real-time components of $A$  and $B$. In the spinor case, the same logic can be applied, since the contour structure is the same as for the spinless case. Having in mind the same product structure in time, space and spins, the product function $C_{\alpha \beta}$

\begin{align}
        C_{\alpha \beta}(1,1')  = \sum_\gamma \int d2 A_{\alpha\gamma}(1,2)B_{\gamma \alpha}(2,1')
\end{align}

obeys the same Langreth rules of the spinless case since each term $\gamma$ of the sum trivially does.

\section*{Appendix C: link to the perturbative Expansion in the time-local Screened Coulomb Interaction}
\label{sec:static_screening}

An alternative point of view that we adopted in Ref.~\cite{Mocatti2026} is to interpret the electron self-energy part originating from the electron-electron interaction in terms of the time-local (or statically) screened interaction $W_s$. Here, I will suppress the local time dependence of $W_s$ for simplicity. The starting point is the exact Dyson equation for the dynamically screened
Coulomb interaction $W(\mathbf{q},\omega)$,
\begin{equation}
  W(\mathbf{q},\omega)
  = v(\mathbf{q})
    + v(\mathbf{q})\,\Pi(\mathbf{q},\omega)\,W(\mathbf{q},\omega),
\end{equation}
where $v(\mathbf{q}) = e^{2}/\varepsilon_{0}q^{2}$ is the bare Coulomb
interaction and $\Pi(\mathbf{q},\omega)$ is the irreducible polarizability.
I decompose the polarizability into its static value and a frequency-dependent remainder:
\begin{equation}
  \Pi(\mathbf{q},\omega)
  = \Pi_{s}(\mathbf{q})
    + \delta\Pi(\mathbf{q},\omega),
  \qquad
  \Pi_{s}(\mathbf{q}) \equiv \Pi(\mathbf{q},0),
\end{equation}
so that $\delta\Pi(\mathbf{q},0)=0$ by construction. The static polarizability $\Pi_{s}$ alone generates its own Dyson equation: 
\begin{equation}
  W_{s}(\mathbf{q})
  = v(\mathbf{q})
    + v(\mathbf{q})\,\Pi_{s}(\mathbf{q})\,W_{s}(\mathbf{q})
  = \frac{v(\mathbf{q})}
         {1 - v(\mathbf{q})\,\Pi_{s}(\mathbf{q})},
\end{equation}
which defines the \emph{statically screened} (time-local) Coulomb interaction $W_{s}(\mathbf{q})$.  Its inverse reads
\begin{equation}
  v^{-1}(\mathbf{q})
  = W_{s}^{-1}(\mathbf{q}) + \Pi_{s}(\mathbf{q}).
\end{equation}

Substituting the decomposition of $\Pi$ back into the full Dyson equation,
\begin{equation}
  W = v + v\bigl(\Pi_{s}+\delta\Pi\bigr)W
    = v + v\Pi_{s}W + v\,\delta\Pi\,W,
\end{equation}
and using $W_{s}^{-1} = v^{-1} - \Pi_{s}$ to identify
$v + v\Pi_{s}W_{s} = W_{s}$, one can write
\begin{equation}
  W = W_{s} + W_{s}\,\delta\Pi(\omega)\,W.
\end{equation}
This is a new Dyson equation with $W_{s}$ playing the role of the ``bare'' propagator and $\delta\Pi$ as the effective interaction kernel. Formally iterating it, I write:
\begin{equation}
  W(\mathbf{q},\omega)
  = W_{s}(\mathbf{q})
    \sum_{n=0}^{\infty}
    \bigl[\delta\Pi(\mathbf{q},\omega)\,W_{s}(\mathbf{q})\bigr]^{n},
\end{equation}
which is an expansion in powers of the dynamic part of the
screening, $\delta\Pi$, with all static screening already contained into
$W_{s}$. The electron self-energy in the $GW$ approximation is $\Sigma = \mathrm{i}GW$.  Replacing $W$ by the expansion
above and retaining terms up to {second order in $W_s$} corresponds
to only keep the first two terms of the  series,
i.e., $W \approx W_{s} + W_{s}\,\delta\Pi\,W_{s}$, so
\begin{align}
  \Sigma^{(1)}(\mathbf{r},\mathbf{r}';\omega)
  &= \mathrm{i}\int\frac{d\omega'}{2\pi}\,
     G(\mathbf{r},\mathbf{r}';\omega-\omega')\,
     W_{s}(\mathbf{r},\mathbf{r}'),
  \label{eq:Sigma1}\\
  \Sigma^{(2)}(\mathbf{r},\mathbf{r}';\omega)
  &= \mathrm{i}\int\frac{d\omega'}{2\pi}\,
     G(\mathbf{r},\mathbf{r}';\omega-\omega')
     \int d^{3}r_{1}\,d^{3}r_{2}\;
     W_{s}(\mathbf{r},\mathbf{r}_{1})\,
     \delta\Pi(\mathbf{r}_{1},\mathbf{r}_{2};\omega')\,
     W_{s}(\mathbf{r}_{2},\mathbf{r}').
  \label{eq:Sigma2}
\end{align}
Equation~\ref{eq:Sigma1} constitutes an instantaneous (static) quasiparticle
energy renormalization, while Eq.~\ref{eq:Sigma2} carries the full
frequency dependence of $\delta\Pi$ and therefore contains retardation
effects and inelastic scattering processes through the collision integrals. I note that no approximation has been made in splitting
$\Pi$ or resumming $\Pi_{s}$.  The advantage is twofold.  First, the
leading-order term $\Sigma^{(1)}$ already contains the dominant long-range
Coulomb screening captured by $W_{s}$, so the perturbative corrections in
$\delta\Pi$ are small in the weakly correlated limit.  Second,
because $\delta\Pi(\mathbf{q},0)=0$, the second-order contribution
$\Sigma^{(2)}$
contributes to finite-frequency processes only. Finally, in practice one replaces the full $\delta\Pi$ by its RPA independent particle form and
evaluates $G$ at the quasiparticle level, obtaining the direct second-Born
self-energy diagram employed here and in Ref.~\cite{Mocatti2026}. In this approximation, the diagram in Eq.~\ref{eq:Sigma2} is the same second order diagram contained in the $GT$ self-energy, with a $T$ matrix built employing a time-local interaction $W$, as in the main text.

\section*{Appendix D: double counting in the GW+GT scheme}

\label{sec:diagrammatic}

The proposed electron self-energy scheme combines two distinct contributions in the scattering integral: an explicit second-Born 
term appearing from the $GW$ self-energy expansion in terms of the static interaction, and which accounts for electron-electron scattering, 
and a $T$ matrix self-energy from which I retain only the coherent magnon-pole 
contribution, explicitly discarding the incoherent 
particle-hole continuum. In this Appendix I discuss if and under what conditions this scheme is physically consistent. A desirable feature of the self-energy is $\Phi$-derivability. A fully $\Phi$-derivable scheme~\cite{Kadanoff2018,Lipavsky1986,Stefanucci_van_Leeuwen_2025} requires that all self-energy contributions be obtained as functional derivatives of a single Luttinger-Ward functional $\Phi[G]$,
\begin{equation}
    \Sigma = \frac{\delta \Phi}{\delta G},
\end{equation}
which guarantees conservation laws and the absence of double-counting by construction. While $\Phi$-derivability is a property of both self-consistent $GW$ or $GT$ schemes~\cite{PhysRev.127.1391}, it is not guaranteed in any scheme that combines them: indeed the $GW+GT$ scheme suffers from the double counting of the second Born direct diagram even when considering a static (or time-local) interaction $W_s$, as mentioned in the main text and discussed in previous Appendix C. The magnon pole arises from  the full resummation of the ladder series, thus representing a non-perturbative object that cannot be generated by any finite-order truncation of the ladder series, and therefore the second-Born diagram does not directly contribute to the pole. Nevertheless, the quantitative values of magnon frequency and electron-magnon coupling do depend on the inclusion of finite-order diagrams in the $T$ matrix. This indirect renormalization is physical and does not constitute double counting. Close to the magnon pole, the T-matrix takes the form

\begin{equation}
    T(\mathbf{q}, \omega\to \Omega_{\mathbf{q}}) \simeq 
    \frac{|m_\mathbf{q}|^2}{\omega - \Omega_\mathbf{q} + i\Gamma_\mathbf{q}},
\end{equation}
where $\Omega_\mathbf{q}$ is the real magnon frequency, and I explicitly separated the 
magnon linewidth acquired due to Landau damping,$\Gamma_\mathbf{q}$  and $m_\mathbf{q}$ is the electron-magnon 
coupling constant. The spectral decomposition into a coherent magnon pole and an incoherent continuum 
is only exact in the limit of vanishing magnon damping, $\Gamma_\mathbf{q} \to 0$,  and constitutes a controlled approximation only when
\begin{equation}
    \frac{\Gamma_\mathbf{q}}{\Omega_\mathbf{q}} \ll 1.
    \label{eq:validity}
\end{equation}
This condition is obviously satisfied in the long-wavelength limit $\mathbf{q} \to 0$, where  the magnon frequency is well below the Stoner continuum, and the magnon is a sharp quasiparticle. 
At intermediate wavevectors, as the magnon dispersion approaches the boundary of  the Stoner continuum, $\Gamma_\mathbf{q}$ grows and the condition~\ref{eq:validity}  is progressively less well satisfied. In this regime the magnon is broadened, and found on top of the continuum background; as a consequence, the pole and continuum contributions to $T$ are no longer spectrally separated, rather they overlap, making the decomposition only approximate. Even when the magnon frequency intersects the continuum, I keep assuming the magnon pole in the $T$ matrix pole decomposition as purely real, and discard the imaginary part, in the spirit of the quasiparticle approximation. This is similar to how phonons are treated in Eliashberg theory~\cite{ALLEN19831}: phonons can in principle decay into electron-hole pairs (damping), yet the theory is formulated as if the phonon is a sharp mode, with the justification that $\Gamma_\mathbf{q} \ll \omega_D$ in the regime of interest.  The error incurred by treating the mode as sharp is of relative order $\Gamma_\mathbf{q}/\Omega_\mathbf{q}$ and is therefore controlled by the same small parameter that defines the quasiparticle regime. I make the same approximation here for magnons, with the understanding that its regime of validity should be controlled for any practical application.

I note that this argument establishes the absence of overlap between the two terms that are retained, and not $\Phi$-derivability of the combined scheme: the
separation achieved is between spectral regions of the $T$ matrix, rather than between disjoint classes of diagrams.

\section*{Appendix E: relation between the $GT$ self-energy and transverse susceptibility}

In order to make an explicit connection between the present treatment of the electron-magnon interaction and the discussions in the context of the Hubbard model~\cite{Hertz_1973}, I now examine the fundamental relationship between the $GT$ self-energy obtained in Eq.~\ref{GTselfen2} and transverse susceptibility obtained in linear response theory. I will consider the collinear case for simplicity. In the second quantization formalism, the transverse spin operators for a certain state $i$ can be defined from the electronic creation ($\hat{c}^\dagger$) and annihilation ($\hat{c}$) operators. Namely, the spin-raising operator $\hat{S}^+_i$ and spin-lowering operator $\hat{S}^-_i$ are defined as:
\begin{equation}
    \hat{S}^+_i = \hat{c}^\dagger_{i\uparrow} \hat{c}_{i\downarrow}, \quad \hat{S}^-_i = \hat{c}^\dagger_{i\downarrow} \hat{c}_{i\uparrow}
\end{equation}
These operators satisfy the commutation relation $[\hat{S}^+_i, \hat{S}^-_j] = 2\delta_{ij} \hat{S}^z_i$, where

\begin{equation}
    \hat{S}^z_i = \frac12 (\hat{c}_{i\uparrow}^\dagger \hat{c}_{i\uparrow}-\hat{c}^\dagger_{i\downarrow}\hat{c}_{i \downarrow})
\end{equation}

In the collinear case, the net spin magnetization of the system along $z$ is the ground-state expectation value of the longitudinal component: $M_z = 2\mu_B\langle \hat{S}^z \rangle = \mu_B\sum_i( f_i^{\uparrow\uparrow} - f_i^{\downarrow\downarrow})$. To make the connection with collective excitations (magnons) with a specific momentum $\mathbf{q}$ in periodic systems, I define the generalized spin-flip operator from the Fourier transform of $\hat{S}_i^{\pm}$. In a multiband system:
\begin{equation}
    \hat{S}^+_{\mathbf{q}} = \sum_{\mathbf{k},n,m} \hat{c}^\dagger_{\mathbf{k}+\mathbf{q}, n \uparrow} \hat{c}_{\mathbf{k}, m \downarrow}
\end{equation}

The magnon state $|S, \mathbf{q}\rangle$ can be seen as a superposition of these transitions, weighted by the magnon amplitudes $A_{S,\uparrow\downarrow}^{nm}(\mathbf{k}, \mathbf{q})$:
\begin{equation}
    A_{S, \uparrow\downarrow}^{nm}(\mathbf{k}, \mathbf{q}) = \langle 0 | \hat{c}^\dagger_{\mathbf{k}+\mathbf{q}, n \uparrow} \hat{c}_{\mathbf{k}, m \downarrow} | S, \mathbf{q} \rangle
\end{equation}

Spin-raising and spin-lowering operators are connected to the transverse spin susceptibility $\chi_{+-}(\mathbf{q}, t)$, which describes the linear response of the spin system and is defined as the retarded correlation function of the spin-raising and spin-lowering operators~\cite{Moriya1973,PhysRevB.85.054305,PhysRevB.103.245110}:
\begin{equation}
    \chi_{+-}(\mathbf{q}, \tau) = -i \theta(t) \langle 0 | [\hat{S}^-_{\mathbf{q}}(\tau), \hat{S}^+_{-\mathbf{q}}(0)] | 0 \rangle 
\end{equation}
where $|0\rangle$ represents the magnetic ground state. I insert a complete set of excited states $I = \sum_S |S, \mathbf{q}\rangle \langle S, \mathbf{q}|$, and consider its Lehmann representation, which in the case of isolated poles is written:
\begin{equation}
    \chi_{+-}(\mathbf{q}, \omega) = \sum_S \frac{|\langle S, \mathbf{q} | \hat{S}^+_{-\mathbf{q}} | 0 \rangle|^2}{\omega - \Omega_{\mathbf{q}}^S + i\eta} \label{chi}
\end{equation}

where $\Omega_{\mathbf{q}}^S$ is the frequency of the $S$-th magnon branch. The numerator represents the spectral weight of the mode, indicating how strongly the external probe couples to that specific collective excitation. In the $T$ matrix framework discussed in the main text, the magnon amplitudes $A_S$ represent transition density between the ground state and the excited state~\cite{PhysRevB.62.4927}. Specifically, the matrix element in the numerator of Eq.~\ref{chi} is the sum over all particle-hole transitions:
\begin{equation}
\begin{gathered}
    \langle S, \mathbf{q} | \hat{S}^+_{-\mathbf{q}} | 0 \rangle = \sum_{\mathbf{k}, n, m} \langle S, \mathbf{q} | \hat{c}^\dagger_{\mathbf{k}-\mathbf{q}, n \uparrow} \hat{c}_{\mathbf{k}, m \downarrow} | 0 \rangle = \\
    \sum_{\mathbf{k}, n, m} [A_{S, \uparrow\downarrow}^{nm}(\mathbf{k}, \mathbf{q})]^* = A^*_S(\mathbf{q})
    \end{gathered}
\end{equation}

In the exact Lehmann representation of Eq.~\ref{chi}, the residue of an isolated delta-peak is simply given by $|A_S(\mathbf{q})|^2$. However, when considering the full response function, poles can intersect the Stoner continuum and, as for the case of the $T$ matrix, it is convenient to separate a magnon pole contribution from the Stoner continuum:

\begin{equation}
    \chi_{+-}(\mathbf{q}, \omega) = \sum_S \frac{|A_{S}(\mathbf{q})|^2 R_{\mathbf{q}}^S}{\omega - \Omega_{\mathbf{q}}^S + i\eta}+ \chi_{+-}^{cont}(\mathbf{q},\omega)
\end{equation}

 A quasiparticle residue $R_{\mathbf{q}}^S  = \sqrt{(\dfrac{d\lambda^S_{\mathbf{q}}}{d \omega})^{-1}}\le 1$ will generally be present at each pole, such that the total pole spectral weight becomes $|A_S(\mathbf{q})|^2 R_{\mathbf{q}}^S$. 

 I discussed in the main text how the $GT$ self-energy in this approximation has the following form in terms of the electronic Green's function $G$ and the magnon propagator $D_{S}$:

\begin{equation}
    \Sigma_{\mathbf{k},\alpha\alpha}^{n_1 n_1}(\tau) = -i \int d\mathbf{q} \sum_{S} G_{\mathbf{k}-\mathbf{q},,\bar{\alpha}\bar{\alpha}}^{n_2 n_2}(\tau) |m^S_{n_1 n_2, \alpha\bar{\alpha}}(\mathbf{k}, \mathbf{q})|^2 D_{\mathbf{q},S}(\tau)
\end{equation}

 In order to form an explicit connection with models of itinerant ferromagnetism~\cite{Hertz_1973}, I assume a $\mathbf{k}$- and band-independent interaction, i.e. $W_{\mathbf{k},\mathbf{k}''}^{n_1n_2m_1m_2}(\mathbf{q},\bar{t}) \approx W(\mathbf{q},\bar{t})$. By identifying the magnon spectral weight $Z_{\mathbf{q}}^S = |A_S(\mathbf{q})|^2 R_{\mathbf{q}}$ (where $R_{\mathbf{q}}$ is the residue), one can relate the internal summation in the self-energy to the transverse susceptibility $\chi_{+-}$:
\begin{equation}
\begin{gathered}
    \sum_S |m^S(\mathbf{k},\mathbf{q})|^2 D_{\mathbf{q},S}( \omega) \approx |W(\mathbf{q},t)|^2 \sum_S \frac{Z_{\mathbf{q}}^S}{\omega - \Omega_{\mathbf{q}}^S} = \\ 
    |W(\mathbf{q},t)|^2 \chi^{poles}_{+-}(\mathbf{q}, \omega)
    \end{gathered}
\end{equation}

Further assuming a time- and $\mathbf{q}$-independent interaction and summing over all magnon modes, one recovers the $GT$ expression of Ref.~\cite{Hertz_1973}:
\begin{equation}
    \Sigma^{GT}_{\mathbf{k}}(\omega) \approx \frac{U^2}{N} \int d\omega'\sum_{\mathbf{q}} G_{\mathbf{k}-\mathbf{q}}(\omega-\omega') \chi_{+-}(\mathbf{q},\omega')
\end{equation}

where $N$ is the number of unit cell, normalizing the discrete $\mathbf{q}$ sum. The connection to magnetization $M$ is rooted in the sum rule for the transverse spectral function $A_{+-}(\mathbf{q}, \omega) = -\frac{1}{\pi} \text{Im} \chi_{+-}(\mathbf{q}, \omega)$. Integrating over all frequencies gives the sum rule~\cite{PhysRevB.103.245110}:

\begin{equation}
    \int_{-\infty}^{\infty} A_{+-}(\mathbf{q}, \omega) d\omega = \langle 0 | [\hat{S}^+_{\mathbf{q}}, \hat{S}^-_{-\mathbf{q}}] | 0 \rangle = 2\langle \hat{S}^z \rangle = \dfrac{M_z^{tot}}{V} \label{sumrule}
\end{equation}

where $M_z^{tot}$ is the magnetization (assuming as usual that the quantization axis is $z$), and noting that the commutator $\hat{S}^+_{\mathbf{q}}$ and $\hat{S}^-_{\mathbf{q}}$ evaluates to: 

\begin{equation}
    [\hat{S}^+_{\mathbf{q}}, \hat{S}^-_{\mathbf{q}'}] = 2\hat{S}^z_{\mathbf{q}-\mathbf{q}'}
\end{equation}

Clearly, both localized poles and the Stoner continuum will generally contribute to the integral in Eq.~\ref{sumrule}, but it is typically reasonable to assume that the pole contribution dominates the integral for localized systems. In this limit, the spectrum of $\chi$ is reduced to a sum of delta functions (or Lorentzians $L$ of linewidth $\eta$):
\begin{equation}
    A_{+-}(\mathbf{q}, \omega) \approx \sum_S Z_{\mathbf{q}}^S L(\omega - \Omega_{\mathbf{q}}^S,\eta)
\end{equation}

Comparing this result to Eq.~\ref{sumrule} one finds, independently from $\mathbf{q}$: 

\begin{equation}
    \sum_S Z_{\mathbf{q}}^S = \dfrac{M_z^{tot}}{V}
\end{equation}

\section*{Appendix F: Symmetry of the Magnon Propagator}

I wish to verify the fundamental symmetry relation for the transverse magnon propagator $D_S^{R,m, \uparrow\downarrow}(\mathbf{q}, \bar{t},\omega)$ in a collinear ferromagnetic system (Eq.~\ref{propD} in the main text) : 

\begin{equation}
     D^{R,m,\uparrow \downarrow}_{\mathbf{q},S}(\bar{t},\omega)) = [D^{R,m,\downarrow \uparrow}_{\mathbf{q},S}(\bar{t},-\omega)]^* 
\end{equation}

I assume the system is in a non-equilibrium state characterized by a slow center-of-mass time $\bar{t}$, and a relative time $\tau = t_1 - t_2$ which has been Fourier transformed into the frequency domain ($\omega$). I thus start from the retarded equation for the $T$ matrix:
\begin{equation}
\begin{gathered}
           T^{R,n_1 n_2 n_3 n_4}_{\alpha \alpha'}(\mathbf{k},\mathbf{k}',\mathbf{q},\bar{t},\omega) =\\ W^{n_1 n_2 n_3 n_4}_{\mathbf{k},\mathbf{k}'}(\mathbf{q},\bar{t})+W^{n_1 n_2 m_1m_2}_{\mathbf{k},\mathbf{k}''}(\mathbf{q},\bar{t})K^{R,m_1 m_2 m_3 m_4}_{\alpha \alpha'}(\mathbf{k}'',\mathbf{q},\bar{t},\omega)T^{R,m_3 m_4 n_3 n_4}_{\alpha \alpha'}(\mathbf{k}'',\mathbf{k}',\mathbf{q},\bar{t},\omega) 
           \end{gathered}
\end{equation}

where the instantaneous, spin-independent screened interaction matrix element is:
\begin{equation}
W_{\mathbf{k},\mathbf{k}'}^{n_1 n_2 n_3 n_4}(\mathbf{q}, \bar{t}) = W_{\mathbf{k},\mathbf{k}',\alpha \alpha'}^{n_1 n_2 n_3 n_4}(\mathbf{q}, \bar{t}) = \int d\mathbf{r} d\mathbf{r}' \, \psi^*_{\mathbf{k},n_1\alpha}(\mathbf{r}) \psi_{ \mathbf{k}',n_3\alpha}(\mathbf{r}) W(\mathbf{r},\mathbf{r}',\bar{t}) \psi^*_{ \mathbf{k}'+\mathbf{q},n_2\alpha'}(\mathbf{r}') \psi_{\mathbf{k}+\mathbf{q},n_4\alpha'}(\mathbf{r}')
\end{equation}

$W(\mathbf{r}-\mathbf{r}',\bar{t})$ is defined in Eq.~\ref{eq:W_recip}, and I recall that in this case the $\{\alpha,\alpha'\}$ dependence is suppressed in $W$ when the interaction is assumed to be proportional to the identity in spinor space. Exchanging $\mathbf{r}$ and $\mathbf{r}'$, one easily verifies that the following relationship holds:


\begin{equation}
  [W_{\mathbf{k},\mathbf{k}'}^{n_1 n_2 n_3 n_4}(\mathbf{q}, \bar{t})]^* =  W_{\mathbf{k}+\mathbf{q},\mathbf{k}'+\mathbf{q}}^{n_4 n_3 n_2 n_1}(-\mathbf{q}, \bar{t}) \label{Wprop}
\end{equation}

Next, I examine the symmetry properties of the retarded non-interacting particle-hole propagator $K_{\alpha\alpha'}^{R,m_1 m_2 m_3 m_4}(\mathbf{k}'', \mathbf{q}, \bar{t}, \omega)$. In the transverse channel, the retarded propagator $K^R$ is constructed from single particle Green's functions with opposite spin:

\begin{equation}
\begin{gathered}
K_{\alpha \alpha'}^{R,m_1 m_2 m_3 m_4}(\mathbf{k}, \mathbf{q}, \bar{t}, \omega) = -i \int \frac{d\omega'}{2\pi} G_{\mathbf{k},\alpha\alpha}^{R,m_1 m_3}( \bar{t}, \omega') G_{\mathbf{k}+\mathbf{q},\alpha'\alpha'}^{<,m_4 m_2}( \bar{t}, \omega'+\omega)\\
-i\int \frac{d\omega'}{2\pi} G_{\mathbf{k},\alpha\alpha}^{<,m_1 m_3}( \bar{t}, \omega') G_{\mathbf{k}+\mathbf{q},\alpha'\alpha'}^{A,m_4 m_2}( \bar{t}, \omega'+\omega)
\end{gathered}
\end{equation}

Nothing that $[G^R]^*=G^A$ and that $[G^<]^*=-G^<$, it is straightforward to verify that the particle-hole retarded propagator possesses the following structural symmetry property:

\begin{equation}
K_{\alpha \alpha'}^{R,m_1 m_2 m_3 m_4}(\mathbf{k}, \mathbf{q}, \bar{t}, \omega) = [K_{\alpha' \alpha}^{R,m_4 m_3 m_2 m_1}(\mathbf{k}+\mathbf{q}, -\mathbf{q}, \bar{t}, -\omega)]^* \label{Kprop}
\end{equation}




By exchanging particle and hole, and consider the right form of the $T$ matrix self-consistent equation, the self-consistent $T$ matrix equation reads:

\begin{equation}
\begin{gathered}
           T^{R,n_4 n_3 n_2 n_1}_{\alpha' \alpha}(\mathbf{k}+\mathbf{q},\mathbf{k}'+\mathbf{q},-\mathbf{q},\bar{t},-\omega) = \\T^{R,n_4 n_3 m_4 m_3}_{\alpha' \alpha}(\mathbf{k}''+\mathbf{q},\mathbf{k}'+\mathbf{q},-\mathbf{q},\bar{t},-\omega) K^{R,m_4 m_3 m_2 m_1}_{\alpha' \alpha}(\mathbf{k}+\mathbf{q},-\mathbf{q},\bar{t},-\omega)W^{m_2 m_1 n_2 n_1}_{\mathbf{k}+\mathbf{q},\mathbf{k}''+\mathbf{q}}(-\mathbf{q},\bar{t}) + W^{n_4 n_3 n_2 n_1}_{\mathbf{k}+\mathbf{q},\mathbf{k}'+\mathbf{q}}(-\mathbf{q},\bar{t})
           \end{gathered}
\end{equation} 

Applying the complex conjugation and employing the properties in Eqs.~\ref{Wprop} and \ref{Kprop}, it is immediate to find the following relationship for the $T$ matrix:

\begin{equation}
\left[T_{\alpha'\alpha}^{R,n_4 n_3 n_2 n_1}(\mathbf{k}+\mathbf{q}, \mathbf{k}'+\mathbf{q}, -\mathbf{q}, \bar{t}, -\omega)\right]^* = T_{\alpha\alpha'}^{R,n_1 n_2 n_3 n_4}(\mathbf{k}, \mathbf{k}', \mathbf{q}, \bar{t}, \omega)
\end{equation}

This relationship, together with the pole decomposition, Eq.~\ref{Tbar_poles2}, finally gives: 

\begin{equation}
\begin{gathered}
       \sum_{S} \dfrac{m^{*S}_{n_2n_1,\alpha'\alpha}(\mathbf{k}+\mathbf{q},-\mathbf{q},\bar{t})  m^{S}_{n_2n_1,\alpha'\alpha}(\mathbf{k'}+\mathbf{q},-\mathbf{q},\bar{t})}{-\omega-\Omega^{S,\alpha'\alpha}_{-\mathbf{q}}(\bar{t})-i\eta} =    \sum_{S} \dfrac{m^{S}_{n_1n_2,\alpha\alpha'}(\mathbf{k},\mathbf{q},\bar{t})  m^{S*}_{n_1n_2,\alpha\alpha'}(\mathbf{k'},\mathbf{q},\bar{t})}{\omega-\Omega^{S,\alpha\alpha'}_{\mathbf{q}}(\bar{t})+i\eta}
       \end{gathered}
\end{equation}

which has to hold separately for each pole to be true at any frequency, thus demonstrating Eq.~\ref{propD}.

\section*{Appendix G: effect of electron-electron interaction in the model calculation}

\begin{figure*}[t]
\includegraphics[width=\textwidth]{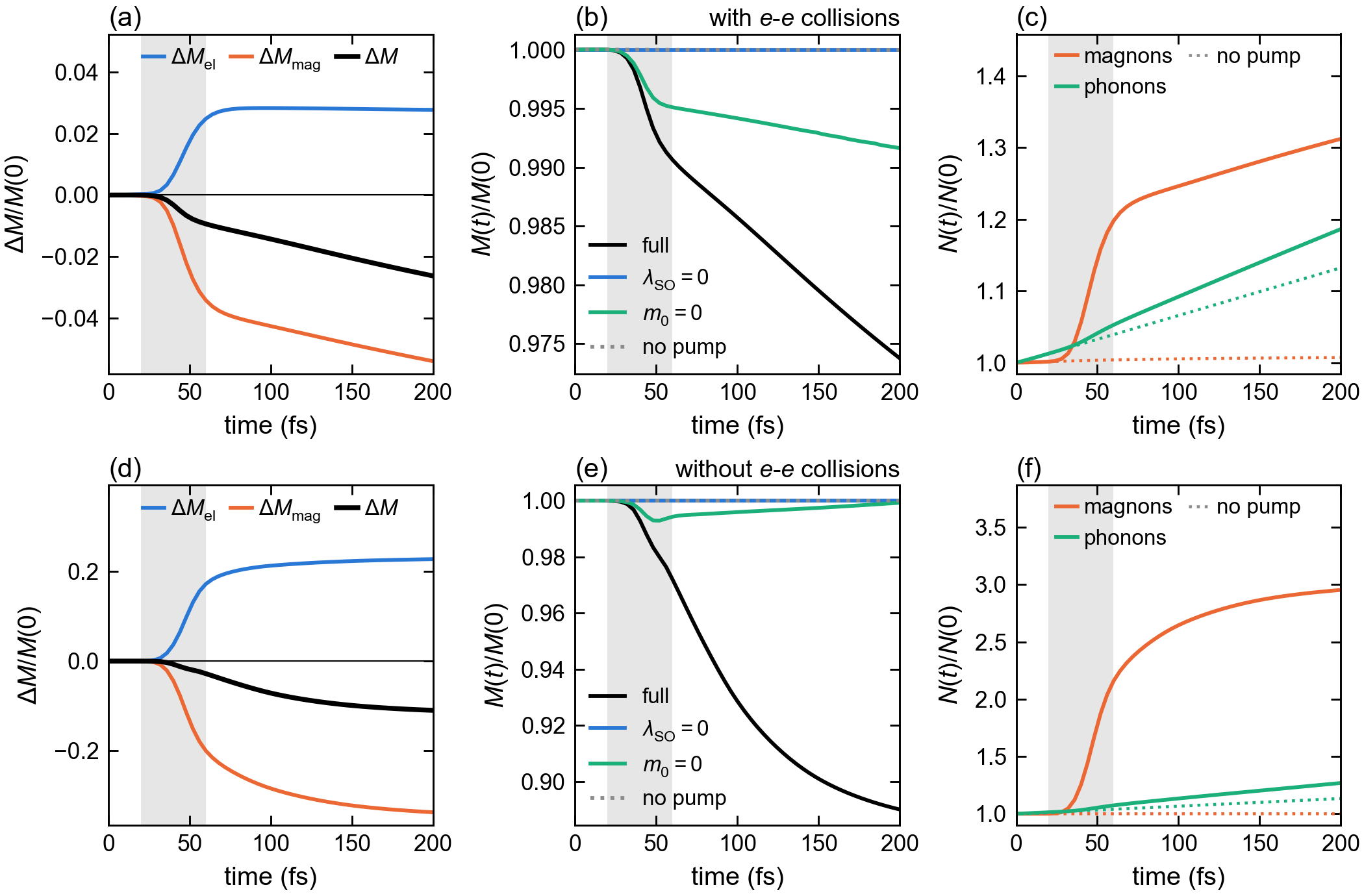}
\caption{Solution of the coupled equations of Fig.~\ref{fig:EOM} for the model
system with $N_k=16$, with (top row) and without (bottom row) electron-electron
collisions. (a),(d) Electronic, magnonic and total change of the magnetization,
normalized to the equilibrium value. (b),(e) Total magnetization for the full
calculation, for $\lambda_{\rm SO}=0$, for $m_0=0$, and for the unpumped system
(dotted). (c),(f) Total magnon and phonon populations normalized to their
equilibrium values; dotted lines are the unpumped baselines. The grey band marks
the pump pulse.}
\label{fig:toy2}
\end{figure*}

The calculations shown in Sec.~\ref{sec:toy} neglected the electron-electron
collision integral, as its evaluation scales as $N_k^3$ and is
incompatible with the sampling required to converge the electron-boson channels. To assess its effect, in this Appendix I compare
calculations performed with and without $I^{el-el}$ on a coarse grid, $N_k=16$, where
both are affordable and the results are still qualitatively comparable to the ones obtained at convergence in the main text. 
 
The comparison is shown in Fig.~\ref{fig:toy2}, where a calculation including electron-electron collisions (top row) is compared to a calculation without it (bottom row). While the inclusion of the electron-electron
collision integral does not qualitatively change the dynamics, it does have an evident quantitative effect, reducing the net demagnetization in the first 200 fs by a factor of four. The mechanism is the following: carrier-carrier scattering redistributes the energy
deposited by the pump within the electronic subsystem on a timescale shorter than
that of magnon emission, so that the photoexcited distribution thermalizes towards a
hot Fermi-Dirac form before a substantial number of magnons can be emitted. The
electron-electron and electron-magnon channels therefore compete for the same
population of hot carriers, and neglecting the former overestimates the amount of
angular momentum transferred to the magnon bath.
 
Two caveats should be kept in mind. First, at $N_k=16$ neither calculation is
converged with respect to the sampling, so that this result is only a qualitative indication. Second, conservation
properties are unaffected: the
electron-electron collision integral conserves particle number and quasiparticle electronic energy
on its own, and, in the absence of spin-orbit coupling, does not contribute to the
spin balance of Eq.~\eqref{eq:spincons}. The result of Sec.~\ref{sec:toy} that the
electron-magnon channel alone cannot demagnetize the system is thus independent of whether $I^{el-el}$ is included. The conclusion is that, while the electron-electron collision integral is
not needed to establish the qualitative mechanism underlying demagnetization, it must be included in any calculation aiming at quantitative predictions of demagnetization amplitudes and timescales.

\end{widetext}

\bibliography{bibliography}
\end{document}